\documentclass[english, 12pt]{article}
\usepackage[T1]{fontenc}
\usepackage[utf8]{inputenc}
\usepackage[english]{babel}
\usepackage{lmodern}
\usepackage[top=1in, bottom=1in, left=1in, right=1in]{geometry}
\usepackage[pdftex]{graphicx}
\usepackage{verbatim}
\usepackage[width=0.8\textwidth]{caption}
\usepackage{amsmath,amssymb,amsopn,dsfont,mathrsfs, bm}
\usepackage{float,placeins,flafter,longtable,array,booktabs}
\usepackage{color}
\usepackage{natbib}
\usepackage{xr}
\usepackage[hidelinks]{hyperref}
\makeatletter

\newtheorem{thm}{Theorem}

\newtheorem{prop}{Proposition}
\newtheorem{lem}{Lemma}
\newtheorem{hyp}{Assumption}
\newtheorem{example}{Example}
\newtheorem{defin}{Definition}

\newcommand{\norm}[1]{\left\|#1\right\|}
\newcommand{\norminf}[1]{\left\Vert#1\right\Vert_\infty}
\newcommand{\abs}[1]{\left\vert#1\right\vert}
\newcommand{\intpart}[1]{\left\lfloor #1\right\rfloor}

\newcommand{\ind}[1]{\mathds{1}\left\{#1\right\}}
\newcommand{\D}{\mathcal{D}}
\newcommand{\E}{E}
\newcommand{\V}{V}
\newcommand{\R}{\mathbb R}
\newcommand{\N}{\mathbb N}

\newcommand{\eps}{\varepsilon}
\newcommand{\deriv}[2]{\partial #1/\partial #2}
\newcommand{\Deriv}[2]{\frac{\partial #1}{\partial #2}}
\newcommand{\convL}{\stackrel{d}{\longrightarrow}}
\newcommand{\M}{\mathcal{M}}

\newcommand{\Supp}{\text{Supp}}
\newcommand{\indep}{\perp \!\!\! \perp}

\newcommand{\interior}{\text{Int }\M}
\newcommand{\frontier}{\partial\M}
\newcommand{\convP}{\stackrel{P}{\longrightarrow}}
\newcommand{\convD}{\stackrel{d}{\longrightarrow}}

\newcommand{\CI}[1]{\text{CI}^{\, #1}_{1-\alpha}}
\newcommand{\lpi}{\Pi_0}

\renewcommand{\section}{\@startsection{section}{2}{0mm}{-1.5\baselineskip}{1\baselineskip}{\normalfont\large\bfseries}}
\renewcommand{\subsection}{\@startsection{subsection}{2}{0mm}{-1.2\baselineskip}{1\baselineskip}{\normalfont\normalsize\bfseries}}
\renewcommand{\subsubsection}{\@startsection{subsubsection}{3}{0mm}{-0.8\baselineskip}{0.4\baselineskip}{\normalfont\normalsize\itshape}}
\allowdisplaybreaks

\date{}

\begin{document}
	
	\title{Identification and Estimation of Average Causal Effects in Fixed Effects Logit Models\thanks{We gratefully acknowledge financial support from the research grants Otelo (ANR-17-CE26-0015-041), ERC POEMH 337665 and ANR-17-EURE-0010 (Investissements d’Avenir program). We would like to thank the editor Francesca Molinari, three anonymous referees, Benjamin Walter for his work at an early stage of this project and Christophe Gaillac and Ma\"el Laoufi for outstanding research assistance. We are also grateful to Irene Botosaru, Iv\'an Fern\'andez-Val, Ulrich M\"uller, Alex Poirier and participants of various conferences and seminars for their feedback.}}
	\author{Laurent Davezies\thanks{CREST-ENSAE, laurent.davezies@ensae.fr} \and Xavier D'Haultf\oe uille\thanks{CREST-ENSAE, xavier.dhaultfoeuille@ensae.fr. This research was partly conducted while Xavier D'Haultf\oe uille was at PSE, which he thanks for its hospitality.} \and Louise Laage\thanks{Georgetown University, louise.laage@georgetown.edu.}}
	
	\maketitle
	
	\begin{abstract}
This paper studies identification and estimation of average causal effects, such as average marginal or treatment effects, in fixed effects logit models with short panels. Relating the identified set of these effects to an extremal moment problem, we first show how to obtain sharp bounds on such effects simply, without any optimization. We also consider even simpler outer bounds, which, contrary to the sharp bounds, do not require any first-step nonparametric estimators. We build confidence intervals based on these two approaches and show their asymptotic validity. Monte Carlo simulations suggest that both approaches work well in practice, the second being typically competitive in terms of interval length. Finally, we show that our method is also useful to measure treatment effect heterogeneity.

\end{abstract}

\section{Introduction} 
\label{sec:introduction}

In this paper, we consider the identification and estimation of average causal effects in the fixed effects  (FE) static binary logit model, with short panels. Estimation of the slope parameters dates back to \cite{rasch1961} and \cite{andersen1970} \citep[see also][]{chamberlain1980analysis} but up to now, there has been no study of the identification and estimation of average causal effects such as average marginal effects (AME) or average treatment effects (ATE) in this model. These parameters are yet of more direct interest than the slope parameter, which only provides information on relative marginal effects. For this reason, and following the influential work of \cite{angrist2001estimation} \citep[see also][]{angrist2008mostly}, many applied economists have turned to using FE linear probability models, as we illustrate in Section \ref{sec:lit_review}.

\medskip
However, this approach can be misleading, for at least two reasons. Firstly, the linearity of the conditional expectation is often implausible with binary outcome. In a simple setup where covariates only include time dummies and an additional binary variable, the ``treatment'' of interest, this means that the so-called parallel trend assumption would typically be violated. In Appendix \ref{sec:pb_FE_linear}, we give an example where, because of this, the FE linear model estimand is negative, even though the true ATE and average treatment effect on the treated (ATT) are positive. Secondly, even if the parallel trend assumption holds, the two-way fixed effect estimator underlying the FE linear probability model with time dummies is in general inconsistent for the ATT if treatment effects are heterogeneous \citep{dCDH2020}. And treatment effects are unlikely to be homogenous with binary outcomes, again because probabilities are constrained to be in $[0,1]$.

\medskip
Unlike the FE linear model, and even if it does impose other restrictions, the FE logit model allows for heterogeneity of treatment effects and does not rely on the parallel trend restriction discussed above. Moreover, we demonstrate in this paper that estimation and inference on average causal effects in this model can be performed simply, without any optimization once the slope parameter is estimated.

\medskip
We first study in Section \ref{sec:identification} the identification of a class of average causal parameters, including the AME and the ATE, in the FE logit model. Such parameters are generally not point identified, but sharp bounds can be obtained by solving an extremal moment problem, that is, maximizing a moment over probability distributions given the knowledge of some other moments. Using existing results on such problems,  we show that the bounds are simple functions of some matrices of moments. One drawback of this approach, yet, is that it depends on functions that need to be estimated nonparametrically in a first step. We then consider even simpler outer bounds avoiding this issue. Importantly, these outer bounds are still informative: the length of the corresponding interval quickly decreases to $0$ as $T$, the number of periods, tends to infinity. We also study identification of heterogeneity measures using the same apparatus.

\medskip
Next, we consider in Section \ref{sec:inference} estimators of the sharp and outer bounds and corresponding confidence intervals on the average causal effects. We establish the root-$n$ consistency of the estimators of the sharp bounds under regularity conditions. The estimators of the bounds are asymptotically normal except if a function of the slope coefficient is zero. We build a confidence interval of the true effect that is asymptotically valid in both cases. We also show asymptotic validity of confidence intervals based on the outer bounds.

\medskip
We study in Section \ref{sec:monte_carlo_simulations} the finite sample properties of our two estimation and inference methods. In line with the theory, they show that the estimated bounds are very informative in practice. Also, the two confidence intervals have coverage close to their nominal level already for moderate sample sizes. Interestingly, we also find in our simulations that the second inference method leads to confidence intervals of similar size as those obtained with the first method. This may seem surprising given that they rely on outer bounds but as it turns out,  for typical sample sizes and number of periods, the difference between the outer and the sharp bounds is small compared to the standard errors of their estimators.

\medskip
Finally, we show in Online Appendix that our method also applies to the (static) ordered and dynamic (binary) FE logit models. Also, we developed with Christophe Gaillac and Ma\"el Laoufi the R package \texttt{MarginalFElogit} and the Stata command \texttt{mfelogit}, which perform inference on the AME and ATE (depending on whether $X$ is continuous or binary) with the two methods considered here, and accommodates the case of an individual-specific number of observations.\footnote{The R package can be found \href{https://github.com/cgaillac/MarginalFElogit}{\texttt{here}}. The more preliminary Stata command is available on the SSC repository.}

\paragraph{Related literature} 
\label{par:related_literature}


\medskip
Our work is related to the literature on the identification and estimation of average causal effects with panel data. Bias-correcting approaches have been developed for panels with large $T$ for both the logit and probit models \citep{fernandez2009fixed,fernandez2016individual}. Few papers have studied average causal effects with fixed $T$ in parametric models with FE, as we do here. In the dynamic FE logit model, \cite{aguirregabiria2021identification} show point identification of some average effects of lagged values of the outcome, and \cite{dobronyi2021} study the partial identification of average causal effects that can be obtained from the knowledge of the slope parameter. We complement these papers by considering other parameters, including average causal effects (AME or ATE) of exogenous covariates in the same dynamic model, though our main focus is on the static, binary FE logit model. 

\medskip
Several papers have studied identification of causal effects in a nonparametric context, see in particular \cite{altonji2005cross,Hoderlein2012,chernozhukov2013average,chernozhukov2015nonparametric,chernozhukov2019,botosaru2017binarization}. Compared to these papers, we consider a more constrained model, with the aim of providing a simple characterization and estimation of the bounds in this set-up. Note that \cite{chernozhukov2013average} describe in their Section 8 a generic method for computing sharp bounds on average causal effects of semiparametric models, which also applies to FE logit models \citep[see also][for a similar contruction specialized to panel dynamic discrete choice models]{honore2006bounds}. It is based on the idea that the distribution of individual effects can be approximated arbitrarily well by a distribution with fixed and finite support, if the number of support points is large enough. Then, bounds can be computed by solving as many linear programming problems as, typically, twice the number of units. In addition to being computationally more costly than our method for the FE logit model, this approach relies on an approximation, and it remains unclear how to quantify the error of this approximation. Thus, even if our approach is much more specialized than theirs, it offers important advantages for logit models.

\medskip
Finally, our work relies on results about moment problems, which have been extensively studied since Chebyshev and Markov. We refer to \cite{schmudgen2017moment} for a recent mathematical exposition and to \cite{dette1997} for applications to various statistical problems. \cite{DR17} use similar results on moment problems to obtain bounds on segregation measures with small units. \cite{dobronyi2021} use other results on moment problems to characterize the identified set of slope parameters in dynamic FE logit models, generalizing the work of \cite{honore2022moment}. Even if the primary goal of \cite{dobronyi2021} differs from ours, in both papers we use similar ideas to rewrite the initial identification problem as a more standard moment problem.


\section{Literature review on applications with FE and binary outcomes} 
\label{sec:lit_review}

To assess how often and in which manner researchers estimate models with FE and binary outcomes, we conducted a literature review of all papers published in 2022 in the five following economic journals: the \textit{American Economic Review} (AER), \textit{Econometrica} (ECTA), the \textit{Journal of Political Economy} (JPE), the \textit{Quarterly Journal of Economics} (QJE) and the \textit{Review of Economic Studies} (RES). That year, these five journals published 402 papers, excluding comments and corrigenda. 74\% of them have at least some empirical content; note that here, we even include mostly theoretical papers. Among those (but excluding their supplementary material), we looked for linear or binary choice models, such as  logit and probit models, that include FEs. We considered ``fixed effect'' as being either individual effects in a standard panel data, or dummies corresponding to a discrete variable whose number of support points is at least 10\% of the number of observations, so that if used in a nonlinear model, the incidental parameter problem becomes potentially important. For papers matching these criteria, we checked which model the authors estimated.

\begin{table}[H]
\begin{center}
	\begin{tabular}{lcccccc}
		\toprule		
		& Total & AER & ECTA & JPE & QJE & RES \\ \midrule
		\# of papers & 402 & 112 & 84 & 68 & 48 & 90 \\  \midrule
		\# of papers with some &  & &  & &  & \\
		empirical content & 296 & 81 & 50 & 49 & 45 & 71 \\  \midrule
		Among these, \# (\%) of papers  & 26 & 12 & 2 & 4 & 4 & 4 \\
		with FE and binary outcomes &  (9\%) & (15\%) & (4\%) & (8\%) & (9\%) & (6\%)\\ \midrule
		Among the latter, \# (\%) of  & 24 & 11 & 2 & 4 & 4 & 3 \\
		papers using LPM only  & (92\%) & (92\%) & (100\%) & (100\%) & (100\%) & (75\%) \\
		\bottomrule
		\multicolumn{7}{p{430pt}}{{\footnotesize Notes: AER, ECTA, JPE, QJE and RES stand respectively for the \textit{American Economic Review}, \textit{Econometrica}, the \textit{Journal of Political Economy}, the \textit{Quarterly Journal of Economics} and the \textit{Review of Economic Studies}. LPM stands for linear probability models.}}
	\end{tabular}
\end{center}
\caption{Models with binary outcomes and FE in papers published in 2022}
\label{tab:lit_review}
\end{table}

The details of our findings are displayed in Table \ref{tab:lit_review}. We identified 26 papers with binary outcomes and FE,  representing 9\% of the set of all papers with some empirical content. In all but two of these papers, only linear probability models (LPM) were used. These findings support that (i) models with FE and binary outcomes are quite common in practice; (ii) in such cases, the common practice is to estimate linear probability models; (iii) even if the FE logit model is probably one of the most well-known nonlinear model with FE, it is seldom used today. One explanation for (iii) could be that inference methods for the  AME and the ATE, which are of primary interest in empirical work, were not available. By developing computationally simple methods for them below, we thus hope that more applied researchers will turn to FE logit models.


\section{Identification} 
\label{sec:identification}

\subsection{The set-up and identification of $\beta_0$}

We consider a panel with $T$ periods and observe binary outcomes $Y_1,...,Y_T$ and for each period $t$, a vector of covariates $X_t:=(X_{t1},...,X_{tp})' \in \R^p$. We let $Y:=(Y_1,...,Y_T)'$, $X:=(X'_1,...,X'_T)'$ and assume in this section that  the joint distribution of $(X,Y)$ is identified. We also rely on the following notation hereafter. For any random variables $A$ and $B$, we let $F_A$ and $F_{A|B=b}$ denote the cumulative distribution function (cdf) of $A$ and its cdf conditional on $B=b$, respectively. We also let $\Supp(A)$ and $\Supp(A|B=b)$ denote the support of $A$ and its support conditional on $B=b$. We also make the following assumption.

\begin{hyp}
We have $Y_t=\ind{X'_t\beta_0 + \alpha + \eps_t\geq 0}$, where the $(\eps_t)_{t=1,...,T}$ are i.i.d., independent of $(\alpha,X)$ and follow a logistic distribution.
\label{hyp:model}
\end{hyp}

Importantly, the individual effect $\alpha$ is allowed to be correlated in an unspecified way with $X$. In this model, $S:=\sum_{t=1}^T Y_{t}$ is a sufficient statistic for $\alpha$ \citep[e.g.][]{andersen1970}. As a result, identification of $\beta_0$ can be achieved by maximizing the expected conditional log-likelihood, conditioning not only on $X$ but also on $S$. For any $y=(y_1,...y_T)\in\{0,1\}^T$, let us define
\begin{align*}
C_k(x, \beta):= & \sum_{(d_1,...,d_T)\in \{0,1\}^T:\sum_{t=1}^T d_t=k} \exp\left(\sum_{t=1}^T d_t x_t'\beta \right), \\	
\ell_c(y|x;\beta) := &\sum_{t=1}^T y_t x_t'\beta - \ln\left[C_{\sum_{t=1}^T y_t}(x,\beta)\right].
\end{align*}
The second term $\ell_c(y|x;\beta)$ is the conditional log-likelihood. To ensure that $\beta_0$ is identified as the unique maximizer of the expected conditional log-likelihood, we impose the following.

\begin{hyp}
$\E[ \sum_{t,t'} (X_t-X_{t'})(X_t-X_{t'})']$ is nonsingular.
\label{hyp:varying_X}
\end{hyp}

Assumption \ref{hyp:varying_X} is necessary and sufficient for the identification of the slope parameter in fixed effects linear models. The following proposition ensures that this is also the case in FE logit models. It must be well-known, but we have not been able to find it in the literature. Its proof and the proofs of other identification results are presented in Appendix \ref{sec:proofs_of_the_results}.

\begin{prop}
\label{prop:beta}	
Suppose that the distribution of $(X,Y)$ is identified, Assumption \ref{hyp:model} holds and for all $t\neq t'$ and $k\in\{1,...,p\}$, $E[(X_{tk} - X_{t'k})^2]<\infty$. Then $\beta_0$ is point identified if and only if Assumption \ref{hyp:varying_X} holds. In this case, $\beta_0 = \arg\max_\beta \E\left(\ell_c(Y|X,\beta)\right)$ and $\mathcal{I}_0=-\E\left(\deriv{{}^2\ell_c}{\beta\partial\beta'}(Y|X;\beta_0)\right)$ is nonsingular.
\end{prop}
The second part of Proposition \ref{prop:beta} shows that $\beta_0$ can be identified as the unique maximizer of the average  expected log-likelihood. Under mild regularity conditions, $\mathcal{I}^{-1}_0$ is the asymptotic variance of the conditional maximum likelihood estimator (CMLE) but also the semiparametric efficiency bound for $\beta_0$ \citep{Hahn1997}.

\subsection{Partial identification of average causal effects}
\label{sub:ident_general}

We now turn to the partial identification of average effects. Specifically, we consider parameters of the form
$$\delta_0 := \E\left[g(X,\alpha,\beta_0)\right],$$
where the function $g$ is known. We will impose an additional restriction on $g$ below but our analysis will cover the  following three key parameters:

\begin{example}[Average marginal effect, AME] Suppose that the distribution of $X_{tk}$ is continuous. Then, the average marginal effect of $X_{tk}$ (considering here the $t$-th period) is the infinitesimal change of $X_{tk}$ on the probability that $Y_t=1$:
$$\delta_0 := E\left[\Deriv{P(Y_t=1|X,\alpha)}{X_{tk}}\right] = \beta_{0k} E\left[\Lambda'(X_t'\beta_0 + \alpha)\right],$$
where $\Lambda(x):=1/(1+\exp(-x))$ denotes the cdf of the logistic distribution. Because $\Lambda'=\Lambda\times (1-\Lambda)$, we also have 
\begin{equation}
\delta_0 = \beta_{0k} E\left[\Lambda(X_t'\beta_0 + \alpha)(1-\Lambda(X_t'\beta_0 + \alpha)) \right].	
	\label{eq:other_expr_AME}
\end{equation}
\end{example}

\begin{example}[Average treatment effect, ATE] Suppose now that $\Supp(X_{tk})=\{0,1\}$. Then, the average treatment effect of $X_{tk}$ is the average effect on $Y_t$ of moving $X_{tk}$ from $0$ to $1$ for all units:
$$\delta_0 := E\left[\Lambda\left(X_t^{(1)}{}'\beta_0+\alpha\right)- \Lambda\left(X_t^{(0)}{}'\beta_0+\alpha\right)\right].$$
There, $X^{(d)}_t$ is such that its $k$-th component is equal to $d\in\{0,1\}$ and any of its other $j$-th component ($j\ne k$) is equal to $X_{tj}$. We can rewrite the ATE as
$$\delta_0 = E\left[(2X_{tk}-1)\left(Y_t - \Lambda\left(X_t^{(1-X_{tk})}{}'\beta_0+\alpha\right)\right)\right],$$
where the term $E[(2X_{tk}-1)Y_t]$ is identified. For convenience, we focus in this section on the remaining term, namely
\begin{equation}
E\left[-(2X_{tk}-1)\Lambda\left(X_t^{(1-X_{tk})}{}'\beta_0+\alpha\right)\right].	
	\label{eq:part_ATE}
\end{equation}
\end{example}

\begin{example}[Average structural function, ASF] This function, evaluated at $\widetilde{x}_t\in \Supp(X_t)$, corresponds to  the counterfactual probability that $Y_t=1$ if everyone had $X_t$ equal to $\widetilde{x}_t$:
	\begin{equation}
\delta_0 := E\left[\Lambda(\widetilde{x}_t'\beta_0+\alpha)\right].		
		\label{eq:ASF}
	\end{equation}
\end{example}

The model does not impose any restriction between $F_{\alpha|X=x}$ and $F_{\alpha|X=x'}$. As a result, to obtain the sharp identified set $\Delta$ for $\delta_0$, it suffices to obtain it on
$$\delta_0(x):= E\left[g(X,\alpha,\beta_0)|X=x\right]$$
and then integrate the corresponding sets over $x$ to obtain the sharp identified set of $\delta_0$ (note that $\delta_0(x)$ could also be of interest by itself). The only restrictions on $F_{\alpha|X=x}$ come from the data, namely from the distribution of $Y|X=x$. Actually, because $S$ is a sufficient statistic for $\alpha$, its (conditional) distribution exhausts all the information available on $\alpha$. Then, the restrictions on $F_{\alpha|X=x}$ reduce to
\begin{equation}
C_k(x,\beta_0) \int \frac{\exp(ka)}{\prod_{t=1}^T [1+\exp(x_t'\beta_0 +a)]} dF_{\alpha|X=x}(a)=P(S=k|X=x), \quad \text{for } k=0,...,T.
\label{eq:constraints}
\end{equation}
In order to obtain the identified set of $\delta_0(x)$, we thus have to find all possible values of $E\left[g(X,\alpha,\beta_0)|X=x\right]$ under the $T+1$ constraints on $F_{\alpha|X=x}$ given by \eqref{eq:constraints}.

\medskip
To derive simple formulas for the sharp and an outer identified set of $\delta_0(x)$ and $\delta_0$, we use two ideas. The first is to reparameterize the problem, so as to replace, under conditions on $g$, the moment function $a\mapsto g(x,a,\beta_0)$ by $a\mapsto a^{T+1}$, and accordingly simplify the moment restrictions \eqref{eq:constraints}. This leads us to consider the identified set of $\theta:=\int u^{T+1} d\mu(u)$ when the unknown positive measure $\mu$ is constrained by moments of the kind $\int u^k d\mu(u)=m_k$ $(k=0,...,T)$. The second idea, then, is to rely on existing results, in particular the theory of moment problems \citep[see, e.g., Chapter 10 in][]{schmudgen2017moment}, to obtain simple expressions for the sharp and an outer identified set for $\theta$. This, in turn, allows us to obtain the sharp and an outer identified set on $\delta_0(x)$ and $\delta_0$.

\subsubsection{Reparametrization of the problem} 
\label{ssub:reparametrization}

Our reparameterization relies on the fact that  for any $v(x,\beta)\in\R$, if we let $U:=\Lambda(v(X,\beta_0)+\alpha)$, the constraints \eqref{eq:constraints} satisfy, for $k=0,...,T$,
\begin{align}
P(S=k|X=x) & = C_k(x,\beta_0) \E\left[\frac{\exp(k\alpha)}{\prod_{t=1}^T [1+\exp(x_t'\beta_0 +\alpha)]} |X=x\right] \nonumber\\
& = \exp(-kv(x,\beta_0)) C_k(x,\beta_0) \E\left[ \frac{U^k(1-U)^{T-k}}{\Omega_{x,\beta_0}(U)} |X=x\right],	
\label{eq:from_alpha_to_U}
\end{align}
with $\Omega_{x,\beta}(u):=\prod_{t=1}^T [u(\exp(x_t'\beta -v(x,\beta)) - 1)+1]$. 
We then assume that $\delta_0(x)$ takes a related form $E[Q(U)/\Omega_{x,\beta_0}(U)|X=x]$, for some polynomial $Q$ with degree at most $T+1$:

\begin{hyp}
For all $(x,\beta)\in\Supp(X)\times \R^p$, there exists $v(x,\beta)\in\R$ such that letting $\Omega_{x,\beta}(u):=\prod_{t=1}^T [u(\exp(x_t'\beta -v(x,\beta)) - 1)+1]$, $u\mapsto g(x,\Lambda^{-1}(u) - v(x,\beta),\beta)\times \Omega_{x,\beta}(u)$ defined on $(0,1)$ is a polynomial of degree at most $T+1$. We let $\lambda_t(x,\beta)$ denote the coefficient of $u^t$ of this polynomial.
\label{hyp:fct_obj_simple}
\end{hyp}

Under Assumption \ref{hyp:fct_obj_simple}, $\delta_0(x)=E[h_{x,\beta_0}(U)|X=x]$ with $U:=\Lambda(v(X,\beta_0)+\alpha)$ and $h_{x,\beta}(u):=g(x,\Lambda^{-1}(u) - v(x,\beta),\beta)$. Then, we have 
	\begin{align}
		\delta_0(x)= & E\left[\frac{h_{x,\beta_0}(U)\Omega_{x,\beta_0}(U)}{\Omega_{x,\beta_0}(U)}|X=x\right]	 \notag \\
		= & \sum_{t=0}^{T+1} \lambda_t(x,\beta_0) E\left[\frac{U^t}{\Omega_{x,\beta_0}(U)}|X=x\right]. \label{eq:Delta_x}
\end{align}
Importantly, Assumption \ref{hyp:fct_obj_simple} holds for the three previous average causal parameters: 

\setcounter{example}{0}
\begin{example}[AME]
If we let $v(x,\beta)=x_t'\beta$, we obtain, in view of \eqref{eq:other_expr_AME}, $h_{x,\beta}(u) = \beta_k u (1-u)$. Then, remark that $\Omega_{x,\beta}$ is of degree at most $T-1$, since its $t$-th term in the product simplifies to one. As a result, $h_{x,\beta}\times \Omega_{x,\beta}$ is a polynomial of degree at most $T+1$, and Assumption \ref{hyp:fct_obj_simple} holds.
\end{example}

\begin{example}[ATE]
Recall that $\delta_0$ is defined by \eqref{eq:part_ATE}, as the remaining part of the ATE is identified. By letting $v(x,\beta)=x^{(1-x_{tk})}_t{}'\beta$, we obtain $h_{x,\beta}(u) = -(2x_{tk}-1) u$. Since $\Omega_{x,\beta}$ is of degree at most $T$, $h_{x,\beta}\times \Omega_{x,\beta}$ is a polynomial of degree at most $T+1$.
\end{example}

\begin{example}[ASF]
If we let $v(x,\beta)=\widetilde{x}_t'\beta$, we obtain, in view of \eqref{eq:ASF}, $h_{x,\beta}(u) = u$. Since $\Omega_{x,\beta}$ is of degree at most $T$, $h_{x,\beta}\times \Omega_{x,\beta}$ is a polynomial of degree at most $T+1$.
\end{example}

We showed in \eqref{eq:from_alpha_to_U} that the probabilities $(P(S=k|X=x))_{k=0,...,T}$ are linear combinations of the expectations $(E[U^t/\Omega_{x,\beta_0}(U)|X=x])_{t=0,...,T}$. Actually, Lemma \ref{lem:optim} below establishes that there is a one-to-one relationship between the two, so that these moments are identifiable. Specifically, we show that \begin{align}
E\left[\frac{U^t}{\Omega_{x,\beta_0}(U)}|X=x\right] &= E[Z_t|X=x],
\label{eq:Z}
\end{align} 
with
$$Z_t(x,s,\beta) := \binom{T-t}{s-t} \frac{\exp(sv(x,\beta))}{C_s(x;\beta)}, \quad Z_t := Z_t(X,S,\beta_0),$$
and where we use the convention $\binom{T-t}{s-t}=0$ if $s<t$. Hence, the first $T+1$ terms of the sum in \eqref{eq:Delta_x} are identified. The last term of the sum, on the other hand, is not identified in general. To write this term in a more convenient way and complete our reparameterization, let us define the following probability measure on $[0,1]$ (remark that $\Omega_{x,\beta_0}(U)>0$ almost surely):
\begin{equation}
\mu_x(A) := \frac{E\left[\ind{U\in A}/\Omega_{x,\beta_0}(U)|X=x\right]}{E\left[1/\Omega_{x,\beta_0}(U)|X=x\right]},	
	\label{eq:def_mu_x}
\end{equation}
for any Borel set $A\subseteq [0,1]$. Then, in view of \eqref{eq:Delta_x}-\eqref{eq:Z}, we obtain
\begin{equation}
\delta_0(x) = \sum_{t=0}^T \lambda_t(x,\beta_0)E[Z_t|X=x] + \lambda_{T+1}(x,\beta_0) E[Z_0|X=x] \int_0^1 u^{T+1}d\mu_x(u).	
	\label{eq:delta_1}
\end{equation}
Moreover, the first $T$ moments of $\mu_x$ are identified: 
\begin{equation}
m_t(x):=\int_0^1 u^t d\mu_x(u) = \frac{E[Z_t|X=x]}{E[Z_0|X=x]}, \quad t=0,...,T.
\label{eq:def_m}
\end{equation}
In other words, $\delta_0(x)$ satisfies \eqref{eq:delta_1} with $\mu_x$ a probability measure whose vector of first moments is equal to $m(x):=(m_0(x),...,m_T(x))$. Lemma \ref{lem:optim} below shows that the converse holds as well: for any probability measure $\mu$ on $[0,1]$ with a vector of first moments equal to $m(x)$ and such that $\mu(\{0,1\})=0$, we prove that 
$$\sum_{t=0}^T \lambda_t(x,\beta_0)E[Z_t|X=x] + \lambda_{T+1}(x,\beta_0) E[Z_0|X=x] \int_0^1 u^{T+1}d\mu(u).$$
is in the identified set of $\delta_0(x)$. To state Lemma \ref{lem:optim}, we denote by $\mathcal{D}$ the set of positive measures on $[0,1]$ and for any $m=(m_0,...,m_T)\in [0,1]^{T+1}$, we let
$$\mathcal{D}(m)=\left\{\mu\in\mathcal{D}: \int u^k d\mu(u)=m_k, \; k=0,...,T\right\}.$$
In words, $\mathcal{D}(m)$ is the subset of positive measures on $[0,1]$ whose vector of first $T+1$ raw moments (including the moment of order 0) is equal to $m$.

\begin{lem}\label{lem:optim}
Suppose that the distribution of $(Y,X)$ is identified and Assumptions \ref{hyp:model}-\ref{hyp:fct_obj_simple} hold. Then, for all $x\in\Supp(X)$, the identified set of $\delta_0(x)$ is
\begin{align}
	&\bigg\{\sum_{t=0}^T \lambda_t(x,\beta_0) E[Z_t|X=x] + \lambda_{T+1}(x,\beta_0) E[Z_0|X=x]\int_0^1 u^{T+1}d\mu(u): \; \notag \\
	& \quad \mu\in \mathcal{D}(m(x)), \; \mu(\{0,1\})=0\bigg\}. \label{eq:other_expr_Delta_x}
\end{align}
\end{lem}

We prove this result by, basically, showing that there is a one-to-one mapping between $F_{\alpha|X=x} $ and $\mu_x$ defined by \eqref{eq:def_mu_x}. This implies that we can express the problem using $\mu_x$ and the $(m_t(x))_{t=0,...,T}$ instead of $F_{\alpha|X=x}$ and the $(P(S=s|X=x))_{s=0,...,T}$, respectively.


\subsubsection{Sharp bounds} 
\label{ssub:sharp_bounds}

Lemma \ref{lem:optim} ensures that finding the sharp identification region of $\delta_0(x)$ reduces to finding the set of values of $\int_0^1 u^{T+1}d\mu_x(u)$ for $\mu_x\in \mathcal{D}(m(x))$ satisfying $\mu_x(\{0,1\})=0$. Without this last constraint, the problem is known as the truncated Hausdorff problem. The solution, accounting for the constraint, is given in Proposition \ref{prop:extension_mom} below. Let us first introduce additional  notation. For any $t\ge 1$, $s\geq t$ and $m=(m_0,...,m_s)\in\R^{s+1}$, we define the Hankel matrices $\underline{\mathbb{H}}_t(m)$ and $\overline{\mathbb{H}}_t(m)$ as
$$\begin{array}{lll}
\underline{\mathbb{H}}_t(m)=\left(m_{i+j-2}\right)_{1\leq i,j\leq t/2+1},& \overline{\mathbb{H}}_t(m)=\left(m_{i+j-1}-m_{i+j}\right)_{1\leq i, j \leq t/2} & \text{if }t\text{ is even,}\\
\underline{\mathbb{H}}_t(m)=\left(m_{i+j-1}\right)_{1\leq i,j \leq (t+1)/2},& \overline{\mathbb{H}}_t(m)=\left(m_{i+j-2}-m_{i+j-1}\right)_{1\leq i,j \leq (t+1)/2} & \text{if }t\text{ is odd.}
\end{array}$$
Then, let $\underline{H}_t(m)=\det \left(\underline{\mathbb{H}}_t(m)\right)$ and $\overline{H}_t(m)=\det(\overline{\mathbb{H}}_t(m))$. Now, for $m\in \mathbb{R}^{T+1}$ and $q\in\R$, consider $\underline{H}_{T+1}(m,q)$. By expanding this determinant along its last column, we see that $q\mapsto \underline{H}_{T+1}(m,q)$ is linear. Then, let $\underline{a}_{T+1}(m)$ and $\underline{b}_{T+1}(m)$ denote the corresponding intercept and slope parameters, so that  $\underline{H}_{T+1}(m,q) = \underline{a}_{T+1}(m)+\underline{b}_{T+1}(m)q$. We define similarly $ \overline{a}_{T+1}(m)$ and $\overline{b}_{T+1}(m)$. Finally, let $|A|$ denote the cardinal of set $A$ and $\lfloor x\rfloor$ denote the integer part of $x\in \R$.

\begin{prop}
Let $\theta_0=\int_0^1 u^{T+1} d\mu(u)$ for some unknown measure $\mu\in \mathcal{D}(m)$ satisfying $\mu(\{0,1\})=0$ and some identified $m\in[0,1]^{T+1}$. Then,  $\underline{H}_T(m)\ge 0$, $\overline{H}_T(m)\ge 0$ and  $\Theta$, the closure of the identified set  of $\theta_0$, satisfies $\Theta=[\underline{q}_T(m),\, \overline{q}_T(m)]$, where the functions $\underline{q}_T$ and $\overline{q}_T$ are such that:
\begin{enumerate}
	\item If $\underline{H}_T(m) \times \overline{H}_T(m)> 0$, then $\underline{q}_T(m)<\overline{q}_T(m)$ with $\underline{q}_T(m)=-\underline{a}_{T+1}(m)/$ $\underline{b}_{T+1}(m)$ and $\overline{q}_T(m)=-\overline{a}_{T+1}(m)/\overline{b}_{T+1}(m)$. Moreover, $\left|\Supp(\mu)\right|> \lfloor T/2 \rfloor$.
	\item If $\underline{H}_T(m) \times \overline{H}_T(m)= 0$, then $\D(m)=\{\mu\}$, $\underline{q}_T(m)=\overline{q}_T(m)=\theta_0$ and $\Theta=\{\theta_0\}$. Moreover, letting $T'=\min\{t\leq T: \underline{H}_t(m) \times \overline{H}_t(m)= 0\}$,  $\underline{q}_T(m)$ is equal to
	\begin{equation}
		\begin{array}{rcl}
			-\underline{a}_{T'}(m_{T-T'+1},...,m_T)/\underline{b}_{T'}(m_{T-T'+1},...,m_T) & \text{ if } \underline{H}_{T'}(m)=0, \\
			-\overline{a}_{T'}(m_{T-T'+1},...,m_T)/\overline{b}_{T'}(m_{T-T'+1},...,m_T) & \text{ if } \overline{H}_{T'}(m)=0.			 	
		\end{array}
		\label{eq:qinf_bord}
	\end{equation} Finally, $\left|\Supp(\mu)\right|\leq \lfloor T'/2 \rfloor$.
\end{enumerate}	
\label{prop:extension_mom}
\end{prop}

Proposition \ref{prop:extension_mom} shows that the sharp bounds of $\theta_0$ are very simple to obtain as rational functions of determinants of some identified matrices. Point 1 follows from classical results in moment theory, see in particular Theorem 10.8 and Proposition 10.15 in \cite{schmudgen2017moment}. The first part of the Point 2 is also well-known. On the other hand, to the best of our knowledge, its second part is new. To understand why we get respectively partial identification and point identification in Point 1 and Point 2, let us assume $T=2$. Since the support of $\mu$ is included in $[0,1]$, we have $m_1^2 \le m_2 \le m_1$. This implies that $\underline{H}_2(m)\ge 0$ and $\overline{H}_2(m)\ge 0$. Moreover, if $m_1^2 = m_2$, so that $\overline{H}_2(m)=0$, the variance of the distribution is 0. Then, $\mathcal{D}(m)=\{\mu\}$ where $\mu$ is the Dirac distribution at $m_1$.\footnote{The case  $m_1 = m_2$ corresponds to a Bernoulli distribution with parameter $m_1$, which is excluded here because $\mu(\{0,1\})=0$.} If, on the other hand, $m_1 > m_2 > m_1^2$, so that $\underline{H}_2(m) \times \overline{H}_2(m)>0$, the variance is strictly positive and there are infinitely many distributions with first two moments equal to $(m_1, m_2)$. More intuition on Proposition \ref{prop:extension_mom} is given at the beginning of its proof.

\medskip
By combining Lemma \ref{lem:optim} with Propositions \ref{prop:beta} and \ref{prop:extension_mom}, we obtain the following result on the closure of the sharp identified sets of $\delta_0(x)$ and $\delta_0$, denoted respectively by $\Delta(x)$ and $\Delta$.

\begin{thm}\label{thm:ident1}
If the distribution of $(X,Y)$ is identified and Assumptions \ref{hyp:model}-\ref{hyp:fct_obj_simple} hold, $\Delta(x)=[\underline{\delta}(x),\, \overline{\delta}(x)]$, with
\begin{equation}
	\begin{array}{rcl}
		\underline{\delta}(x) & := & \sum_{t=0}^T \lambda_t(x,\beta_0) \E[Z_t|X=x]  + \lambda_{T+1}(x,\beta_0) \E[Z_0|X=x] \big( \underline{q}_T(m(x)) \\[2mm]
		& & \hspace{1.2cm} \ind{\lambda_{T+1}(x,\beta_0)\geq 0}  +  \overline{q}_T(m(x)) \ind{\lambda_{T+1}(x,\beta_0)< 0} \big), \\[2mm]
		\overline{\delta}(x) & := & \sum_{t=0}^T \lambda_t(x,\beta_0) \E[Z_t|X=x]  + \lambda_{T+1}(x,\beta_0) \E[Z_0|X=x] \big( \overline{q}_T(m(x)) \\[2mm]
		& & \hspace{1.2cm} \ind{\lambda_{T+1}(x,\beta_0)\geq 0}  +  \underline{q}_T(m(x)) \ind{\lambda_{T+1}(x,\beta_0)< 0} \big).
	\end{array}
	\label{eq:bounds_Delta}
\end{equation}
Moreover,  $\Delta =[\underline{\delta},\, \overline{\delta}]$, with $\underline{\delta}=E(\underline{\delta}(X))$, $\overline{\delta}=E(\overline{\delta}(X))$. $\delta_0$ is point identified if and only if
$$P\left(\big\{\lambda_{T+1}(X,\beta_0)=0\big\} \cup \big\{|\Supp(\alpha|X)|\leq \lfloor T/2\rfloor \big\}\right)=1.$$
\end{thm}

From \eqref{eq:bounds_Delta}, point identification of $\delta_0(x)$ holds if and only if $\lambda_{T+1}(x_t,\beta_0)=0$ or $\underline{q}_T(m(x))=\overline{q}_T(m(x))$. The first case occurs for the AME and ATT if $\beta_{0k}=0$. It also holds for the AME at $t$ if $\min_{s\ne t} |(x_s-x_t)'\beta_0|=0$. This could be expected, as \cite{Hoderlein2012} show that average marginal effects are nonparametrically identified on ``stayers'', namely individuals for whom $X_{it}$ remains constant between two periods. Finally, point identification can also be achieved if $|\Supp(\alpha|X)|\le \intpart{T/2}$.\footnote{Such identification is achieved using the logit structure: as a complement to \cite{Hoderlein2012}, \cite{chernozhukov2019} show that average marginal effects are not identified nonparametrically for non-stayers.} This corresponds to the second case described in Proposition \ref{prop:extension_mom}. Intuitively, if $\alpha|X$ has few points of support, its full distribution is characterized by its first moments. Then, given these moments, the higher moments  are fully determined. As an illustration, assume that $T=2$ and $\alpha|X$ is degenerate and equal to $\alpha_0$. Then, some algebra shows that $m(X)=(1,\Lambda(\alpha_0+X_T'\beta_0),\Lambda(\alpha_0+X_T'\beta_0)^2)'$ which implies that the variance of any distribution in $\mathcal{D}(m(X))$ is zero. Thus, $\mathcal{D}(m(X))$ reduces to the Dirac distribution at $\Lambda(\alpha_0+X_T'\beta_0)$, implying $\underline{q}_2(m(X))=\overline{q}_2(m(X))= \Lambda(\alpha_0+X_T'\beta_0)^3$.

\subsubsection{Outer bounds} 
\label{ssub:outer_bounds}

A drawback of the previous approach, in terms of estimation, is that even if we are interested in $\delta_0$ rather than $\delta_0(x)$, we need to estimate nonparametrically the functions $m_k(x)$ for all $x$ observed in the data, because they enter nonlinearly into the bounds. This requires regularity conditions and the choice of tuning parameters. Also, we need to account for the constraint that $m(x)$ is a vector of moments. An estimator  $\widehat{m}(x)$ need not be a vector of moments and, for instance, need not satisfy the variance constraint $\widehat{m}_2(x)\ge \widehat{m}^2_1(x)$. 
If not, the set $\D(\widehat{m}(x))$ is empty and the bounds $(\underline{q}_T(\widehat{m}(x)),\overline{q}_T(\widehat{m}(x)))$ are not defined anymore. We suggest in Section \ref{sub:definition_of_the_estimators} below a constrained estimator which is a vector of moments but this complicates the estimation of the bounds.

\medskip
As an alternative, we consider simple outer bounds that, when focusing on the parameter $\delta_0$, do not require nonparametric estimation of $m_k(x)$. The idea is to find good approximations, in a sup-norm sense on $[0,1]$, of $u\mapsto u^{T+1}$ by a polynomial of degree $T$. Let us define, as in Proposition \ref{prop:extension_mom}, $\theta_0:=\int_0^1 u^{T+1}d\mu(u)$ for some $\mu\in\mathcal{D}(m)$ and suppose that there exists $K>0$ and $(b_0,...,b_T)$  such that
$$\sup_{u\in[0,1]} \left|u^{T+1}-\sum_{k=0}^T b_k u^k\right|\le K.$$
By Jensen's inequality, this implies that $|\theta_0 - \sum_{k=0}^T b_k m_k| \le K$. As a result, we obtain that $[\sum_{k=0}^T b_k m_k - K,\; \sum_{k=0}^T b_k m_k +K]$ is an outer set for $\theta_0$. Moreover, an optimal solution to the uniform approximation problem is very simple, and the corresponding $K$ is small, as the following proposition shows. Let $\mathbb{T}^c_{T+1}$ denote the Chebyshev polynomial of degree $T+1$, renormalized so that its leading coefficient is equal to 1.\footnote{Recall that the unnormalized Chebyshev polynomials are defined by $\mathbb{T}^u_0(x)=1$, $\mathbb{T}^u_1(x)=x$ and $\mathbb{T}^u_{k+1}(x)=2x\mathbb{T}^u_{k}(x)-\mathbb{T}^u_{k-1}(x)$ for any $k\ge 1$.} Then, let  $\mathbb{T}_{T+1}(u):= 2^{-T-1}\mathbb{T}^c_{T+1}(2u-1)$ and let $-b^*_{k,T}$ denote the coefficient of degree $k$ of $\mathbb{T}_{T+1}$. Finally, recall that $\Theta$ denotes the closure of the identified set of $\theta_0$.

\begin{prop}
We have $(b^*_{0,T},...,b^*_{T,T})=\arg\min_{(b_0,...,b_T)\in\R^{T+1}} \sup_{u\in[0,1]} \left|u^{T+1}-\sum_{k=0}^T b_K u^k\right|$. Moreover, under the conditions of Proposition \ref{prop:extension_mom}, we have $\Theta \subseteq \Theta^o$, with
\begin{align*}
	\Theta^o & := \left[\sum_{k=0}^T b^*_{k,T}m_k - \frac{1}{2\times 4^T} ,\,\sum_{k=0}^T b^*_{k,T}m_k + \frac{1}{2\times 4^T}\right].
\end{align*}
Finally, we may have $\Theta=\Theta^o$.
\label{prop:outer_ident}
\end{prop}

\begin{figure}[H]
\begin{center}
	\includegraphics[trim=10mm 115mm 0mm 110mm,scale=0.85, clip=true]{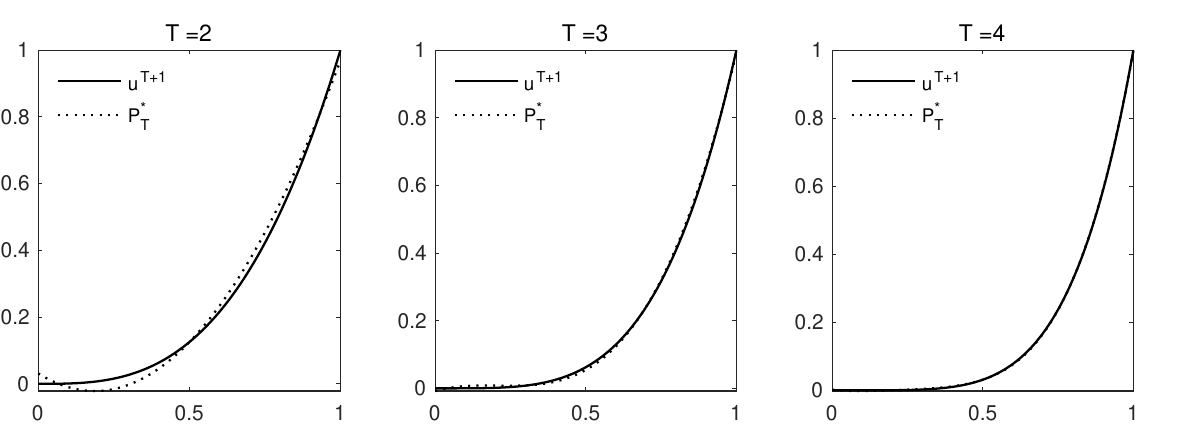}
\end{center}
\caption{Approximation of $u\mapsto u^{T+1}$ by $P^*_T$.}
\label{fig:Cheb_approx}
\end{figure}

Figure \ref{fig:Cheb_approx} displays $u\mapsto u^{T+1}$ and its best uniform approximation $P^*_T(u):=\sum_{k=0}^T b^*_{k,T} u^k$ for $T=2, 3$ and 4. As we can see, the approximation is already good for $T=2$, and the two functions become almost indistinguishable for $T=4$. This could be expected as the bound $1/(2\times 4^T)$ decreases very quickly with $T$.

By combining Lemma \ref{lem:optim} with Propositions \ref{prop:beta} and \ref{prop:outer_ident}, we obtain the following outer identified sets for $\delta_0(x)$ and $\delta_0$.
Recall that $\Delta(x)$ and $\Delta$ denote the closure of their identified sets.

\begin{thm}\label{thm:ident2}
If the distribution of $(X,Y)$ is identified and Assumptions \ref{hyp:model}-\ref{hyp:varying_X} hold, we have $\Delta(x)\subseteq \Delta^o(x) :=[\tilde{\delta}(x)\pm\overline{b}(x)]$ and $\Delta \subseteq \Delta^o:=[\tilde{\delta}\pm \overline{b}]$, with 
\begin{align*}
	\tilde{\delta}(x) & := \sum_{t=0}^T \left(\lambda_t(x,\beta_0) + b^*_{t,T}\lambda_{T+1}(x,\beta_0)\right) E(Z_t|X=x), \\
	\overline{b}(x) & :=  \frac{1}{2\times 4^T} |\lambda_{T+1}(x,\beta_0)| E\left[Z_0|X=x\right],\\
	\tilde{\delta} & := 	E\bigg[\sum_{t=0}^T \left(\lambda_t(X,\beta_0) + b^*_{t,T}\lambda_{T+1}(X,\beta_0)\right)Z_t \bigg],\\
	\overline{b} & := \frac{1}{2\times 4^T} E\left[|\lambda_{T+1}(X,\beta_0)|\times Z_0\right].		
\end{align*}
Moreover, we may have $\Delta^o(x)= \Delta(x)$ and $\Delta^o= \Delta$.
\end{thm}

As mentioned above, the outer bounds on $\delta_0$ are very simple and do not require nonparametric estimation of $m(X)$. Moreover, because $[\underline{\delta},\overline{\delta}] \subseteq [\tilde{\delta}-\overline{b},\tilde{\delta}+\overline{b}]$, we  obtain the following bound for the length of the identified set of $\delta_0$:
\begin{equation}
\overline{\delta} - \underline{\delta} \leq \frac{1}{4^T} E\left[|\lambda_{T+1}(X,\beta_0)|\times Z_0\right].	
\label{eq:upper_bd_length}
\end{equation}
For some distributions of $X$, this inequality yields an upper bound on the rate of decrease of the size of the identified set as $T$ increases. Specifically, assume that for all $t\le T$, $P[|X_t'\beta_0-v(X,\beta_0)|\leq \ln(2)]=1$. 
Then, additional algebra shows that $E\left[|\lambda_{T+1}(X,\beta_0)|\times Z_0\right]\leq 1$, which in turn implies
$$\overline{\delta} - \underline{\delta} \leq \frac{1}{4^T}.$$
Similarly if for all $t\le T$, $P(|(X_t'\beta_0-v(X,\beta_0)|\leq c)=1$ for some $c\in [\ln(2), \ln(5))$ then $\overline{\delta} - \underline{\delta} \leq (e^c-1)^{T-1}/4^T \leq K^{T}$ for $K=(e^c-1)/4< 1$. This discussion may convey the impression that for $X$ with larger support, possibly unbounded, the identified set could be large. However, recall that the right-hand side of \eqref{eq:upper_bd_length} is only an upper bound on the true length $\overline{\delta} - \underline{\delta}$. In Appendix \ref{sec:illustr_bds}, we show how the sharp bounds evolve when varying the support of $X$, for a specific distribution of $\alpha|X$. We find that even if $X$ has large support, the bounds remain extremely informative and their length quickly decreases with $T$.

\medskip
\cite{chernozhukov2013average} also obtain, in their Theorem 4, an exponential rate of decrease on the length of the identified set of the ASF. Actually, their result imposes substantially weaker conditions on the distribution of $Y_t|X_t,\alpha$. On the other hand, it only holds for finitely supported $X$ and imposes additional restrictions on the distribution of $(X,\alpha)$.


\subsection{Measuring the heterogeneity of treatment effects} 
\label{sub:het_TE}

Whereas in the linear probability model, $(x,\alpha) \mapsto g(x,\alpha,\beta_0)$ is constant when considering the AME or ATE, this function varies in the FE logit model. Understanding this heterogeneity is important for, e.g., policy design \citep[see, e.g.,][]{manski2004statistical}. We have shown above how to get sharp and outer bounds for $\delta_0(x)=E[g(X,\alpha,\beta_0)|X=x]$. However, $X$ has dimension $pT$, so even if the $\alpha_i$ were observed, $\delta_0(x)$ would typically be inaccurately estimated due to the curse of dimensionality. On the other hand, if $\widetilde{X}$ is a low-dimensional subvector of $X$ (e.g., $\widetilde{X}=X_{tj}$), we can measure how $g(X,\alpha,\beta_0)$ varies with $\widetilde{X}$ by partially identifying $$\delta_{\widetilde{X}}(\widetilde{x}):=E[g(X,\alpha,\beta_0)|\widetilde{X}=\widetilde{x}]=E[\delta_0(X)|\widetilde{X}=\widetilde{x}].$$
One may also be interested by heterogeneity with respect to factors not included in $X$, e.g. time-invariant features. For instance, if one considers participation to the labor market as a function of the number of children, we may be interested by heterogeneity with respect to education. We then consider the parameter $\delta_W(w):= E[g(X,\alpha,\beta_0)|W=w]$. This parameter has a similar interpretation as $\delta_{\widetilde{X}}(\widetilde{x})$, even if $W$ is not a function of $X$, provided that the following conditional independence holds:
\begin{equation}
(\eps_1,...,\eps_T) \indep W | X,\alpha.	
\label{eq:condit_indep_W}
\end{equation}
Then, if we consider for instance the AME, we have $g(x,\alpha,\beta_0)=\deriv{P(Y_T=1|X=x,W=w,\alpha)}{x_{tk}}$, and $\delta_W(w)$ is the average marginal effect of $X_k$ on the subpopulation $W=w$. When $W$ is multivariate, $W=(W_1,...,W_p)$, one may also want to learn about the main drivers of heterogeneity, by considering the coefficient $\varsigma_j$ of $W_j \ (j\in\{1,...,p\})$ in the best linear prediction of $g(X,\alpha,\beta_0)$ by $W$.

\medskip
We obtain sharp bounds on $\delta_W(w)$ by using $\delta_W(w)= E[\delta_0(W,X)|W=w]$, with $\delta_0(w,x)=E[g(X,\alpha,\beta_0)|W=w, X=x]$, and remarking that Lemma \ref{lem:optim} holds when conditioning on $(X,W)$ instead of $X$ only. Then, the same reasoning as for obtaining the sharp bounds on $\delta_0(x)$ applies to $\delta_0(w,x)$. For instance, its sharp lower bound satisfies
\begin{align*}
\underline{\delta}(w,x) = & \sum_{t=0}^T \lambda_t(x,\beta_0) \E[Z_t|W=w, X=x]  + \lambda_{T+1}(x,\beta_0)  \E[Z_0|W=w, X=x] \\
& \hspace{0.45cm} \times \big( \underline{q}_T(m(w,x)) \ind{\lambda_{T+1}(x,\beta_0)\geq 0}  +  \overline{q}_T(m(w,x)) \ind{\lambda_{T+1}(x,\beta_0)< 0} \big),
\end{align*}
where $m_t(w,x):=\E[Z_t|W=w, X=x]/\E[Z_0|W=w, X=x]$. Then the sharp lower bound on $\delta_W(w)$ is simply $E[\underline{\delta}(W,X)|W=w]$, and similarly for its upper bound. Also, following \cite{bontemps2012set}, the sharp lower bound on $\varsigma_j$ is
$$\underline{\varsigma}_j = \frac{E\left[W^r_j\left(\underline{\delta}(W,X)\ind{W^r_j>0}+ \overline{\delta}(W,X)\ind{W^r_j<0}\right)\right]}{E[W^{r2}_j]},$$
where $\overline{\delta}(W,X)$ is the sharp upper bound on $\delta_0(W,X)$ and $W^r_j$ is the residual of the theoretical regression of $W_j$ on 1 and the $(W_k)_{k\ne j}$.

\medskip
The sharp bounds above rely on the nonparametric functions $\E[Z_t|W=w, X=x]$, with $(W,X)$ possibly of high dimension. The following proposition shows that we can avoid this by considering, again, outer bounds. Its proof is identical to that of Theorem \ref{thm:ident2} and therefore omitted.

\begin{prop}
If the distribution of $(W,X,Y)$ is identified, Assumptions \ref{hyp:model}-\ref{hyp:varying_X}  and \eqref{eq:condit_indep_W} holds, an outer identified set on $\delta_W(w)$ (resp. on $\varsigma_j$) is $[\tilde{\delta}_W(w)\pm \overline{b}_W(w)]$ (resp. $[\tilde{\varsigma}_j\pm \overline{b}_{\varsigma_j}]$), with
\begin{align*}
	\tilde{\delta}_W(w) & = E\left[\sum_{t=0}^T (\lambda_t(X,\beta_0) + b^*_{t,T} \lambda_{T+1}(X,\beta_0)) Z_t \big| W=w\right], \\
	\overline{b}_W(w) & = \frac{1}{2\times 4^T} E\left[|\lambda_{T+1}(X,\beta_0)| Z_0 \big| W=w \right],\\
	\tilde{\varsigma}_j & = \frac{E\left[W^r_j\sum_{t=0}^T \left(\lambda_t(X,\beta_0) + b^*_{t,T} \lambda_{T+1}(X,\beta_0)\right)Z_t\right]}{E[W^{r2}_j]}, \\
	\overline{b}_{\varsigma_j} & = \frac{E\left[\left|W^r_j\lambda_{T+1}(X,\beta_0) \right|\times Z_0 \right]}{E[W^{r2}_j]}.
\end{align*}
\label{prop:heterog}
\end{prop}
\vspace{-1cm}



\section{Estimation and inference} 
\label{sec:inference}

We turn to the estimation of bounds on $\delta_0$, and inference on this parameter, using a sample $(Y_i,X_i)_{i=1,...,n}$. We first consider sharp bounds, before considering outer bounds. Though we omit details to save space, the methodology developed below can be applied to estimate bounds on the heterogeneity parameters $\delta_W(w)$ and $\theta_j$.

\subsection{Estimation and inference based on the sharp bounds} 
\label{sub:estimation_and_inference_based_on_the_sharp_bounds}

\subsubsection{Definition of the estimators} 
\label{sub:definition_of_the_estimators}

By Theorem \ref{thm:ident1} and the law of iterated expectations, we have $\underline{\delta} = E[\underline{h}(X,S)]$, with
\begin{align*}
\underline{h}(X,S) = & r(X,S,\beta_0) +  \lambda_{T+1}(X,\beta_0) \E(Z_0 | X) \big( \underline{q}_T(m(X)) \ind{\lambda_{T+1}(X,\beta_0)\geq 0}  \\[2mm]
&  +  \overline{q}_T(m(X)) \ind{\lambda_{T+1}(X,\beta_0)< 0} \big),
\end{align*}
with $r(x,s,\beta) := \sum_{t=0}^T \lambda_t(x,\beta) Z_t(x,s,\beta)$, so that $r(X,S,\beta_0) =  \sum_{t=0}^T \lambda_t(X,\beta_0)Z_t$.\footnote{Recall that for the ATE, we actually focused so far on a part of it, defined by \eqref{eq:part_ATE}. To estimate the ATE itself, we thus need to replace $r(x,s,\beta)$ by $(2x_{kt}-1)y_t+ \sum_{t=0}^T Z_t(x,s,\beta) \lambda_t(x,\beta)$.} Also, $\overline{\delta}= E[\overline{h}(X,S)]$, with $\overline{h}(X,S)$ defined similarly as $\underline{h}(X,S)$. We estimate $\underline{h}$ and $\overline{h}$, and in turn $\underline{\delta}$ and $\overline{\delta}$ by plug-in, in four steps:
\begin{enumerate}
\item Estimation of $\beta_0$ by the conditional maximum likelihood estimator: $$\widehat{\beta}:=\arg\max_{b\in B}\sum_{i=1}^n\ell_c(Y_i|X_i,b).$$
\item Initial estimation of $m=(m_0,...,m_T)$:\\
Let $\gamma_{0j}(x):=P \left(S=j|X=x\right)$ for $j=0,...,T$. By definition, $m_t(x)=c_t(x)/c_0(x)$, with
\begin{equation}
	c_t(x):=E(Z_t|X=x)=\sum_{j=t}^T\binom{T-t}{j-t}\frac{\gamma_{0j}(x)\exp(j v(x,\beta_0))}{C_{j}(x,\beta_0)}.	 	
	\label{eq:otherdef_m}
\end{equation}
We first estimate the $(\gamma_{0j})_{j=0,...,T}$ nonparametrically. We consider a local polynomial estimator $\widehat{\gamma}_j$ of order $\ell$, with kernel function $K$ and bandwidth $h_n$. Then, let $\widehat{c}_t(x)$ be the plug-in estimator of $c_t(x)$ based on \eqref{eq:otherdef_m}, and let $\widetilde{m}_t(x)=\widehat{c}_t(x)/\widehat{c}_0(x)$ for $t=0,...,T$. Remark that by construction, $\widetilde{m}_0(x)=1$. \label{item:other_terms}
\item \label{step3} Constrained estimation of $m$: \\	
$\widetilde{m}$ has  an important drawback: because of sampling uncertainty, it is not necessarily a vector of moments. Formally, if we let $\M_T:=\left\{m\in [0,1]^{T+1}: \mathcal{D}(m)\ne \emptyset\right\}$, we may have $\widetilde{m}\not\in\M_T$, in which case $\underline{q}_T(\widetilde{m}(X_i))$ and $\overline{q}_T(\widetilde{m}(X_i))$ are not defined.\footnote{In our simulations, this already occurs with $T=3$ and $n=1,000$, even if $m(x)$ is in the interior of $\M_T$.} We thus construct an estimator $\widehat{m}$ such that for all $i=1,...,n$, $\widehat{m}(X_i)\in\M_T$, by exploiting Proposition \ref{prop:extension_mom}. For any $(m_t)_{t\geq 0}$ and $t\in\{0,...,T\}$, let $m_{\rightarrow t}:=(m_0,...,m_t)$. The idea of the estimator is to use the first elements of $\widetilde{m}(x)$, until $\widetilde{m}_t(x) \not\in [\underline{q}_{t-1}(\widetilde{m}_{\rightarrow t-1}(x)), \overline{q}_{t-1}(\widetilde{m}_{\rightarrow t-1}(x))]$, or (for technical reasons) $\widetilde{m}_t(x)$ is close to $\underline{q}_{t-1}(\widetilde{m}_{\rightarrow t-1}(x))$ or $\overline{q}_{t-1}(\widetilde{m}_{\rightarrow t-1}(x))$. In such a case, we simply replace $\widetilde{m}_t(x)$ by $\underline{q}_{t-1}(\widetilde{m}_{\rightarrow t-1}(x))$ or $\overline{q}_{t-1}(\widetilde{m}_{\rightarrow t-1}(x))$. We finally complete the vector using the second part of Proposition \ref{prop:extension_mom}.

\medskip
Specifically, let $c_n$ be a sequence tending to 0 at a rate specified later and define
\begin{equation}
\widehat{I}(x):= \max\left\{t\in \{1,...,T\}: \;\forall s\le t,\; \underline{H}_s(\widetilde{m}_{\rightarrow s}(x))\times \overline{H}_s(\widetilde{m}_{\rightarrow s}(x)) > c_n\right\},
\label{eq:I_hat}
\end{equation}
with the convention that $\max \emptyset = 0$. We then let $\widehat{m}_{\rightarrow \widehat{I}(x)}(x):= \widetilde{m}_{\rightarrow \widehat{I}(x)}(x)$. If $\widehat{I}(x)=T$, $\widehat{m}(x)$ is fully defined. Otherwise, we complete $\widehat{m}(x)$ by first letting
$$\widehat{m}_{\widehat{I}(x)+1}(x):= \left|\begin{array}{cl}
	\underline{q}_{\widehat{I}(x)}(\widetilde{m}_{\rightarrow \widehat{I}(x)}(x)) & \text{ if } \underline{H}_{\widehat{I}(x)+1}(\widetilde{m}_{\rightarrow \widehat{I}(x)+1}(x)) < c_n^{1/2}, \\
	\overline{q}_{\widehat{I}(x)}(\widetilde{m}_{\rightarrow \widehat{I}(x)}(x)) & \text{ otherwise.}
\end{array}\right.$$
Next, if $\widehat{I}(x)+1<T$, we have, by construction, 
$$\underline{H}_{\widehat{I}(x)+1}(\widehat{m}_{\rightarrow \widehat{I}(x)+1}(x))\times \overline{H}_{\widehat{I}(x)+1}(\widehat{m}_{\rightarrow \widehat{I}(x)+1}(x)) = 0. $$
Then, Part 2 of Proposition \ref{prop:extension_mom} shows that there are unique moments $\widehat{m}_{\widehat{I}(x)+2},...,\widehat{m}_{T}$ that are compatible with $\widehat{m}_{\rightarrow \widehat{I}(x)+1}(x)$. We construct them by induction using this proposition. In the end, the corresponding vector $\widehat{m}(x)$ satisfies $\widehat{m}(x)\in\M_T$.

\item Plug-in estimation of $\underline{h}$, $\overline{h}$ and the sharp bounds:\\
We compute the estimator $\widehat{\underline{\delta}}  = \frac{1}{n}\sum_{i=1}^n \widehat{\underline{h}}(X_i,S_i)$, with
\begin{align}
\widehat{\underline{h}}(x,s) & = r(x,s,\widehat{\beta}) + \widehat{c}_0(x) \lambda_{T+1}(x, \widehat{\beta}) \bigg[\overline{q}_T(\widehat{m}(x)) \ind{\lambda_{T+1}(x,\widehat{\beta})\geq 0}  \nonumber \\
	&  + \underline{q}_T(\widehat{m}(x)) \ind{\lambda_{T+1}(x,\widehat{\beta}) < 0}  \bigg]. \label{eq:def_estim_bds}
\end{align}
We define $\widehat{\overline{\delta}}$ and $\widehat{\overline{h}}(x,s)$ similarly.
\end{enumerate}

Two remarks on this estimation method are in order. First, we produce estimators of the bounds that never cross, even if the model is misspecified. In particular, misspecification can induce $m(x)\not\in\mathcal{M}_T$ for some $x$, but still $\widehat{m}(x)\in \M_T$ in this case.\footnote{\label{foot:misspecif} In theory, one could exploit the constraint that $m(X)\in\mathcal{M}_T$ almost surely (a.s.) to derive a specification test for the FE logit model. However, we would recommend using only the conditional moment restrictions $E[\deriv{\ell_c}{\beta}(Y|X,\beta_0)|X]=0$. This leads to a much simpler test and it is unclear whether adding the former constraint would lead to large power gains.} Second, we can define in an exact similar way estimators of the sharp bounds of parameters related to the heterogeneity of treatment effects described in Subsection \ref{sub:het_TE}.


\subsubsection{Asymptotic properties of the estimated bounds} 
\label{sub:asymptotic_properties}

We first derive the asymptotic distribution on the estimated bounds $(\widehat{\underline{\delta}}, \widehat{\overline{\delta}})$. To this end, we rely on Assumptions \ref{hyp:model}-\ref{hyp:fct_obj_simple} and the assumptions below. We refer to Theorem \ref{thm:consistency} in the Online Appendix for a consistency result on $(\widehat{\underline{\delta}}, \widehat{\overline{\delta}})$ under weaker conditions than those below. To state some regularity conditions below, we let (rearranging terms here) $X=(X^{c \, \prime},X^{d \, \prime})'$ with $X^c$ a vector of $p_cT$ continuous regressors and $X^d$ a vector of $(p-p_c)T$ discrete regressors.

\begin{hyp}
	\
\begin{enumerate}
	\item \label{assn:iid} The variables $(X_i, \alpha_i, \eps_{i1},...,\eps_{iT})$  are i.i.d across $i$.
	\item \label{assn:Xborne} $\Supp(X)$ is a compact set and $\beta_0$ belongs to the interior of $B$, a convex compact set.
\end{enumerate}
\label{hyp:iid_bdX}
\end{hyp}

\begin{defin}[Regularity Condition $Reg(j)$]\
	A function $f$ from $\Supp(X)\times \mathbb{R}^p$ to $\mathbb{R}$ is $Reg(j)$ if $\beta\in \mathbb{R}^{p}\mapsto f(x,\beta)$ admits continuous derivatives of order $j$ for any $x\in \Supp(X)$ and
	$x\in \Supp(X)\mapsto \partial^{|k|} f(x,\beta)/(\partial^{k_1} \beta_{1}... \partial^{k_p} \beta_{p})$ is continuous for any $\beta\in \mathbb{R}^{p}$ and any $(k_1,...,k_p)\in \mathbb{N}^p$ such that $|k|:=\sum_{r=1}^p k_r\leq j$.
\end{defin}

\newcounter{counterhypprod}
\setcounter{counterhypprod}{\value{hyp}}
\begin{hyp}
	The functions $v$, $\lambda_{0},...,$ and $\lambda_{T}$ are $Reg(2)$. Moreover, there exist $k$, $a$ and $\rho$ such that $\lambda_{T+1}(x,\beta)=a(\beta_k)\rho(x,\beta)$, $\rho$ is $Reg(2)$, $a$ is twice continuously differentiable and either $a(\beta_{0k}) = 0$ or $P\left(\rho(X,\beta_0)=0\right)=0$.
	\label{hyp:lambda}
\end{hyp}

\begin{hyp}~
\begin{enumerate}
	\item The support of $X^d$ is finite and $X^c|X^d=x^d$ admits a density $f_{X^c|X^d=x^d}$ with respect to the Lebesgue measure on $\R^{p_cT}$. $f_{X^c|X^d=x^d}$ is $C^{\ell+2}$ and bounded away from $0$ on $\Supp(X^c|X^d=x^d)$, which is convex. Also, $\ell \geq p_cT/2$.\footnote{If $p_c=p$, this condition and Assumption \ref{hyp:np_for_consistency}.\ref{assn:smoothgamma} should be understood replacing $X^d$ by a constant variable. If $p_c=0$, all the conditions on $X^c|X^d=x^d$ should be removed.} 
	\label{assn:densityX}
	\item $x^c \mapsto \gamma_0(x^c,x^d)$ is $C^{\ell+1}$ on $\Supp(X^c|X^d=x^d)$ for all $x_d$. \label{assn:smoothgamma} 
	\item $K$ is a density  on $\R^{pT}$ with compact support bounded away from 0 in a neighborhood of 0. $K$ is $C^{\ell+2}$ on $\R^{pT}$. \label{assn:K}
		\item $h_n=C_h n^{-\xi_h}, c_n=C_c n^{-\xi_c}$ with $0<\xi_c<\xi_h(\ell+1)$, $\frac{1}{4(\ell+1)}<\xi_h<\frac{1}{p_cT+2(\ell+1)}$ and $C_c,C_h>0$. \label{assn:hn_cn}
	\end{enumerate}
	\label{hyp:np_for_consistency}
\end{hyp}

\begin{hyp}
	\label{hyp:cushion_m} Either $|\Supp(\alpha|X=x)|>\intpart{T/2}$ for all $x \in \Supp(X)$, or the function $x\mapsto |\Supp(\alpha|X=x)|$ is constant on $\Supp(X)$.
\end{hyp}

Note that the decomposition of $\lambda_{T+1}(x,\beta)$ in Assumption \ref{hyp:lambda} may not be unique. Also, this assumption holds for the three average parameters introduced in Section \ref{sub:ident_general}, under regularity conditions on $X$:

\setcounter{example}{0}
\begin{example}[AME]
	In view of \eqref{eq:other_expr_AME}, $\lambda_{T+1}(x,\beta)=-\beta_k \prod_{s\neq t}\left(\exp[(x_s-x_t)'\beta]-1\right)$. Let $a(\beta_{k}) := \beta_{k}$ and $\rho(x,\beta) =-\prod_{s \neq t}\left(\exp\left[(x_s-x_t)'\beta\right]-1\right)$. Assumption \ref{hyp:lambda} then holds if either $\beta_{0k}=0$ or $(X_s - X_t)'\beta_0 \neq 0$ a.s. for all $s\ne t$.
\end{example}

\begin{example}[ATE]
	Recall that $\delta_0$ is defined by \eqref{eq:part_ATE}, as the remaining part of the ATE is identified and $\lambda_{T+1}(x,\beta)=-(2x_{tk} - 1) \prod_{s=1}^T\left(\exp[(x_s-x_t^{(1-x_{tk})})'\beta]-1\right)$. Let $a(\beta_k)=1-\exp(\beta_k)$ and $\rho(x,\beta)=\exp\left((x_{tk}-1)\beta_k\right)\prod_{s \neq t} \left(\exp[(x_s -x_t^{(1 - x_{tk})})'\beta]-1 \right)$. Assumption \ref{hyp:lambda} then holds if either $\beta_{0k}=0$ or $(X_s-X_{t}^{(1-X_{tk})})'\beta_0 \neq 0$ a.s.  for all $s\ne t$.
\end{example}

\begin{example}[ASF]
In view of \eqref{eq:ASF}, $\lambda_{T+1}(x,\beta)=\prod_{s=1}^T\left(\exp[(x_s-\tilde{x}_t)'\beta]-1\right)$. Let $a(\beta_k):=1$ and $\rho(x,\beta)=\prod_{s=1}^T\left(\exp[(x_s-\tilde{x}_t)'\beta]-1\right)$. Assumption \ref{hyp:lambda} then holds if the distribution of $X_s'\beta_0$ is continuous for all $s\in\{1,...,T\}$.
\end{example}

Assumption \ref{hyp:np_for_consistency} is a standard regularity condition when estimating, as here, parameters involving first-step nonparametric estimators. It ensures that $\widehat{\gamma}$  converges uniformly  to $\gamma_0$ over the support of $X$ at a rate of at least $\eta_n:= (\ln n/(n h_n^{p_cT}))^{1/2} + h_n^{\ell+1}=o(n^{-1/4})$. We also impose restrictions on the $c_n$ used in our constrained estimation of $m(x)$. In particular, the restrictions on $\xi_h$ and $\xi_c$, when combined with Assumption \ref{hyp:cushion_m}, ensure that $\widehat{I}(x)$ defined in \eqref{eq:I_hat} converges to $I(x):=\max\left\{t\in \{1,...,T\}: \;\forall s\le t,\; \underline{H}_s(m_{\rightarrow s}(x))\times \overline{H}_s(m_{\rightarrow s}(x)) > 0\right\}$. The restrictions on $\xi_h$ and $\xi_c$ also ensure that the derivatives of $\widehat{\gamma}(x)$ remain bounded in probability. 

\medskip
Assumption \ref{hyp:cushion_m} is less standard than the other conditions. It restricts the number of support points of $\alpha$ given $X=x$. This number should either be large enough, in which case it may vary with $x$, or remain constant as a function of $x$. We impose this condition  because $\underline{q}_T$ and $\overline{q}_T$ are not regular everywhere for $T\geq 3$: whereas they  are infinitely differentiable on the relative interior of the moment space $\M_T$, they may not even be directionally differentiable on the boundary of $\M_T$.\footnote{See \cite{DR17} for a proof of the first statement. Regarding the second, one can show that, e.g., $m_1\mapsto \underline{q}_3(m_0,m_1,m_2,m_3)$ is not differentiable at $m=(1,m_1, m_1^2,m_1^3)$.}

\medskip
Before presenting the asymptotic distribution of $\left(\widehat{\underline{\delta}}, \widehat{\overline{\delta}}\right)$, we introduce additional notation. Let $\phi=\mathcal{I}_0^{-1}\partial \ell_c(Y|X,\beta_0)/\partial \beta$ denote the influence function of $\widehat{\beta}$ and $\Gamma = (\ind{S=0},...,$ $\ind{S=T})'$. In general, the estimated bounds are asymptotically normal. Their asymptotic variance is $\Sigma:=V[\underline{\psi}, \overline{\psi}]$, with
\begin{align*}
\underline{\psi} & := \underline{h}(X, S) -  \underline{\delta} + \underline{v}_\beta' \phi + \underline{v}_\gamma(X,S)' [\Gamma - \gamma_0(X)],\\
\overline{\psi} & := \overline{h}(X, S) -  \overline{\delta} + \overline{v}_\beta' \phi + \overline{v}_\gamma(X,S)' [\Gamma - \gamma_0(X)],
\end{align*}
where the vectors $\underline{v}_\beta$ and $\underline{v}_\gamma(x,s)$ (and similarly $\overline{v}_\beta$ and $\overline{v}_\gamma(x,s)$) are, respectively, the gradient of $\E(\underline{h}(X, S))$ with respect to $\beta$ and the gradient of $\underline{h}(x, s)$ with respect to $\gamma(x)$. The exact expressions of these four vectors are given in Online Appendix \ref{sub:proof_prop_asnormality_old}. The term $\underline{v}_\beta' \phi$ captures the effect of estimating $\beta_0$, whereas $\underline{v}_\gamma(X,S)' [\Gamma - \gamma_0(X)]$ captures the effect of estimating the nonparametric function $\gamma_0$.

\medskip
In some cases, the estimated bounds are not asymptotically normal. Then, their asymptotic distribution depends on $\Omega$, the variance matrix of $(r(X,S,\beta_0),\phi)$ and on the functions $\underline{K}(u):=E[\underline{k}(X,S,u)]$ and $\overline{K}(u):=E[\overline{k}(X,S,u)]$, with $u\in \mathbb{R}^p$ and where
\begin{align*}
	\underline{k}(x,s,u)=& E\left(\Deriv{r}{\beta}(X_1,S_1,\beta_0)\right)'u+c_0(x) \min\bigg(\underline{q}_T(m(x)) \; \Deriv{\lambda_{T+1}}{\beta}(x,\beta_0)'u, \\
	& \overline{q}_T(m(x)) \Deriv{\lambda_{T+1}}{\beta}(x,\beta_0)'u\bigg).
\end{align*}
$\overline{k}$ is defined similarly, with the minimum replaced by the maximum.

\begin{thm}\label{thm:as_normality}
	Suppose that Assumptions \ref{hyp:model}-\ref{hyp:cushion_m} hold and let $(\underline{N},\overline{N})\sim \mathcal{N}(0,\Sigma)$ and $(N_1,N_2)\sim \mathcal{N}(0, \Omega)$. Then:
$$\sqrt{n} \left(\widehat{\underline{\delta}}  - \underline{\delta}, \widehat{\overline{\delta}}  - \overline{\delta}\right) \convD \left|\begin{array}{ll} (\underline{N},\overline{N}) & \text{if } a(\beta_{0k})\ne 0, \\
\left(N_1+\underline{K}(N_2),N_1+\overline{K}(N_2)\right) & \text{otherwise.} \end{array}
\right. $$	
\end{thm}

The proof of Theorem \ref{thm:as_normality}, as the proofs of other estimation results, is in Section \ref{sec:proofs_of_asymptotic_results} of the Online Appendix.  The main difficulty in this proof is to deal with the nonlinear terms $\underline{q}_T(\widehat{m}_T)$ and $\overline{q}_T(\widehat{m}_T)$, given that $\underline{q}_T$ and $\overline{q}_T$ may not be differentiable. The construction of $\widehat{m}_T$ above and Assumption \ref{hyp:cushion_m} are key for dealing with this issue.

\medskip
If $a(\beta_{0k})\ne 0$ (equivalently, $\beta_{0k}\ne 0$ for the AME and the ATE), the limit distribution is normal. But this is generally not the case when $a(\beta_{0k})=0$, as the functions $\underline{K}$ and $\overline{K}$ are not linear. We still get asymptotic normality of the AME and the ATE, however, if the whole vector $\beta_0$ is equal to 0: then, the functions $\underline{K}$ and $\overline{K}$ are equal and linear. Note that because $a(\beta_{0k})=1$ for the ASF, the estimated bounds are always asymptotically normal.


\subsubsection{Inference on $\delta_0$} 
\label{ssub:inference_on_delta}

With Theorem \ref{thm:as_normality} at hand, we can construct confidence intervals on $\delta_0$ that are asymptotically valid, whether or not the estimated bounds are asymptotically normal. For simplicity, we focus hereafter on the AME, the ATE and the ASF. First, we estimate $\Sigma$ by $\widehat{\Sigma}=\frac{1}{n}\sum_{i=1}^n(\widehat{\underline{\psi}}_i,\widehat{\overline{\psi}}_i)' (\widehat{\underline{\psi}}_i,\widehat{\overline{\psi}}_i)$, where
$$\widehat{\underline{\psi}}_i := \widehat{\underline{h}}(X_i, S_i) -  \widehat{\underline{\delta}} + \widehat{\underline{v}}_\beta' \widehat{\phi}_i + \widehat{\underline{v}}_\gamma' [\Gamma_i - \widehat{\gamma}(X_i)],$$
with $\widehat{\phi}_i:=-\left[\frac{1}{n}\sum_{j=1}^n \deriv{{}^2 \ell_c}{\beta \partial \beta'}(Y_j|X_j;\widehat{\beta})\right]^{-1}\deriv{\ell_c}{\beta}(Y_i|X_i;\widehat{\beta})$ and $\widehat{\underline{v}}_\beta$ and $\widehat{\underline{v}}_\gamma$ are estimators of $\underline{v}_\beta$ and $\underline{v}_\gamma$ defined in Section \ref{sub:proposition_ref_prop_ci1} of the Online Appendix. $\widehat{\overline{\psi}}_i$ is defined similarly.

\medskip
Next, let $\varphi_\alpha=1$ for the ASF and otherwise, let $\varphi_\alpha=1$ denote a consistent test of asymptotic level $\alpha$ of $\beta_{0k}=0$. For instance we can use $\varphi_\alpha=1$ if a Wald test rejects $\beta_{0k}=0$, $\varphi_\alpha=0$ otherwise. Following \cite{imbens2004confidence}, let $c_\alpha$ denote the unique solution to
\begin{equation}
	\Phi\left(c_\alpha+\frac{n^{1/2}\left(\widehat{\overline{\delta}}-\widehat{\underline{\delta}}\right)}{\max\left(\widehat{\Sigma}^{1/2}_{11}, \widehat{\Sigma}^{1/2}_{22}\right)}\right) - \Phi(-c_\alpha)=1-\alpha,	
	\label{eq:def_c_alpha}
\end{equation}
with $\Phi$ the cdf of a standard normal distribution and $\Sigma_{ij}$ the $(i,j)$ term of $\Sigma$.  Then, we define $\CI{1}$ as
$$\CI{1}:=\left|\begin{array}{ll}
	\left[\widehat{\underline{\delta}} - c_\alpha (\widehat{\Sigma}_{11}/n)^{1/2}, \; \widehat{\overline{\delta}} + c_\alpha (\widehat{\Sigma}_{22}/n)^{1/2} \right] & \quad \text{ if } \varphi_\alpha=1, \\[2mm]
	\left[\min\left(0, \widehat{\underline{\delta}} - c_\alpha (\widehat{\Sigma}_{11}/n)^{1/2}\right), \; \max\left(0, \widehat{\overline{\delta}} + c_\alpha (\widehat{\Sigma}_{22}/n)^{1/2}\right)\right] & \quad \text{ if } \varphi_\alpha=0.
\end{array}\right.$$
The following proposition shows that $\CI{1}$ is pointwise valid as $n\to\infty$.

\begin{prop}
	\label{prop:CI1}
	Suppose that $\delta_0$ is the AME, the ATE or the ASF, Assumptions \ref{hyp:model}-\ref{hyp:cushion_m} hold and $\min(\Sigma_{11},\Sigma_{22})>0$. Then $\liminf_n \inf_{\delta_0\in[\underline{\delta},\overline{\delta}]} P(\delta_0\in\CI{1})\geq 1-\alpha$, with equality if  $a(\beta_{0k})\neq 0$.
\end{prop}

Intuitively, $\CI{1}$ asymptotically reaches its nominal level when $a(\beta_{0k})\neq 0$ because it includes $[\widehat{\underline{\delta}} - c_\alpha
(\widehat{\Sigma}_{11}/n)^{1/2}, \; \widehat{\overline{\delta}} + c_\alpha (\widehat{\Sigma}_{22}/n)^{1/2}]$, and the latter interval has asymptotic coverage $1-\alpha$, by Theorem \ref{thm:as_normality}. 
When $a(\beta_{0k})=0$, which applies only to the AME and the ATE, 
the asymptotic coverage of $\CI{1}$ is also at least $1-\alpha$ because $\delta_0=0\in\CI{1}$ as soon as $\varphi_\alpha=0$. Finally,  for simplicity we did not consider other parameters, for which $\delta_0$ may be unknown when $a(\beta_{0k})=0$, but we can still adapt $\CI{1}$ to such cases. We simply have to replace (i) $c_\alpha$ and $\varphi_\alpha$ by $c_{\alpha_1}$ and $\varphi_{\alpha_1}$ for some $\alpha_1\in (0,\alpha)$; (ii) $0$ in $\CI{1}$ by the lower and upper bounds of confidence intervals with nominal level $\alpha-\alpha_1$ (for some $\alpha_1\in (0,\alpha)$) on $E[r(X,S,\beta_0)]$, since $\delta_0=E[r(X,S,\beta_0)]$ when $a(\beta_{0k})=0$.

\medskip
The interval $\CI{1}$ may have a uniform coverage over an appropriate set of data generating processes (DGPs). 
Establishing this formally would however require to establish the uniform convergence in distribution of $(\widehat{\underline{\delta}}, \widehat{\overline{\delta}})$, a multistep estimator with a nonparametric first step. We leave this issue for future research. 



\subsection{Estimation and inference based on the outer bounds} 
\label{sec:simple_method}

We now turn to the estimation of the outer bounds, and inference on $\delta_0$ based on these outer bounds. First, by Theorem \ref{thm:ident2} and the law of iterated expectations, these outer bounds are $\tilde{\delta} \pm \overline{b}$, with $\tilde{\delta} =  E[p(X, S,\beta_0)]$, and\footnote{Again, in the case of the ATE, we must add $(2x_{kt}-1)y_t$ to $p(x,s,\beta)$.}
\begin{align*}
	p(x,s,\beta) & :=\sum_{t=0}^s \left(\lambda_t(x,\beta)+b^*_{t,T}\lambda_{T+1}(x,\beta)\right)\binom{T-t}{s-t} \frac{\exp(sv(x,\beta))}{C_s(x,\beta)} , \\
	\overline{b} & := \frac{1}{2\times 4^T} E\left[\left|\lambda_{T+1}(X,\beta_0)\right| \binom{T}{S}\frac{\exp(Sv(X,\beta_0))}{C_S(X,\beta_0)} \right].	
\end{align*}
We estimate $\tilde{\delta}$ and $\overline{b}$ by plug-in: $\widehat{\tilde{\delta}}  =  \sum_{i=1}^n p(X_i,S_i,\widehat{\beta})/n$ and
$$\widehat{\overline{b}} = \frac{1}{2\times 4^T} \frac{1}{n}\sum_{i=1}^n \left|\lambda_{T+1}(X_i,\widehat{\beta})\right| \binom{T}{S_i}\frac{\exp(S_i v(X_{i},\widehat{\beta}))}{C_{S_i}(X_i,\widehat{\beta})}.$$
Lemma \ref{lem:as_nor_triv} in the Online Appendix implies that $\widehat{\tilde{\delta}} - \widehat{\overline{b}}$ (resp. $\widehat{\tilde{\delta}} + \widehat{\overline{b}}$) is a consistent estimator of the outer bound $\tilde{\delta} - \overline{b}$ (resp. $\tilde{\delta} + \overline{b}$).

\medskip
To define confidence intervals on $\delta_0$ based on $\widehat{\tilde{\delta}}$ and $\widehat{\overline{b}}$, we introduce additional notation.  Let
\begin{align*}
	\psi_i = & p(X_i,S_i,\beta_0) - \E\left[p(X,S,\beta_0)\right] + \E\left[\Deriv{p}{\beta}(X,S,\beta_0)\right]'\phi_i, \\
	\widehat{\psi}_i = & p(X_i,S_i,\widehat{\beta}) - \frac{1}{n} \sum_{i=1}^n p(X_i,S_i,\widehat{\beta})   + \left(\frac{1}{n} \sum_{i=1}^n \Deriv{p}{\beta}(X_i,S_i,\widehat{\beta})\right)' \widehat{\phi}_i.
\end{align*}
Then, we define $\sigma^2=\V(\psi)$  and $\widehat{\sigma}^2 = \sum_{i=1}^n \widehat{\psi}_i^2/n$. The first confidence interval we consider is
$$\CI{2} = \left[\widehat{\tilde{\delta}} \pm q_\alpha\left(\frac{n^{1/2}\widehat{\overline{b}}}{\widehat{\sigma}}\right) \frac{\widehat{\sigma}}{n^{1/2}}\right],$$
where $q_\alpha(b)$ denotes the quantile of order $1-\alpha$ of a $|\mathcal{N}(b,1)|$. The intuition for the asymptotic validity of this confidence interval is as follows. We prove in Lemma \ref{lem:as_nor_triv} in the Online Appendix that
\begin{equation}
	n^{1/2}\left(\widehat{\tilde{\delta}} - \tilde{\delta}\right) \convL \mathcal{N}(0,\sigma^2)	
	\label{eq:asnor_triv}
\end{equation}
and $\widehat{\sigma}\convP \sigma$. For simplicity, let us assume that $\widehat{\sigma}=\sigma$, $\widehat{\overline{b}}=\overline{b}$ and the asymptotic approximation \eqref{eq:asnor_triv}  is exact. Then
\begin{equation}
	n^{1/2}\frac{\widehat{\tilde{\delta}} - \delta_0}{\widehat{\sigma}} \sim \mathcal{N}\left(n^{1/2}\frac{\tilde{\delta} - \delta_0}{\widehat{\sigma}}, 1\right).
	\label{eq:nor_triv}
\end{equation}
It is not difficult to show that $b \mapsto q_\alpha(b)$ is symmetric and increasing on $[0,\infty)$. Then, because $|\tilde{\delta} - \delta_0| \le \overline{b}$, we have
\begin{equation}
	P\left(n^{1/2}\left|\frac{\widehat{\tilde{\delta}} - \delta_0}{\widehat{\sigma}}\right|\leq q_\alpha\left(\frac{n^{1/2}\widehat{\overline{b}}}{\widehat{\sigma}}\right)\right)\geq 1- \alpha.	
	\label{eq:ineg_for_CI}
\end{equation}

Theorem \ref{thm:CI_triv} below shows that \eqref{eq:ineg_for_CI} holds asymptotically even if $\widehat{\sigma}\ne \sigma$, $\widehat{\overline{b}}\ne \overline{b}$ and \eqref{eq:asnor_triv} is not exact, as soon as $|\delta_0 - \tilde{\delta}| < \overline{b}$. The only difference between $\CI{2}$ and a standard confidence interval is that because of the possible bias, we consider $q_\alpha\left(n^{1/2}\widehat{\overline{b}}/\widehat{\sigma}\right)$ instead of the usual normal quantile $q_\alpha\left(0\right)$. This difference is important however: it implies that $\CI{2}$ converges to the outer set $[\tilde{\delta} \pm \overline{b}]$, rather than to $\{\tilde{\delta}\}$.

\begin{thm}
Suppose that Assumptions \ref{hyp:model}-\ref{hyp:iid_bdX} hold, $v$, $\lambda_{0}, \, ...,\lambda_{T+1}$ are $Reg(1)$, $\sigma^2>0$  and either $\overline{b}=0$ or $\abs{\delta_0-\tilde{\delta}}\neq\overline{b}$. Then:
		$$\liminf_{n\to\infty} P\left(\delta_0\in \CI{2}\right) \geq 1-\alpha.$$
\label{thm:CI_triv}
\end{thm}

\vspace{-1cm}
Hence, $\CI{2}$ is pointwise asymptotically conservative under the condition that $\overline{b}=0$ or $|\delta_0-\tilde{\delta}|<\overline{b}$. This condition  is very weak, as the following lemma shows:

\begin{lem}\label{lem:cond_CI2}
	Suppose that Assumptions \ref{hyp:model}-\ref{hyp:iid_bdX} hold. Then, $|\delta_0-\tilde{\delta}|=\overline{b}>0$ implies \
\begin{align}
	& P(\lambda_{T+1}(X,\beta_0)\ne 0)>0 \text{ and either } P\left(\Supp(U|X)\subseteq \mathcal{R}_{T,X} | \lambda_{T+1}(X,\beta_0)\ne 0\right)=1 \notag \\ & \text{or } P\left(\Supp(U|X)\subseteq \mathcal{R}'_{T,X} | \lambda_{T+1}(X,\beta_0)\ne 0\right)=1, 
	\label{eq:eg_R_Rsup}
\end{align}
where $\mathcal{R}_{T,x}$ is defined as
\begin{equation}
\mathcal{R}_{T,x}=\left|\begin{array}{ccl} \arg\max_{u\in[0,1]} \mathbb{T}_{T+1}(u) & \text{if} & \lambda_{T+1}(x,\beta_0)>0 \\
\arg\min_{u\in[0,1]} \mathbb{T}_{T+1}(u) & \text{if} & \lambda_{T+1}(x,\beta_0)<0
\end{array}
\right.	
\label{eq:def_set_R}	
\end{equation}
and $\mathcal{R}'_{T,X}$ is defined similarly, simply switching the min and the max in \eqref{eq:def_set_R}. 
\end{lem}

Condition \eqref{eq:eg_R_Rsup} strongly restricts the support of $\alpha|X$. First, because $\mathbb{T}_{T+1}$ has at most $\intpart{T/2}+1$ maxima and minima, one must have $|\Supp(\alpha|X)|\le \intpart{T/2}+1$. Second, \eqref{eq:eg_R_Rsup} imposes that the support points of $\alpha$ change discontinuously around any $x_0$ for which $\lambda_{T+1}(x_0,\beta_0)=0$ and $\deriv{\lambda_{T+1}(x_0,\beta_0)}{x}\neq 0$. In any case, our simulations below suggest that $\CI{2}$ still has good coverage when \eqref{eq:eg_R_Rsup} holds. Another potential issue with $\CI{2}$ is that because it does not account for the variability of $\widehat{\overline{b}}$, it may not be asymptotically uniformly valid for the AME and ATE for sequences of DGPs such that $\beta_{0k}$ tends to 0 at the rate $n^{1/2}$. We consider in Appendix \ref{app:uniform_CI} another confidence interval that is uniformly valid on a set of DGPs allowing for such sequences of $\beta_{0k}$. 


	
\section{Monte Carlo simulations} 
\label{sec:monte_carlo_simulations}

We now study the estimators of the bounds on the AME, and inference based on them, through simulations.\footnote{For an application of our methodology to a real dataset with both continuous and discrete regressors, see the \href{https://github.com/cgaillac/MarginalFElogit/blob/master/MarginalFElogit.pdf}{\texttt{documentation}} of our R package \texttt{MarginalFELogit}.} We first compare our two methods. Then, we compare our approach with FE linear probability models. Finally, we study measures of heterogeneous effects.

\subsection{Comparison of the two inference methods} 
\label{sub:comparison_of_the_two_inference_methods}

We first compare  the finite sample performances of our two methods. We consider $T\in\{2,3\}$ and $n\in\{250; 500;1,000\}$ and four DGPs. In all of them, we assume that $(X_1,...,X_T)$ are i.i.d., with $X_t\in \R$, uniformly distributed on $[-1/2,1/2]$ and $\beta_0=1$. We also suppose that $\alpha = - X_T'\beta_0 +\eta$. Then, we consider different distributions for $\eta|X$. In DGP1, we let $\eta=0$. By Theorem \ref{thm:ident1}, $\delta_0$ is point identified for all $T\ge 2$ in this case. In DGP2, we let $\eta|X\sim \mathcal{N}(0,1)$. Again by Theorem \ref{thm:ident1}, $\delta_0$ is partially identified in this case for all $T\ge 2$. We also consider DGP$3_T$ where $\eta|X$ is uniformly distributed over $\Lambda^{-1}(\mathcal{R}_{T,X})$, where $\mathcal{R}_{T,X}$ is defined in \eqref{eq:def_set_R}. Note that we index DGP3 by $T$ because contrary to DGP1 and DGP2, this DGP actually varies with $T$. Table \ref{tab:prop_DGP} shows the true parameter, the sharp bounds and the outer bounds $\underline{\delta}^o := \tilde{\delta}-\overline{b}$ and $\overline{\delta}^o:= \tilde{\delta}+\overline{b}$ for $T\in\{2,3\}$. In the partially identified case of DGP2, the sharp bounds are very informative, even with $T=2$. The outer bounds are also very informative in all DGPs.

\begin{table}[H]
	\begin{center}
	\begin{tabular}{lc|cc|cc}
		\toprule
		 & True & \multicolumn{2}{c|}{$\Delta=[\underline{\delta},\, \overline{\delta}]$} & \multicolumn{2}{c}{$\Delta^o=[\underline{\delta}^o,\, \overline{\delta}^o]$} \\
		DGP & parameter & $T=2$ & $T=3$ &  $T=2$ & $T=3$ \\ \midrule
		 1 & 0.25 & [0.25,\, 0.25] & [0.25,\, 0.25] & $[0.2398,\, 0.2602]$ & $[0.2497, 0.2515]$ \\
		 2 & 0.2066 & [0.2006,\, 0.2124] & [0.2059,\, 0.2069] & $[0.1971,\, 0.2177]$ & $[0.2058,\, 0.2076]$ \\
		 3$_2$ & 0.1875 & [0.1875,\, 0.1875] & --- & $[0.1692,\, 0.1875]$ & --- \\
		 3$_3$ & 0.1667 & --- & [0.1652,\, 0.1667] & --- & $[0.1647,\, 0.1667]$ \\
		 \bottomrule
		\multicolumn{6}{p{440pt}}{{\footnotesize Notes: $(\underline{\delta}^o, \overline{\delta}^o)$ denote the outer bounds, $\underline{\delta}^o := \tilde{\delta}-\overline{b}$ and $\overline{\delta}^o:= \tilde{\delta}+\overline{b}$. The sharp and outer bounds are computed in closed form for DGP1 and by simulations (with a sample of size $10^6$) in other cases.}}
	\end{tabular}
	\end{center}
	\caption{Value of the true parameter and its bounds in the DGPs}
	\label{tab:prop_DGP}
\end{table}

For each of the DGPs above, we consider and perform 3,000 simulations for each such $(T,n)$. We then compute the estimators of the sharp bounds, $\widehat{\underline{\delta}}$ and $\widehat{\overline{\delta}}$, those of the outer bounds, $\widehat{\underline{\delta}^o} :=\widehat{\tilde{\delta}} - \widehat{\overline{b}}$ and $\widehat{\overline{\delta}^o} :=\widehat{\tilde{\delta}} + \widehat{\overline{b}}$, and $\text{CI}^1_{0.95}$ and $\text{CI}^2_{0.95}$. To estimate nonparametrically $\gamma_0$, we use local linear estimators with a Gaussian product kernel. We use data-driven bandwidths $h_n$ and thresholds $c_n$, on which further details are given in Section \ref{sub:details_on_the_simulations} of the Online Appendix. 

\medskip
Table \ref{tab:prop_est} displays the properties of the estimators underlying the two methods. The estimators of the bounds appear to have a small bias in all cases, except perhaps with DGP3$_2$. With this DGP, the distribution of $\eta|X=x$ does not vary in a smooth way with $x$: $\eta=\Lambda^{-1}(1/4)$ when $x_1 \le x_2$ while $\eta=\Lambda^{-1}(3/4)$ otherwise. As a result, the regularity condition we impose on $\gamma_0$ (see Assumption \ref{hyp:np_for_consistency}.\ref{assn:smoothgamma}) is actually violated, which could explain the larger bias.

\begin{table}[H]
	\begin{center}
		{\small
			\begin{tabular}{ccc|cccc|cccc}
				\toprule		
				& & & \multicolumn{4}{c|}{First method} &  \multicolumn{4}{c}{Second method}  \\
				DGP & T & n & $\sigma(\widehat{\underline{\delta}})$ & Bias$(\widehat{\underline{\delta}})$ & $\sigma(\widehat{\overline{\delta}})$ & Bias$(\widehat{\overline{\delta}})$ & $\sigma(\widehat{\underline{\delta}^o})$ & Bias$(\widehat{\underline{\delta}^o})$ & $\sigma(\widehat{\overline{\delta}^o})$ & Bias$(\widehat{\overline{\delta}^o})$\\ 
				\midrule
1 &2 &250 &0.113 &-0.001 &0.116 &0.003 &0.107 &0$^*$ &0.126 &0.005 \\
 & &500 &0.081 &0$^*$ &0.082 &0.002 &0.076 &0.001 &0.089 &0.004 \\
 & &1000 &0.056 &0.002 &0.057 &0.003 &0.053 &0.002 &0.062 &0.004 \\
 &3 &250 &0.079 &-0.002 &0.079 &-0.002 &0.080 &0$^*$ &0.082 &0$^*$ \\
 & &500 &0.056 &0.001 &0.056 &0.001 &0.056 &0.002 &0.057 &0.002 \\
 & &1000 &0.041 &-0.001 &0.041 &-0.001 &0.041 &0$^*$ &0.042 &0$^*$ \\ \midrule
2 &2 &250 &0.103 &0.002 &0.115 &0.005 &0.099 &0$^*$ &0.123 &0.008 \\
 & &500 &0.070 &0.002 &0.078 &0.004 &0.068 &0.001 &0.083 &0.005 \\
 & &1000 &0.049 &0.001 &0.054 &0.002 &0.047 &0$^*$ &0.058 &0.002 \\
 &3 &250 &0.071 &0.001 &0.072 &0.001 &0.071 &0.001 &0.073 &0.002 \\
 & &500 &0.050 &0$^*$ &0.051 &0$^*$ &0.050 &0$^*$ &0.051 &0$^*$ \\
 & &1000 &0.036 &0$^*$ &0.036 &0$^*$ &0.036 &0$^*$ &0.037 &0$^*$ \\ \midrule
3$_2$ &2 &250 &0.126 &0.016 &0.130 &0.021 &0.099 &0.004 &0.119 &0.01 \\
 & &500 &0.083 &0.008 &0.086 &0.013 &0.067 &0.001 &0.080 &0.004 \\
 & &1000 &0.060 &0.006 &0.062 &0.011 &0.048 &0.001 &0.058 &0.003 \\
3$_3$ &3 &250 &0.068 &0.006 &0.069 &0.005 &0.064 &0.001 &0.068 &0.002 \\
 & &500 &0.045 &0.003 &0.046 &0.003 &0.043 &-0.001 &0.045 &0$^*$ \\
 & &1000 &0.032 &0.004 &0.033 &0.004 &0.031 &0.001 &0.033 &0.001 \\
\bottomrule
\multicolumn{11}{p{440pt}}{{\footnotesize Notes: Results based on 3,000 simulations. $^*$: on absolute values, smaller than $0.0005$.}}
		\end{tabular}}
	\end{center}
	\caption{Properties of the estimators}
	\label{tab:prop_est}
\end{table}

On the other hand, the estimated outer bounds exhibit very little bias. Also, in DGP3$_2$, the estimated outer bounds are  more precise than the estimated sharp bounds. This already suggests that the corresponding inference may be more precise than that based on the sharp bounds.

\medskip
Table \ref{tab:sims_CI} presents the coverage rate and length of both confidence intervals. The second confidence interval shows a very good coverage, always greater than 94\%. This is the case even with DGP3$_T$, for which Theorem \ref{thm:CI_triv} does not provide any guarantee. Hence, neglecting the variability of $\widehat{\overline{b}}$ does not seem to lead to undercoverage here. The first method also leads to coverage very close to the 95\% nominal level. 

\begin{table}[H]
	\begin{center}
		{\small\begin{tabular}{ccc|cc|cc}
\toprule
& & & \multicolumn{2}{c|}{$\text{CI}^1_{0.95}$} & \multicolumn{2}{c}{$\text{CI}^2_{0.95}$} \\ 
DGP & T & n & Coverage & Avg. length & Coverage & Avg. length \\ 
\midrule
1 &2 &250 &0.95 &0.448 &0.96 &0.462 \\
 & &500 &0.95 &0.314 &0.96 &0.326 \\
 & &1000 &0.94 &0.221 &0.96 &0.232 \\
 &3 &250 &0.95 &0.306 &0.95 &0.316 \\
 & &500 &0.95 &0.217 &0.95 &0.223 \\
 & &1000 &0.94 &0.155 &0.94 &0.158 \\ \midrule
2 &2 &250 &0.94 &0.410 &0.95 &0.421 \\
 & &500 &0.94 &0.286 &0.96 &0.297 \\
 & &1000 &0.95 &0.202 &0.96 &0.211 \\
 &3 &250 &0.95 &0.281 &0.95 &0.284 \\
 & &500 &0.95 &0.197 &0.95 &0.200 \\
 & &1000 &0.94 &0.139 &0.94 &0.141 \\ \midrule
3$_2$ &2 &250 &0.95 &0.483 &0.95 &0.422 \\
 & &500 &0.95 &0.333 &0.95 &0.295 \\
 & &1000 &0.95 &0.234 &0.95 &0.210 \\
3$_3$ &3 &250 &0.96 &0.280 &0.94 &0.249 \\
 & &500 &0.96 &0.185 &0.95 &0.175 \\
 & &1000 &0.95 &0.129 &0.95 &0.124 \\
\bottomrule
\multicolumn{7}{p{300pt}}{{\footnotesize Notes: Results based on 3,000 simulations.}}
		\end{tabular}}
	\end{center}
	\caption{Coverage and average length of CI$^1_{0.95}$ and CI$^2_{0.95}$.}
	\label{tab:sims_CI}
\end{table}

In terms of length of the confidence intervals, the first method delivers slightly shorter intervals for DGP1 and DGP2, but the difference is very small, for all $T$ and $n$. This was not obvious: when $n\to\infty$, the length of $\text{CI}^1_{0.95}$ becomes smaller than that of $\text{CI}^2_{0.95}$ because $\overline{\delta}- \underline{\delta}<\overline{\delta}^o- \underline{\delta}^o$. Hence, at least with our four DGPs, the difference in the length of the sharp and outer identified sets is small compared to the standard errors of $\widehat{\underline{\delta}}$, $\widehat{\overline{\delta}}$ and $\widehat{\tilde{\delta}}$, even with $n=1,000$.\footnote{Additional simulations show that, as expected, $\text{CI}^1_{0.95}$ becomes shorter than $\text{CI}^2_{0.95}$ as $n$ increases. For $n=10^4$ for instance, we observe a gain of around 10\%-15\% for DGP1 and DGP2 with $T=2$. Nonetheless, the gain remains negligible for this sample size with $T=3$.} The second method actually leads to shorter intervals with DGP3$_2$ and DGP3$_3$, especially for small $n$. 


\subsection{Comparison with the linear probability model estimator} 
\label{sub:comparison_with_the_linear_probability_model}

Next, we compare our confidence interval $\text{CI}^2_{0.95}$ with $\text{CI}^{\text{LPM}}_{0.95}$, obtained using the linear probability model (LPM) estimator and the usual standard error accounting for clustering at the individual level. We consider DGP1 but also two incorrectly specified models.\footnote{\label{foot:MC_misspec} Remark that the estimated outer bounds never cross, even if the model is misspecified.} In DGP4, the $(\eps_t)_{t=1,...,T}$ still marginally follow a logistic distribution, so that $\delta_0$ is the same as in DGP1, but these variables are autocorrrelated: their copula is gaussian with a variance matrix $(\Sigma_{s,t})_{1\le s,t\le T}$ satisfying $\Sigma_{s,t}=1/2^{|s-t|}$. In DGP5, we assume instead that the $\eps_t$ are independent over time but $\eps_t\sim \mathcal{N}(0,8/\pi)$. We chose this variance so that again, the AME 
is the same as in DGP1. We consider $T\in\{2,3,4\}$, $n=1,000$ and two possible values of $\beta_0$, namely $\beta_0=1$ and $\beta_0=2$.

\medskip
Table \ref{tab:sims_compar} displays the results. It first shows that in the two misspecified DGPs we consider, our confidence interval still performs very well, with a coverage very close to or above 95\%. Inference based on the linear probability model estimator also works well when $\beta_0=1$, with a coverage always larger than 93\%. However, when $\beta_0=2$, its performance deteriorates, especially for larger $T$. Note that this sensitivity on $\beta_0$ may be more exacerbated with fixed effects. To see this, we consider the same DGP as DGP1 but with $\alpha=0$. Then, the coverage rate of $\text{CI}^{\text{LPM}}_{0.95}$ only decreases from 95\% when $\beta_0=1$ to 89\% when $\beta_0=2$ with $T=4$, as opposed to the decrease from 94\% to 52\% that we observe with DGP1.

\begin{table}[H]
	\begin{center}
		{\small\begin{tabular}{ccc|cc|cc}
				\toprule
				& & & \multicolumn{2}{c|}{$\text{CI}^2_{0.95}$} & \multicolumn{2}{c}{$\text{CI}^{\text{LPM}}_{0.95}$}  \\
				DGP & $\beta_0$ & T & Coverage & Length & Coverage & Length  \\
				\midrule
1 &1 &2 &0.960 &0.232 &0.947 &0.209 \\
 & &3 &0.941 &0.158 &0.942 &0.148 \\
 & &4 &0.948 &0.128 &0.941 &0.120 \\
 &2 &2 &0.977 &0.300 &0.821 &0.198 \\
 & &3 &0.954 &0.179 &0.655 &0.139 \\
 & &4 &0.950 &0.140 &0.520 &0.113 \\ \midrule
4 &1 &2 &0.971 &0.214 &0.943 &0.173 \\
 & &3 &0.956 &0.137 &0.948 &0.127 \\
 & &4 &0.953 &0.113 &0.938 &0.106 \\
 &2 &2 &0.994 &0.379 &0.743 &0.167 \\
 & &3 &0.973 &0.183 &0.574 &0.122 \\
 & &4 &0.954 &0.137 &0.427 &0.101 \\ \midrule
5 & 1 &2 &0.957 &0.232 &0.945 &0.209 \\
 & &3 &0.953 &0.158 &0.948 &0.148 \\
 & &4 &0.951 &0.128 &0.946 &0.120 \\
 & 2 &2 &0.978 &0.304 &0.854 &0.197 \\
 & &3 &0.943 &0.18 &0.742 &0.138 \\
 & &4 &0.941 &0.141 &0.637 &0.112 \\
				\bottomrule	
				\multicolumn{7}{p{290pt}}{{\footnotesize Notes: DGP4 as DGP1, but with autocorrrelated $(\eps_t)_{t=1,...,T}$. DGP5 as DGP1, but with $\eps_t\sim \mathcal{N}(0,8/\pi)$. In the three DGPs, $\delta_0=0.25\beta_0$.
				Results based on 3,000 simulations.}}
		\end{tabular}}
	\end{center}
	\caption{Comparison with the linear probability model}
	\label{tab:sims_compar}
\end{table}


\subsection{Measuring heterogeneous effects} 
\label{sub:heterogeneity_measures}

Finally, we study the estimator of $\delta_W(w)$, one of the parameter considered in Section \ref{sub:het_TE} above. We focus on DGP2, defining $W$ as $W=\ind{|\eta+\nu|>1}$, with $\nu|\eta,X,\eps \sim\mathcal{N}(0,0.3^2)$. In this setup, we have $\delta_W(0)\simeq 0.2317$ and $\delta_W(1)\simeq 0.1575$. We consider below the coverage of the confidence intervals on $\delta_W(0)$ and $\delta_W(1)$ and the length of these confidence intervals. We also consider the test of $\delta_W(1)=\delta_W(0)$, by considering the critical region $\{|T_h|>q^d_\alpha\}$, where the test statistic for homogeneity $T_h$ satisfies
$$T_h= n^{1/2} \frac{\widehat{\tilde{\delta}}_W(1)-\widehat{\tilde{\delta}}_W(0)}{\left(\widehat{\sigma}_0^2 +\widehat{\sigma}_1^2\right)^{1/2}}, \quad q^d_\alpha=q_\alpha\left(n^{1/2}\frac{\widehat{\overline{b}}_0+\widehat{\overline{b}}_1}{\left(\widehat{\sigma}_0^2+\widehat{\sigma}_1^2\right)^{1/2}}\right),$$
where $\widehat{\sigma}_d^2$ is a consistent estimator of the asymptotic variance of $\widehat{\tilde{\delta}}_W(d)$. Reasoning as in Section \ref{sec:simple_method}, we obtain, under the null hypothesis, that $\lim_{n\to\infty} P(|T_h|>q^d_\alpha)\le \alpha$, so this test is valid albeit asymptotically conservative.

\medskip
The results are displayed in Table \ref{tab:MC_het}. As one could expect, larger sample sizes are necessary to detect heterogeneous treatment effects. But already with $n=2,000$, our test of $\delta_W(1)=\delta_W(0)$ has good power when $T=3$.

\begin{table}[H]
	\begin{center}
		{\small\begin{tabular}{cc|cc|cc|c}
				\toprule
				& & \multicolumn{2}{c|}{Coverage (95\% CIs)} & \multicolumn{2}{c|}{Length of the CIs} & Power of the test \\
				T & n & $\delta_W(0)$& $\delta_W(1)$  & $\delta_W(0)$& $\delta_W(1)$ & of $\delta_W(0)=\delta_W(1)$ \\ \midrule
2&1000 &0.960 &0.966 &0.235 &0.170 &0 \\
 &2000 &0.954 &0.967 &0.168 &0.123 &0.009 \\
 &4000 &0.967 &0.976 &0.122 &0.091 &0.204 \\ \midrule
3&1000 &0.947 &0.950 &0.159 &0.113 &0.102 \\
 &2000 &0.944 &0.953 &0.113 &0.080 &0.650 \\
 &4000 &0.949 &0.957 &0.080 &0.056 &1 \\
				\bottomrule
				\multicolumn{7}{p{350pt}}{{\footnotesize Notes: DGP2, with $W=\ind{|\eta+\nu|>1}$ and $\nu|\eta,X,\eps \sim\mathcal{N}(0,0.3^2)$. Results based on 3,000 simulations}}
		\end{tabular}}
	\end{center}
	\caption{Inference on $\delta_W(w)$, $w\in\{0,1\}$.}
	\label{tab:MC_het}
\end{table}



\section{Conclusion} 
\label{sec:conclusion}

We have shown that in the FE logit model, we can partially identify a class of average causal effects including the AME, the ATE and ASF in a simple way, without requiring any optimizaton. Inference based on the outer bounds, in particular, is computationally cheap, does not require any tuning parameter and is shown to fare very well compared to inference based on the sharp bounds. For these reasons, we recommend this approach in practice.

\medskip
The theory is simple here because only raw moments are involved; but similar results hold with other moments, provided that the corresponding functions form a so-called Chebyshev system \citep[See, e.g.,][for a mathematical exposition]{krein_1977}. Results on these systems have already been applied to the optimal design of experiments \citep[see][]{dette1997} and the measure of segregation with small units \citep{DR17}. By drawing further attention on these tools, we hope that this paper will contribute to their use in econometrics.


\linespread{1}\selectfont
\bibliography{biblio}
\newpage
\linespread{1.3}\selectfont
\appendix

\section{Potential pitfalls of using FE linear models} 
\label{sec:pb_FE_linear}

We illustrate here the point made in the introduction that linear two-way fixed effect estimators corresponding to FE linear probability models may be misleading because of the violation of the parallel trends assumption. Suppose that potential outcomes $Y_t(d)$ satisfy
$$Y_{t}(d)=\ind{\alpha+ \ind{t=2} + d+\eps_{t}\geq 0}, \quad t\in\{1,2\},$$
where $\eps_2, \eps_2$ are i.i.d. and follow a logistic distribution. We observe $Y_t:=Y_t(D_t)$, where the binary treatment satisfies $D_1=0$ a.s., whereas $P(D_2=1)=0.5$. Assume further that $\alpha=-0.5+1.5D_2$. On the other hand, the FE linear model yields the simple difference-in-difference estimand:
\begin{align*}
	\delta_{\text{lin}} & = E[Y_2 - Y_1 |D_2 = 1] - E[Y_2 - Y_1 |D_2 = 0] \simeq -0.02.
\end{align*}
In this example, violations of the parallel trends leads to a negative estimand, even though every unit weakly benefits from the treatment ($Y_2(1)\ge Y_2(0)$) and thus both the ATE and the ATT are positive (and equal respectively to 0.13 and 0.07). 


\section{Numerical illustration on the bounds} 
\label{sec:illustr_bds}

We investigate here how $\overline{\delta} - \underline{\delta}$ varies with the distribution of $X$, when $\alpha|X\sim\mathcal{N}(-X_T'\beta_0, 1)$, as in DGP2 of the simulations. Specifically, we first consider $(X_t)_{t=1,...,T}$ i.i.d. uniform on $[-1/2, 1/2]$, then $(X_t)_{t=1,...,T}$ i.i.d. standard normal and finally $X_t \sim \mathcal{N}(0,1)$ following an AR(1): $X_t=(X_{t-1}+\xi_t)/\sqrt{2}$, $t\ge 2$, with the $(\xi_t)_{t=1,...,T}$ i.i.d. standard normal and $X_1\sim \mathcal{N}(0,1)$.

\medskip
We report in Table \ref{tab:bounds} the bounds on the AME for $T\in\{2,3,4\}$ and the ratio of the length of the identified set between $T=2$ and $T=6$. The marginal distributions of the $(X_t)_{t=1,...,T}$ have an influence on the bounds and the rate at which they tend to each other. The case with $X_t$ uniform on $[-1/2,1/2]$ is the most favourable. The normal case leads to larger bounds that also tend to each other more slowly. Dependence also matters: when $X_t\sim\mathcal{N}(0,1)$ but follows an AR(1), the bounds are tighter. This could be expected since in this case the $|(X_t- X_T)'\beta_0|$, and in turn $\lambda_{T+1}(X,\beta_0)$, are on average smaller. Over the four DGP, the ratio of the decrease of the length of the identified set is at least 219. This number is close to $4^4=256$, which would be what we obtain if the length could be written as $C/4^T$ for some $C>0$ (as \eqref{eq:upper_bd_length} could suggest). The ratio is much larger than $256$ for the two uniform cases, which suggests that the rate of decrease is larger than $4^T$ in such cases.

\begin{table}[H]
	\begin{center}
		\begin{tabular}{lcccc}
			\toprule
			Distribution of $X$ & $T=2$ & $T=3$ & $T=4$ & Ratio length\\ \midrule
			i.i.d. unif. on $[-\frac{1}{2},\frac{1}{2}]$ & $[0.2006,\, 0.2124]$ & $[0.2059,\,  0.2069]$ & $[0.2066,\,  0.2067]$ & 22,923 \\
			i.i.d. unif. on $[-1, 1]$ & $[0.1948,\, 0.2177]$ & $[0.2040,\,  0.2078]$ & $[0.2064,\,  0.2069]$ & 1,598\\
			i.i.d. normal & $[0.1878,\,  0.2236]$ & $[0.2001,\,  0.2099]$ & $[0.2054,\,  0.2077]$ & 235\\
			AR(1) normal & $[0.1958,\, 0.2168]$ & $[0.2033,\,  0.2080]$ & $[0.2061,\,  0.2071]$ & 219 \\ \bottomrule
			\multicolumn{5}{p{460pt}}{{\footnotesize Notes: ``Ratio length'' is the ratio of the length of the identified set between $T=2$ and $T=6$. In the four cases, the true parameter is 0.2066.}}
		\end{tabular}
	\end{center}
	\caption{Evolution of the sharp bounds on the AME in DGP2}
	\label{tab:bounds}
\end{table}


\section{Uniformly valid confidence interval based on outer bounds} 
\label{app:uniform_CI}

We consider here a confidence interval that is slightly wider than $\CI{2}$ but has the advantage of being uniformly valid among a large class of DGP. To this end, we account for the randomness of $\widehat{\overline{b}}$. A complication we face is that $\widehat{\overline{b}}$ may not be asymptotically normal:  $\overline{b}= E[|\lambda_{T+1}(X,\beta_0)|Z_0]/(2\times 4^T)$ and  with the AME or ATE, $\lambda_{T+1}(X,\beta_0)=0$ a.s. but $\deriv{\lambda_{T+1}}{\beta}(X,\beta_0)\ne 0$ when $\beta_{0k}=0$. 
This implies that we cannot apply the results of, e.g., \cite{imbens2004confidence} or \cite{stoye2009more} to construct uniformly valid confidence intervals. Instead, we rely on union bounds, by first estimating an upper bound $\widehat{\overline{b}}_{\alpha_1}$ on $\overline{b}$ that exceeds $\overline{b}$ with asymptotic probability of at least $1-\alpha_1$ for some $\alpha_1\in (0,\alpha)$, and then compute
$$\CI{3}  = \left[\widehat{\tilde{\delta}} \pm \frac{\widehat{\sigma}}{n^{1/2}} q_{\alpha-\alpha_1}\left(\frac{n^{1/2}\widehat{\overline{b}}_{\alpha_1}}{\widehat{\sigma}}\right) \right].$$
We now detail our construction of $\widehat{\overline{b}}_{\alpha_1}$. We rely on the assumption below, which is similar to Assumption \ref{hyp:lambda}:
\newcounter{counterhyp}
\setcounter{counterhyp}{\thehyp}
\setcounter{hyp}{\thecounterhypprod}
\renewcommand{\thehyp}{\arabic{hyp}'}
\begin{hyp}
	$\lambda_{T+1}(x,\beta)=a(\beta_k)\rho(x,\beta)$ where $\rho$ is is $Reg(1)$, $a$ is continuously differentiable. Moreover, either $|a'(\beta_{k})|> 0$ for all $\beta \in B$ or $a(\beta_k)=1$ for all $\beta \in B$.
\label{hyp:lambda2}
	\end{hyp}
\renewcommand{\thehyp}{\arabic{hyp}}
\setcounter{hyp}{\thecounterhyp}

Note that the decomposition of $\lambda_{T+1}(x,\beta)$ is not unique, but Theorem \ref{thm:CI_triv2} below holds if $E(|\rho(X,\beta_0)|)> 0$, which imposes constraints on which pair $(a,\rho)$ we can consider. For the AME, letting $a(\beta_k)=\beta_k$ as above leads to $E(|\rho(X,\beta_0)|)> 0$ provided that the weak condition $P(\min_{s\ne t} |(X_s-X_t)'\beta_0| >0)>0$ holds. For the ATE, $a(\beta_k)=\left(1-\exp(\beta_k)\right) $ works as soon as $P(\min_{s\ne t} |(X_s-X_t^{1 - X_{tk}})'\beta_0| >0)>0$. For the ASF at $\tilde{x}_t$, $a(\beta_k)=1$ leads to $E(|\rho(X,\beta_0)|)> 0$ if $P(\min_{s=1,...,T} |(X_s-\tilde{x}_t)'\beta_0|> 0)>0$.


\medskip
Define $\overline{R} := \E[\abs{\rho(X,\beta_0)}Z_0/(2\times 4^T)]$, so that $\bar{b} = \abs{a(\beta_{0k})} \overline{R}$. Let $\widehat{\overline{R}}$ denote the plug-in estimator of $\overline{R}$. Since $a(\cdot)$ is differentiable, $a(\widehat{\beta}_k)$ is asymptotically normal by the delta method and we can consistently estimate its asymptotic standard deviation $\sigma_a$. We denote by $\widehat{\sigma}_a$ such an estimator. Then, our estimator of the upper bound $\widehat{\overline{b}}_{\alpha_1}$ on $\overline{b}$ is
$$\widehat{\overline{b}}_{\alpha_1} = \left(|a\left(\widehat{\beta}_k\right)|+ z_{1-\alpha_1} n^{-1/2}\widehat{\sigma}_a\right)\widehat{\overline{R}},$$
where $z_{1-\alpha_1}$ is the quantile of order $1-\alpha_1$ of a $\mathcal{N}(0,1)$. 

\medskip
We now define classes of DGPs for which $\CI{3}$ has uniform guarantees in terms of asymptotic coverage. In view of \eqref{eq:Delta_x}, \eqref{eq:Z} and the definition of $\tilde{\delta}$, we have $|\delta_0 - \tilde{\delta}| = |a(\beta_{0k})| R$, where
$$R := \left|E\left[\rho(X,\beta_0)\frac{\mathbb{T}_{T+1}(U)}{\Omega_{X,\beta_0}(U)}\right]\right|,$$
and we recall that $U=\Lambda(v(X,\beta_0)+\alpha)$ (note that $R\le \overline{R}$).

\medskip
Now, fix $\overline{M}>0$, $\underline{\sigma}> 0$, $\omega > 0$, $\zeta > 0$ and $\underline{A}$ a symmetric positive definite matrix. Define the following subset of probability distributions:
\begin{align}
	\mathcal{P}:=& \left\{P: \,\text{Assumption }\ref{hyp:model} \text{ holds, }  P(\|X\|\leq \overline{M})=1,\, \mathcal{I}_{0P}>>\underline{A}, \, \sigma^2_P\geq\underline{\sigma}^2, \right.\nonumber\\
	&\left. \qquad \ E_P\left(|\rho(X,\beta_0)|\right) > \omega \text{ and } \overline{R}_P > (1+\zeta)R_P\right\},	\label{eq:def_P}
\end{align}
where $\|\cdot\|$ denotes the Euclidean norm, $B>>A$ means that $B-A$ is symmetric positive definite and we index $\mathcal{I}_0$, $\sigma^2$, $R$ and $\overline{R}$ by $P$ to underline their dependence in $P$. The first three restrictions ensure that $\widehat{\beta}$ is asymptotically linear in a uniform sense, see Lemma \ref{lem:conv_unif_P} in the supplementary material. The condition $E_P(|\rho(X,\beta_0)|)>\omega$ has already been discussed above. Finally, by a straightforward adaptation of Lemma \ref{lem:cond_CI2}, we have $R=\overline{R}$ only if \eqref{eq:eg_R_Rsup} holds, with $\lambda_{T+1}(X,\beta_0)$ therein replaced by $\rho(X,\beta_0)$. Thus, $\overline{R}_P > (1+\zeta)R_P$ basically excludes DGPs close to the peculiar DGPs for which this modified version of \eqref{eq:eg_R_Rsup} holds. 

\begin{thm}
Suppose that Assumptions \ref{hyp:model}-\ref{hyp:iid_bdX} and \ref{hyp:lambda2} hold. Then:	
$$\liminf_{n\to\infty}\inf_{P\in\mathcal{P}} P\left(\delta_0\in \CI{3}\right) \geq 1-\alpha.$$
	\label{thm:CI_triv2}
\end{thm}


\section{Proofs of the identification results} 
\label{sec:proofs_of_the_results}

\subsection{Proposition \ref{prop:beta}}

First, suppose that Assumption \ref{hyp:varying_X} does not hold. Then, there exists $\lambda\neq 0$ such that $\lambda'\E\left(\sum_{1\leq t\neq t'\leq T}(X_t-X_{t'})(X_t-X_{t'})'\right)\lambda=0$. As a result,
$$\sum_{1\leq t\neq t'\leq T}\E\left((X_t'\lambda-X_{t'}'\lambda)^2\right)=0.$$
Because the quantity inside the expectation is nonnegative, we obtain $X_1'\lambda = ... = X_T' \lambda$ almost surely (a.s.). For any $v\in\R$, let $\alpha'=\alpha- v X'_t\lambda$ and $\beta = \beta_0 + v\lambda$. Then $Y_t =\ind{X'_t\beta + \alpha' +\eps \geq 0}$. This model satisfies Assumption \ref{hyp:model}. Thus, $\beta_0$ is not identified.

\medskip
Now, assume that Assumption \ref{hyp:varying_X} holds. By the concavity of the logarithm and Jensen's inequality, $\E\left(\ell_c(Y|X;\beta)\right)\leq \E\left(\ell_c(Y|X;\beta_0)\right)$ with equality if and only if $\ell_c(Y|X;\beta)=\ell_c(Y|X;\beta_0)$ a.s. Assume that the latter holds. Then, a.s.,
\begin{equation}
	\exp[\ell_c(Y|X;\beta)]\mathds{1}\{S=1\}=\exp[\ell_c(Y|X;\beta_0)]\mathds{1}\{S=1\}.	
	\label{eq:eg_loglik}
\end{equation}
Let us define $P_t(\beta) := \exp(X_t'\beta)/\sum_{s=1}^T \exp(X_s'\beta)$. Equality \eqref{eq:eg_loglik} is equivalent to
$$\sum_{t=1}^T (P_t(\beta) - P_t(\beta_0)) Y_t\prod_{s\neq t}(1-Y_s)=0 \quad \text{a.s.}$$
Because at most one of the variables $(Y_t\prod_{s\neq t}(1-Y_s))_t$ is equal to 1, we have, for all $t$,
$$(P_t(\beta) - P_t(\beta_0)) Y_t\prod_{s\neq t}(1-Y_s)=0 \quad \text{a.s.}$$
By taking the expectation with respect to $X$ and noting that $P(Y_t\prod_{s\neq t}(1-Y_s) =1 |X)>0$
a.s., we get, a.s., $P_t(\beta) = P_t(\beta_0)$. This in turn implies that $X_t'(\beta-\beta_0)$ does not depend on $t$. Hence, a.s., $$\sum_{t,s}\left[(X_t-X_s)'(\beta-\beta_0)\right]^2=0.$$
Taking the expectation, this implies that
$$(\beta-\beta_0)'\E\left[\sum_{t,s}(X_t-X_s)(X_t-X_s)'\right](\beta-\beta_0)=0.$$
By Assumption \ref{hyp:varying_X}, $\beta=\beta_0$. Hence, $\beta_0$ is identified and $\beta_0 = \arg\max_\beta \E\left(\ell_c(Y|X,\beta)\right)$.

\medskip
We finally turn to the last result. If $S=1$, we have $\deriv{\ell_c}{\beta}(Y|X;\beta_0) =
\sum_{t=1}^T X_t \left(Y_t  -  P_t(\beta_0)\right)$. Then, conditional on $S=1$, because $\sum_{s=1}^T P_s(\beta_0)=1$ we have
\begin{align*}
	\Deriv{{}^2\ell_c}{\beta\partial \beta'}(Y|X;\beta_0)& =-\sum_{t=1}^T X_t P_t(\beta_0)\sum_{s=1}^T (X_t -X_s)' P_s(\beta_0) \\ 	
	& = -\frac{1}{2} \sum_{s,t} P_s(\beta_0) P_t(\beta_0)\left(X_t -X_s\right)\left(X_t -X_s\right)'.
\end{align*}
Let $\lambda$ be such that $\lambda'\mathcal{I}_0\lambda=0$. Because $-\deriv{{}^2\ell_c}{\beta\partial \beta'}$ is positive semidefinite, we have
\begin{align*}
	\lambda'\mathcal{I}_0\lambda & \geq \lambda'\E\left[-\Deriv{{}^2\ell_c}{\beta\partial \beta'}(Y|X;\beta_0)\ind{S=1}\right]\lambda \\
	& = \frac{1}{2} \sum_{s,t} \E\left[P_s(\beta_0) P_t(\beta_0) \ind{S=1} \lambda'\left(X_t -X_s\right)\left(X_t -X_s\right)' \lambda \right] \\
	& = \frac{1}{2} \sum_{s,t} \E\left[P_s(\beta_0) P_t(\beta_0) P(S=1|X)\left[(X_t -X_s)' \lambda\right]^2\right].
\end{align*}
Hence, for all $(s,t)$, $P_s(\beta_0) P_t(\beta_0) P(S=1|X)\left[(X_t -X_s)' \lambda\right]^2=0$ almost surely. Since $P(S=1|X)>0$, we have
$(X_t -X_s)' \lambda=0$ almost surely. In turn, this implies that
$$\lambda'\E\left[\sum_{s,t}(X_t-X_s)(X_t-X_s)'\right]\lambda=0.$$
Thus, by Assumption \ref{hyp:varying_X}, $\lambda=0$, proving that $\mathcal{I}_0$ is nonsingular.

\subsection{Lemma \ref{lem:optim}} 

First, note that $\beta_0$ is identified by Proposition \ref{prop:beta}. Now, let $\mathcal{I}(x)$ be the set defined in \eqref{eq:other_expr_Delta_x}. We prove the result by proving the two inclusions in turn.

\paragraph{We have $\Delta(x)\subseteq \mathcal{I}(x)$.} 
\label{par:_delta_x_subset_mathcal_i_x}
Consider $\tilde{\delta}(x)$ an arbitrary element of $\Delta(x)$. There exists some random variable $\widetilde{\alpha}$ 
such that
\begin{align}
P(S=k|X=x) & = C_k(x,\beta_0) \int\frac{\exp(k a)}{\prod_{t=1}^T [1+\exp(x_t'\beta_0 +a)]} dP_{\widetilde{\alpha}|X=x}(a), \label{eq:rationalized_PS} \\
\int g(a,x,\beta_0) dP_{\widetilde{\alpha}|X=x}(a) & = \widetilde{\delta}(x).
\label{eq:rationalized_delta}
\end{align}
Consider $\widetilde{U}=\Lambda(v(X,\beta_0)+\widetilde{\alpha})$.
For all $t=0,...,T$,
\begin{align}
\E\left[Z_t|X=x\right] & = \sum_{k=t}^T \frac{\binom{T-t}{k-t} \exp(kv(x,\beta_0))}{C_k(x,\beta_0)} P(S=k|X=x) \notag \\
& =  E\left[\frac{\sum_{k=t}^T \binom{T-t}{k-t} \widetilde{U}^k(1-\widetilde{U})^{T-k}}{\Omega_{x,\beta_0}(\widetilde{U})}\bigg|X=x\right] \notag \\
& = E\left[\frac{\widetilde{U}^t}{\Omega_{x,\beta_0}(\widetilde{U})}\bigg|X=x\right], \label{eq:mom_Z}
\end{align}
where the last line follows since for all $u\in\R$ and $t\in\{0,...,T\}$, $\sum_{k=t}^T \binom{T-t}{k-t} u^{k} (1-u)^{T-k}=u^t$. Observe that for all $x\in\Supp(X)$, $\inf_{u\in[0,1]}\Omega_{x,\beta_0}(u)>0$. Then, define $\widetilde{\mu}_x$ by $d\widetilde{\mu}_x/dP_{\widetilde{U}|X=x}=1/[\Omega_{x,\beta_0}\int(1/\Omega_{x,\beta_0}(u)dP_{\widetilde{U}|X=x}(u))]$. Since $\E\left[Z_0|X=x\right]=E[1/\Omega_{x,\beta_0}(\widetilde{U})|X=x]$, we obtain
\begin{equation}
m_t(x)=\frac{\E\left[Z_t|X=x\right]}{\E\left[Z_0|X=x\right]} = \int_0^1 u^t d\widetilde{\mu}_x(u), \quad t=0,...,T.	
	\label{eq:constr_bis}
\end{equation}
Thus, $\widetilde{\mu}_x\in\mathcal{D}(m(x))$. Moreover, because $P(U\in\{0,1\}|X=x)=0$, $\widetilde{\mu}_x(\{0,1\})=0$. 
By Assumption \ref{hyp:fct_obj_simple}, we have
\begin{align}
	\widetilde{\delta}(x)= &  \sum_{t=0}^{T+1} \lambda_t(x, \beta_0)  E\left[\frac{\widetilde{U}^t}{\Omega_{x,\beta_0}(\widetilde{U})}|X=x\right] \notag \\
	= & \sum_{t=0}^{T} \lambda_t(x, \beta_0)  E[Z_t|X=x] + \lambda_{T+1}(x, \beta_0) E[Z_0|X=x] \int_0^1 u^{T+1} d\widetilde{\mu}_x(u), \label{eq:AME_bis}
\end{align}
where the first equality is \eqref{eq:Delta_x} and the second uses \eqref{eq:mom_Z} and the definition of $\widetilde{\mu}_x$:
$$\int_0^1 u^{T+1} d\widetilde{\mu}_x(u) = \frac{E\left[\widetilde{U}^{T+1}/\Omega_{x,\beta_0}(\widetilde{U})|X=x\right]}{E[Z_0|X=x]}.$$
Equation \eqref{eq:AME_bis}, together with $\widetilde{\mu}_x\in\mathcal{D}(m(x))$ and $\widetilde{\mu}_x(\{0,1\})=0$, imply $\Delta(x)\subseteq \mathcal{I}(x)$.


\paragraph{We have $\mathcal{I}(x) \subseteq \Delta(x)$.} 
\label{par:_mathcal_i_x_subset_delta_x}

Let $\mu_x\in\mathcal{D}(m(x))$ and $\mu_x(\{0,1\})=0$. To prove the result, we  let
$$\widetilde{\delta}(x) = \sum_{t=0}^{T} \lambda_t(x, \beta_0)  E[Z_t|X=x] + E[Z_0|X=x] \lambda_{T+1}(x, \beta_0) \int_0^1 u^{T+1} d\mu_x(u),$$ and we construct 
a probability measure $P_{\widetilde{\alpha}|X=x}$ 
such that \eqref{eq:rationalized_PS} and \eqref{eq:rationalized_delta} hold.

First, let  $P_{\widetilde{U}|X=x}$ be such that $dP_{\widetilde{U}|X=x}/d\mu_x = \Omega_{x,\beta_0}/\int_0^1 \Omega_{x,\beta_0}d\mu_x$. Then, let $$F_{\widetilde{\alpha}|X=x}(a)=\int_{-\infty}^{\Lambda(v(x,\beta_0)+a)} dP_{\widetilde{U}|X=x}(u).$$
The function $F_{\widetilde{\alpha}|X=x}$ is increasing, c\'adl\'ag and since $P_{\widetilde{U}}(\{0,1\})=0$,
$\lim_{a\to-\infty} F_{\widetilde{\alpha}|X=x}(a)=0$ and $\lim_{a\to\infty} F_{\widetilde{\alpha}|X=x}(a)=1$. Hence, $F_{\widetilde{\alpha}|X=x}$ is a cdf. Let $P_{\widetilde{\alpha}|X=x}$ be its corresponding probability measure. Remark that for any function $q$,
\begin{equation}
	\int q(u)dP_{\widetilde{U}|X=x}(u) = \int q \circ \Lambda(v(x,\beta_0)+a) dP_{\widetilde{\alpha}|X=x}(a).
	\label{eq:link_alpha_U}
\end{equation}
We now show that \eqref{eq:rationalized_delta} holds. One obtains with some algebra that
 \begin{align*}
 	\Omega_{x,\beta_0}(u)&=\sum_{t=0}^Tu^T\left[\sum_{k=0}^t(-1)^{t-k}\binom{T-k}{t-k}C_k(x,\beta_0)\exp\left(-k v(x,\beta_0)\right)\right].
 	\end{align*}
It follows from $\mu_x\in \mathcal{D}(m(x))$ and $m_t(x)=\E(Z_t|X=x)/E(Z_0|X=x)$ that:
\begin{align}
	&E[Z_0|X=x]\int \Omega_{x,\beta_0}(u)d\mu_x(u) \notag\\
&=\sum_{t=0}^TE[Z_t|X=x] \sum_{k=0}^t(-1)^{t-k}C_k(x,\beta_0)\binom{T-k}{t-k}\exp\left(-k v(x,\beta_0)\right) \notag\\
&=\sum_{k=0}^TP(S=k|X=x) \left[\sum_{k'=0}^{k}\frac{C_{k'}(x,\beta_0)}{C_{k}(x,\beta_0)}\exp((k-k')v(x,\beta_0))\sum_{t=k'}^k\binom{T-t}{k-t}\binom{T-k'}{t-k'}(-1)^{t-k'}\right] \notag\\
&=\sum_{k=0}^TP(S=k|X=x)=1, \label{eq:rel_Omega_Z0}
\end{align}
where the last equality follows from
\begin{equation}
\sum_{t=k'}^k\binom{T-t}{k-t}\binom{T-k'}{t-k'}(-1)^{t-k'}=\binom{T-k'}{T-k}\sum_{j=0}^{k-k'}\binom{k-k'}{j}(-1)^j=\mathds{1}\{k=k'\}.
\label{eq:inv_B}	
\end{equation}
Then using again  $m_t(x)=\E(Z_t|X=x)/E(Z_0|X=x)$, we have
\begin{align*}
\widetilde{\delta}(x)  & = E[Z_0|X=x] \int \sum_{t=0}^{T+1} \lambda_t(x,\beta_0) u^t d\mu_x(u) \\
& = E[Z_0|X=x] \left(\int \frac{\sum_{t=0}^{T+1} \lambda_t(x,\beta_0) u^t}{\Omega_{x,\beta_0}(u)}dP_{\widetilde{U}|X=x}(u)\right)\int \Omega_{x,\beta_0}(u)d\mu_x(u) \\
& = \int h_{x,\beta_0}(u) dP_{\widetilde{U}|X=x}(u) \\
& = \int g(x,a,\beta_0) dP_{\widetilde{\alpha}|X=x}(a),	
\end{align*}
where the last line follows using \eqref{eq:link_alpha_U} and $g(x,a,\beta_0)=h_{x,\beta_0}(\Lambda(v(x,\beta_0)+a))$. Thus, \eqref{eq:rationalized_delta} holds. To prove \eqref{eq:rationalized_PS}, first note that by definition of $Z_t$,
$\bm{Z} = B \bm{q}$ where $B$ is the $(T+1)\times (T+1)$ upper triangular matrix of term $(i,j)$ equal to $\binom{T-i}{j-i}$ for $0\leq i\leq j\leq T$,  $\bm{Z}=(E(Z_0|X=x),...,E(Z_T|X=x))'$ and
$$\bm{q}=\left(\frac{P(S=0|X=x)e^{0v(x,\beta_0)}}{C_0(x,\beta_0)},...,\frac{P(S=T|X=x)e^{Tv(x,\beta_0)}}{C_T(x,\beta_0)}\right)'.$$
By \eqref{eq:inv_B}, $B$ is invertible and the $(i,j)$ term of its inverse is $(-1)^{j-i}\binom{T-i}{j-i}$ (for $0\leq i\leq j\leq T$). Then, by using again \eqref{eq:link_alpha_U} and \eqref{eq:rel_Omega_Z0}, we obtain
\begin{align*}
\int\frac{\exp(k a)}{\prod_{t=1}^T [1+\exp(x_t'\beta_0 +a)]} dP_{\widetilde{\alpha}|X=x}(a) & = e^{-kv(x,\beta_0)}\int \frac{u^k(1-u)^{T-k}}{\Omega_{x,\beta_0}(u)} dP_{\widetilde{U}|X=x}(u) \\
& = e^{-kv(x,\beta_0)}\left(\int u^k(1-u)^{T-k}d\mu_x(u)\right)/ \int \Omega_{x,\beta_0}(u) d\mu_x(u) \\
& = e^{-kv(x,\beta_0)} \sum_{t=k}^{T} \binom{T-k}{t-k}  (-1)^{t-k} E[Z_t|X=x] \\
& = \frac{P(S=k|X=x)}{C_k(x,\beta_0)},
\end{align*}
where the last line follows since $\sum_{t=k}^{T} \binom{T-k}{t-k}  (-1)^{t-k} E[Z_t|X=x]$ is the $k$-th line of $B^{-1}\bm{Z}$. To conclude, note that if \eqref{eq:rationalized_PS} and \eqref{eq:rationalized_delta} hold for a random variable $\widetilde{\alpha}$, then $\widetilde{\delta}(x)\in\Delta(x)$ because $S$ is an exhaustive statistic of the fixed effects.



\subsection{Proposition \ref{prop:extension_mom}}
\label{par:proof_of_eq_eq_caract_M}

As we rely on the theory of moments and we do not prove the corresponding results, we first give some intuition on why $\underline{\mathbb{H}}_T(m)$ and $\overline{\mathbb{H}}_T(m)$ appear here. First, a given vector  $m=(m_0,...,m_{T})'$ is a vector of raw moments (namely, $m=(\int_0^1 u^0d\nu(u),...,\int_0^1 u^{T}d\nu(u))'$ for $\nu\in\mathcal{D}$) if and only if for all polynomials $P(x):=\sum_{j=0}^{T} \lambda_j x_j$ that are non-negative on $[0,1]$, we have $\sum_{j=0}^{T} \lambda_j m_j\ge 0$. The ``only if'' part simply follows from $\int_0^1 P(u)d\nu(u)\ge 0$. The ``if'' part is basically due to the Hahn-Banach theorem. Second, by Markov-Luk\'acs theorem \citep[see, e.g., Corollary 3.24 in][]{schmudgen2017moment}, polynomials of degree $T$ that are non-negative on $[0,1]$ are of the form $x\mapsto P_1^2(x)+ x(1-x) P_2^2(x)$ for some polynomials $P_1$ and $P_2$ when $T$ is even, and of the form $x\mapsto x P_1^2(x)+ (1-x) P_2^2(x)$ when $T$ is odd. Now, if  $T$ is even and the polynomial is of the form $x\mapsto\left(\sum_{t=0}^{T/2} \lambda_{t} x^t\right)^2$,
$$\int_0^1\left(\sum_{t=0}^{T/2} \lambda_{t} x^t\right)^2 d\nu(x) = \sum_{s=0}^{T/2} \sum_{t=0}^{T/2} \lambda_s \lambda_t  m_{s+t} = \lambda'\underline{\mathbb{H}}_T(m)\lambda,$$
with $\lambda=(\lambda_0,...,\lambda_{T/2})'$. Similarly, for any polynomial of the form $x\mapsto x(1-x)\left(\sum_{t=0}^{T/2-1} \lambda_{t} x^t\right)^2$,
$$\int_0^1 x(1-x)\left(\sum_{t=0}^{T/2-1} \lambda_{t} x^t\right)^2 d\nu(x) = \sum_{s=0}^{T/2-1} \sum_{t=0}^{T/2-1} \lambda_s \lambda_t  (m_{s+t+1} - m_{s+t+2}) = \lambda'\overline{\mathbb{H}}_T(m)\lambda.$$
This explains why $m$ is a valid vector of moments if and only if $\underline{\mathbb{H}}_T(m)$ and $\overline{\mathbb{H}}_T(m)$ are positive semidefinite matrices (the same holds when $T$ is odd). We refer to Theorems 10.1 and 10.2 in \cite{schmudgen2017moment} for the formal result and its proof.

\medskip
Now, we prove Proposition \ref{prop:extension_mom}. By what precedes, $\underline{H}_T(m)\geq 0$ and $\underline{H}_T(m)\geq 0$. Let $\underline{q}_T(m)=\inf_{\nu\in\D(m)} \int_0^1 u^{T+1}d\nu(u)$ and $\overline{q}_T(m)=\sup_{\nu\in\D(m)} \int_0^1 u^{T+1}d\nu(u)$, so that the identified set $\widetilde{\Theta}$ of $\theta_0$ satisfies $\widetilde{\Theta}\subseteq [\underline{q}_T(m), \overline{q}_T(m)]$. To prove the other statements of the proposition, let us introduce $\mathcal{M}_t$, the set of vectors of size $(t+1)$ corresponding to the first $(t+1)$ moments of a non-negative measure on $[0,1]$:
$$\M_t=\left\{m\in\R^{t+1}:\exists \nu \in \mathcal{D}:\; \int_0^1 u^j d\nu(u)=m_j \text{ for } j=0,...,t\right\}.$$
For any set $A$, let Int$~A$ and $\partial A$ denote the interior and boundary of $A$, respectively.

\paragraph{Point 1.} 
\label{par:point_1}
Since $\underline{H}_T(m) \overline{H}_T(m) > 0$ and $m\in\mathcal{M}_T$, by Theorem 10.8 in \cite{schmudgen2017moment}, $m\in\interior_T$. By his Theorem 10.7 and since $\mu(\{0,1\})=0$, $|\Supp(\mu)|>\lfloor N/2\rfloor$. Moreover, by Proposition 10.15 in \cite{schmudgen2017moment}, $\underline{q}_T(m)<\overline{q}_T(m)$ and $\underline{q}_T(m)$ and $\overline{q}_T(m)$ are the unique solutions of the linear equations $\underline{H}_{T+1}(m,\underline{q}_T(m))=0$ and $\overline{H}_{T+1}(m,\overline{q}_T(m))=0$.

\medskip
We now prove that $(\underline{q}_T(m),\overline{q}_T(m))\subseteq \widetilde{\Theta}$. The result follows since we also have $\widetilde{\Theta}\subseteq [\underline{q}_T(m), \overline{q}_T(m)]$ and $\Theta$ is the closure of $\widetilde{\Theta}$. Fix $\theta\in (\underline{q}_T(m),\overline{q}_T(m))$. By Proposition 10.15 in \cite{schmudgen2017moment}, $v:=(m,\theta)\in\interior_{T+1}$. Then, associated to $v$ are two measures in $\mathcal{D}(v)$, the so-called lower and upper principal representing measures  \citep[see, e.g.,][Definition 10.14 and Theorem 10.17]{schmudgen2017moment}. Moreover, if $T+1$ is odd, the support of the lower principal representing measure is in $(0,1)$ \citep[][eq. (10.17)]{schmudgen2017moment}. Thus, if $T+1$ is odd, there exists $\mu_\theta\in\D(m)$ such that $\int_0^1 u^{T+1}d\mu_\theta(u)=\theta$ and $\mu_\theta(\{0,1\})=0$. If $T+1$ is even, let $m_{T+2}\in (\underline{q}_{T+1}(v),\overline{q}_{T+1}(v))$. Again by Proposition 10.15 in \cite{schmudgen2017moment}, $(v,m_{T+2})\in\interior_{T+2}$. Since $T+2$ is odd, the support of the lower principal representing measure corresponding to $(v,m_{T+2})$ is in $(0,1)$. By construction, this distribution $\mu_\theta$ belongs to $\D(m)$ and satisfies $\int_0^1 u^{T+1}d\mu_\theta(u)=\theta$. Hence, in both cases, we have shown that $\theta\in \widetilde{\Theta}$.


\paragraph{Point 2.} 
\label{par:point_2}

If $\underline{H}_T(m) \overline{H}_T(m) = 0$, Theorem 10.7 in \cite{schmudgen2017moment} implies that $m\in\frontier_T$ and $\D(m)$ is a singleton. Since $\mu\in\D(m)$, we have $\D(m)=\{\mu\}$. Note that $\underline{q}_T(m)=\overline{q}_T(m)=\int_0^1 u^{T+1}d\mu(u)$. Suppose first that $T'$ is even and $\underline{H}_{T'}(m)=0$. Then, there exists a vector $\lambda=(\lambda_1,..., \lambda_{T'/2+1})'\ne 0$ such that $\underline{\mathbb{H}}_{T'}(m)\lambda=0$. Hence, for all $i\in\{1,...,T'/2+1\}$, $\sum_{j=1}^{T'/2+1} \lambda_j m_{i+j-2}=0$. Thus, for all $i\in\{0,...,T'/2\}$,
$$\int_0^1 u^i\left(\sum_{j=0}^{T'/2} \lambda_{j+1} u^j\right) d\mu(u)=0.$$
This implies that $\int_0^1 \left(\sum_{j=0}^{T'/2} \lambda_{j+1} u^j\right)^2 d\mu(u)=0$ and in turn $\sum_{j=0}^{T'/2} \lambda_{j+1} u^j=0$ $\mu$-almost everywhere. This ensures that $|\Supp(\mu)|\leq T'/2=\lfloor T'/2\rfloor$ (otherwise $\lambda=0$). It also implies that for all $k\geq 1$, and letting $m_k:=\int_0^1 u^kd\mu(u)$ for $k>T$, we have $\sum_{j=1}^{T'/2+1} \lambda_{j} m_{j+k-2}=0$. Since this holds for $k\in\{T+2-T',...,T+2-T'/2\}$, we have $\underline{\mathbb{H}}_{T'}(m_{T+1-T'},...,m_{T+1})\lambda=0$. Therefore, $\underline{H}_{T'}(m_{T+1-T'},...,m_{T+1})=0$, with $\underline{q}_T(m)=\overline{q}_T(m)=m_{T+1}$.

\medskip
The reasoning is the same if $T'$ is odd and still $\underline{H}_{T'}(m)=0$, with just one difference. Instead of having $\int_0^1\left(\sum_{j=0}^{T'/2-1} \lambda_{j+1} u^j\right)^2d\mu(u)=0$, we have
$$\int_0^1 u\bigg(\sum_{j=0}^{(T'-1)/2-1} \lambda_{j+1} u^j\bigg)^2d\mu(u)=0.$$
But since $u\geq 0$, this still implies $u\left(\sum_{j=0}^{(T'-1)/2-1} \lambda_{j+1} u^j\right)^2= 0$ $\mu$-a.e., and since $0\not\in\Supp(\mu)$, $\left|\Supp(\mu)\right|\leq (T'-1)/2-1=\lfloor T'/2\rfloor-1$. The rest of the proof is as above.

\medskip
When instead $\overline{H}_{T'}(m)=0$ and $T'$ is even, we have instead
$$\int_0^1 u(1-u)\bigg(\sum_{j=0}^{T'/2-1} \lambda_{j+1} u^j\bigg)^2d\mu(u)=0,$$ implying again
$u(1-u)\left(\sum_{j=0}^{T'/2-1} \lambda_{j+1} u^j\right)^2= 0$ $\mu$-a.e., $\left|\Supp(\mu)\right|\leq T'/2-1=\lfloor T'/2\rfloor-1$ and $\overline{H}_{T'}(m_{T+1-T'},...,m_{T+1})=0$, with $\underline{q}_T(m)=\overline{q}_T(m)=m_{T+1}$. Finally, when $\overline{H}_{T'}(m)=0$ and $T'$ is odd, we have
$$\int_0^1 (1-u)\left(\sum_{j=0}^{(T'-1)/2-1} \lambda_{j+1} u^j\right)^2d\mu(u)=0,$$
implying again $(1-u)\left(\sum_{j=0}^{(T'-1)/2-1} \lambda_{j+1} u^j\right)^2= 0$ $\mu$-almost everywhere, $\left|\Supp(\mu)\right|\leq (T'-1)/2-1=\lfloor T'/2\rfloor-1$ and $\overline{H}_{T'}(m_{T+1-T'},...,m_{T+1})=0$, with $\underline{q}_T(m)=\overline{q}_T(m)=m_{T+1}$.


\subsection{Theorem \ref{thm:ident1}}

It follows directly from Lemma \ref{lem:optim}, Proposition \ref{prop:extension_mom} and the definition of $\underline{\delta}(x)$  and $\overline{\delta}(x)$ that $\Delta(x)=[\underline{\delta}(x),\overline{\delta}(x)]$. Turning to $\Delta$, note that the proof of Proposition \ref{prop:extension_mom} implies that the identified set $\widetilde{\Delta}(x)$ of $\delta_0(x)$ (not its closure) satisfies, since $\delta_0(x)\subseteq \widetilde{\Delta}(x)$,
$$(\underline{\delta}(x),\overline{\delta}(x)) \cup \{\delta_0(x)\} \subseteq \widetilde{\Delta}(x)\subseteq [\underline{\delta}(x),\overline{\delta}(x)].$$
Because the distribution of $\alpha|X=x$ is only constrained by \eqref{eq:constraints}, the identified set of $x\mapsto \delta_0(x)$ is $\{f(\cdot): \; \forall x\in\Supp(X), \, f(x)\in \widetilde{\Delta}(x)\}$. Besides, the identified set $\widetilde{\Delta}$ of $\delta_0$ satisfies $\widetilde{\Delta}=\left\{E[f(X)]: \; \forall x\in\Supp(X), \, f(x)\in \widetilde{\Delta}(x)\right\}$. This implies that
$$(E[\underline{\delta}(X)],E[\overline{\delta}(X))]) \cup \{\delta_0\} \subseteq \widetilde{\Delta}\subseteq [E[\underline{\delta}(X)],E[\overline{\delta}(X))]].$$
Since $\Delta$ is the closure of $\widetilde{\Delta}$, we obtain $\Delta= [\underline{\delta},\overline{\delta}]$.

\medskip
Finally, $\delta_0$ is point identified if and only if $\delta_0(x)$ is point identified for almost all $x\in\Supp(X)$. Because $E(Z_0|X=x)>0$, $\delta_0(x)$ is point identified if and only if $\lambda_{T+1}(x,\beta_0)(\overline{q}_T(m(x))-\underline{q}_T(m(x)))=0$. By Proposition \ref{prop:extension_mom}, $\overline{q}_T(m(x))=\underline{q}_T(m(x))$ if and only if  $\left|\Supp(\mu_x)\right|\leq \intpart{T/2}$, where $\mu_x$ is the single distribution in $\mathcal{D}(m(x))$. The last result follows since by the proof of Lemma \ref{lem:optim}, $\left|\Supp(\mu_x)\right| = \left|\Supp{(\alpha|X=x)}\right|$.

\subsection{Proposition \ref{prop:outer_ident}}

For any given $T \geq 0$, among the polynomials of degree $T+1$ with leading coefficient equal to 1, the normalized Chebyshev polynomial $\mathbb{T}_{T+1}^c(u)$ has minimal supremum norm on the interval $[-1,1]$  \citep[see, e.g.][Section 3.3]{mason2002chebyshev}. Because $\mathbb{T}_{T+1}^c(u)=\cos(n\, \text{arccos}\, u)/2^T$ \citep[see, e.g.,][Eq. (1.1) and p.3]{mason2002chebyshev}, this maximal absolute value is $2^{-T}$, when $u=\cos\left(k\pi/(T+1)\right)$ for $0\leq k\leq T+1$.

\medskip
Thus, for any $u\in[0,1]$, $\left|\mathbb{T}_{T+1}(u)\right|\leq 1/(2\times 4^T)$ with equality for $u=\left(1+\cos\left(k\pi/(T+1)\right)\right)/2$ for $0\leq k\leq T+1$. Next for any $u\in[0,1]$:
$$\sum_{k=0}^{T}b_k^{\ast} u^k-\frac{1}{2\times 4^T}\leq u^{T+1}\leq \sum_{k=0}^{T}b_k^{\ast} u^k+\frac{1}{2\times 4^T}.$$
Integration with respect to any $\mu\in\D(m)$ implies, since $\mu([0,1])=m_0=1$, that
$$\sum_{k=0}^{T}b_k^{\ast} m_k-\frac{1}{2\times 4^T}\leq \theta\leq \sum_{k=0}^{T}b_k^{\ast} m_k+\frac{1}{2\times 4^T}.$$
As a result, $\widetilde{\Theta}\subseteq\Theta^o$ and thus $\Theta\subseteq\Theta^o$. The length of $\Theta^o$ is $4^{-T}$. From \cite{Skibinsky1967}, we have $\overline{q}_T(m)-\underline{q}_T(m)=\prod_{k=1}^Tp_k(1-p_k)$, with $$p_j=\frac{m_j-\underline{q}_{j-1}(m_{\to j-1})}{\overline{q}_{j-1}(m_{\to j-1})-\underline{q}_{j-1}(m_{\to j-1})}.$$
If we consider the first moments of the arcsine distribution, namely $m_k=\prod_{r=0}^{k-1}(2r+1)/(2r+2)$ for $k\leq T$, Theorem 3 in \cite{Skibinsky69} implies that $p_j=1/2$ for any $j\leq T$. In this case, the length of $\Theta$ is also $4^{-T}$, which implies that $\Theta=\Theta^o$.

\subsection{Theorem \ref{thm:ident2}}

Lemma \ref{lem:optim} combined with  Proposition \ref{prop:outer_ident} imply that $\widetilde{\Delta}(x) \subseteq [\tilde{\delta}(x) \pm \overline{b}(x)]$ and thus $\Delta(x) \subseteq \Delta^o(x)$. Since $\widetilde{\Delta}= \{E[\delta_0(X)]:\,\forall x\in\Supp(X), \delta_0(x)\in\widetilde{\Delta}(x) \}$, $\widetilde{\Delta}\subseteq [\tilde{\delta}\pm \overline{b}]$ and thus $\Delta\subseteq \Delta^o$.


\newpage
\pagenumbering{arabic}

\begin{center}
	{\huge Online Appendix}
	
\end{center}

\setcounter{section}{0}

\section{Extensions}
\label{sec:extensions}

We consider three extensions to our main results. First, we consider FE logit models with ordered, non-binary outcomes. Second, we consider dynamic FE logit models (with binary outcomes). Third, we show how to adapt results when $T$ varies per individual. 

\subsection{FE ordered logit models}
\label{sub:ordered_logit}

We now consider a model where the outcome is ordered and takes $J\geq 2$ values.

\begin{hyp}
	We have $Y_t=\sum_{k=1}^{J-1} k\mathds{1}\{\gamma_{k} \leq X_t'\beta_0+\alpha+\varepsilon_t< \gamma_{k+1}\}$ with $\gamma_1=0<...<\gamma_J=+\infty$ and $(\varepsilon_t)_{t=1,...,T}$ are i.i.d., independent of $(\alpha,X)$ and follow a logistic distribution.
	\label{hyp:model2}
\end{hyp}

The condition $\gamma_1=0$ is a mere normalization: only the differences $\gamma_j-\gamma_{j'}$ are identified since the location of the distribution of $\alpha$ is left unrestricted. To identify $\theta_0 := (\beta_0,\gamma_2,...,\gamma_{J-1})$, we follow \cite{Muris2017}. Let $\Pi$ be the set of functions from $\{1,...,T\}$ into $\{1,...,J-1\}$ and for $\pi\in \Pi$, let $Y^{\pi}_t=\ind{Y_t\geq \pi(t)}$. By conditioning on $S^{\pi}=\sum_t Y^{\pi}_t$, we get the conditional log-likelihood
\begin{align*}
	\ell^{\pi}_c(y|x;\beta,\gamma_2,...,\gamma_{J-1}) := &\sum_{t=1}^T y_t (x_t'\beta-\gamma_{\pi(t)}) - \ln\left[C^{\pi}_{\sum_{t=1}^T y_t}(x,\beta,\gamma)\right], \\
	\text{with } C^{\pi}_k(x, \beta,\gamma):= & \sum_{(d_1,...,d_T)\in \{0,1\}^T: \sum_{t=1}^Td_t=k} \exp\left(\sum_{t=1}^T d_t (x_t'\beta -\gamma_{\pi(t)})\right).	
\end{align*}
The parameter $\theta_0$ is then identified by stacking, over all $\pi\in\Pi$, the first-order conditions $\E[\partial \ell^{\pi}_c/\partial \theta(Y|X;\theta_0)]=0$ of the conditional log-likelihood maximization.

\medskip
We now turn to the identification of average effects. As in the main body of the paper, we consider parameters of the form
\begin{equation}
	\delta_0 := \E\left[g(X,\alpha,\theta_0)\right],	
	\label{eq:def_delta_ordered}
\end{equation}
where the function $g$ is known. This encompasses the average marginal or treatment effects of $X_{kt}$ (say) on the probability that $Y_t\geq j_0$ for some $j_0$. For instance, the corresponding average marginal effect is
$$\delta_0 =\E\left[\Deriv{P\left(Y_t\geq j_0|X,\alpha\right)}{X_{tk}}\right]=\beta_{0k}\E\left[\Lambda'(\gamma_{j_0} +  X'\beta_0+\alpha)\right].$$
Equation \eqref{eq:def_delta_ordered} also includes average structural functions, namely the counterfactual probability that $Y_t\geq j_0$ if $X_t$ is fixed to $\widetilde{x}_t$.

\medskip
We impose the following  restrictions on $g$, similar to Assumption \ref{hyp:fct_obj_simple}.

\begin{hyp}
	For all $x\in\Supp(X)$ and $\theta \in \R^{p + J-2}$, there exists $v(x,\theta)\in\R$ such that if we let $\rho(j,t,x):=\exp(x_{t}'\beta_0-\gamma_{j} - v(x,\theta))-1$ and $\Omega_{x,\theta}(u):=\prod_{\substack{1\leq j \leq J-1\\ 1\leq t\leq T}}(1+u\rho(j,t,x))$, the function $u\mapsto g(x,\Lambda^{-1}(u) - v(x,\theta),\theta)\times \Omega_{x,\theta}(u)$ defined on $(0,1)$ is a polynomial of degree at most $(J-1)T+1$. We let $\lambda_t(x,\theta)$ denote the coefficient of $u^t$ of this polynomial.
	\label{hyp:fct_obj_simple_ordered}
\end{hyp}
As with the FE binary logit model, this assumption holds with the aforementioned average marginal effects, average treatment effects and average structural functions.

\medskip
Now, let $\delta_0(x):= E\left[g(X,\alpha,\theta_0)|X=x\right]$, $U:=\Lambda(v(x,\theta)+\alpha)$ and $h_{x,\theta}(u):=g(x,\Lambda^{-1}(u) - v(x,\theta),\theta)$, so that $\delta_0(x)=E[h_{x,\theta_0}(U)|X=x]$. Under Assumption \ref{hyp:fct_obj_simple_ordered},  we obtain
\begin{align}
	\delta_0(x)= & E\left[\frac{h_{x,\theta_0}(U)\Omega_{x,\theta_0}(U)}{\Omega_{x,\theta_0}(U)}|X=x\right]	 \notag \\
	= & \sum_{t=0}^{(J-1)T+1} \lambda_t(x,\theta_0) E\left[\frac{U^t}{\Omega_{x,\theta_0}(U)}|X=x\right]. \label{eq:Delta_x_ordered}
\end{align}
Moreover, we show in the proof of Theorem \ref{thm:ident1_ordered} below that there exist identified variables $(Z_t)_{t=0,...,(J-1)T}$ such that $E[U^t/\Omega_{x,\theta_0}(U)|X=x] = E[Z_t|X=x]$. Thus, as in the FE binary logit model, only the last term of the sum in \eqref{eq:Delta_x_ordered}  is partially identified in general. As in Section \ref{ssub:reparametrization}, let us define
$$m_t(x) := \frac{E[Z_t|X=x]}{E[Z_0|X=x]},$$
and let $m(x):=(m_0(x),...,m_{(J-1)T}(x))$. We obtain the following result on the closure of the sharp identified sets of $\delta_0(x)$ and $\delta_0$, denoted respectively by $\Delta(x)$ and $\Delta$.

\begin{thm}\label{thm:ident1_ordered}
	If the distribution of $(X,Y)$ is identified and Assumptions \ref{hyp:model2}-\ref{hyp:fct_obj_simple_ordered} hold, $\Delta(x)=[\underline{\delta}(x),\, \overline{\delta}(x)]$, with
	\begin{equation}
		\begin{array}{rcl}
			\underline{\delta}(x) & = & \sum_{t=0}^{(J-1)T} \E[Z_t|X=x] \lambda_t(x,\theta) + \E[Z_0|X=x] \lambda_{(J-1)T+1}(x,\theta)\big( \underline{q}_{(J-1)T}(m(x)) \\[2mm]
			& & \hspace{1.2cm} \ind{\lambda_{(J-1)T+1}(x,\theta)\geq 0}  +  \overline{q}_{(J-1)T}(m(x)) \ind{\lambda_{(J-1)T+1}(x,\theta)< 0} \big), \\[2mm]
			\overline{\delta}(x) & = & \sum_{t=0}^{(J-1)T} \E[Z_t|X=x] \lambda_t(x,\theta) + \E[Z_0|X=x] \lambda_{(J-1)T+1}(x,\theta)\big( \overline{q}_{(J-1)T}(m(x)) \\[2mm]
			& & \hspace{1.2cm} \ind{\lambda_{(J-1)T+1}(x,\theta)\geq 0}  +  \underline{q}_{(J-1)T}(m(x)) \ind{\lambda_{(J-1)T+1}(x,\theta)< 0} \big)
		\end{array}
		\label{eq:bounds_Delta_ordered}
	\end{equation}
	and where the variables $(Z_t)_{t=0,...,(J-1)T}$ are identified functions of $(X,Y)$. Moreover,  $\Delta =[\underline{\delta},\, \overline{\delta}]$, with $\underline{\delta}=E(\underline{\delta}(X))$, $\overline{\delta}=E(\overline{\delta}(X))$. $\delta_0$ is point identified if and only if
	$$P\left(\big\{\lambda_{(J-1)T+1}(X,\theta)=0\big\} \cup \big\{|\Supp(\alpha|X)|\leq \lfloor (J-1)T/2\rfloor \big\}\right)=1.$$
\end{thm}

The main difference between this result and Theorem \ref{thm:ident1} in the paper is that the bounds are related to moments of order $(J-1)T+1$ of distributions for which the first $(J-1)T$ raw moments are known. Hence, the bounds are tighter than in the binary case, and substantially more so given the discussion around Equation \eqref{eq:upper_bd_length}.


\subsection{Dynamic logit model}
\label{app:dyn}

We show here that our methodology also applies to dynamic logit models. 
\begin{hyp}
	For $t\ge 1$, we have $Y_t=\mathds{1}( Y_{t-1} \gamma_0 + X_{t}'\beta_0 + \alpha + \epsilon_t \geq 0)$ where the $(\eps_t)_{t=1,...,T}$ are i.i.d., independent of $(\alpha,X, Y_0)$ and follow a logistic distribution.
	\label{hyp:model3}
\end{hyp}
We assume that the econometrician observes $Y_0$. Identification results on $\theta_0:=(\gamma_0,\beta_0)$ can be found in, e.g., \cite{honore2000panel}, \cite{honore2022moment} and \cite{dobronyi2021}. We assume here that these parameters are point identified. A sufficient condition for this, when $T=3$, is basically that the density of $X_3 - X_2$ is continuous and strictly positive in a neighborhood of $0$.\footnote{See Theorem 1 in \cite{honore2000panel} on the consistency of their estimator for a full list of assumptions.}  See also Theorem 1 in \cite{honore2022moment} for more general conditions. Our interest lies in the average effects of changes in the value of $X_{kt}$ on the probability that $Y_t = 1$. We focus on the case $t=1$ and leave the generalization to $t > 1$ for future research. We thus consider parameters of the form
\begin{align*}
	\delta_0 := E\left[g(X,Y_0, \alpha, \theta_0)\right].
\end{align*}
As for the static model, examples of parameters are the average marginal effect of a continuous covariate, the average treatment effect of a discrete covariate and the average structural function corresponding to fixing $X$ to some $x$. Thus, our result below complements those of \cite{aguirregabiria2021identification}, who show point identification results on average effects of lagged values of $Y$, and \cite{dobronyi2021}, who study effects that can be obtained from the knowledge of $\theta_0$.

\medskip

As for the static model, we focus on conditional effects where we also condition on the value of $Y_0$ since  the model does not impose restrictions between $F_{\alpha|X=x,Y_0=y_0}$ and $F_{\alpha|X=x',Y_0=y_0'}$. Thus we consider
$$\delta_0(x,y_0):= E\left[g(X,Y_0, \alpha, \theta_0)|X=x, Y_0 = y_0\right].$$
We impose the following restrictions on $g$, similar to Assumption \ref{hyp:fct_obj_simple}.
\begin{hyp}
	For all $x\in\Supp(X)$, $y_0 \in \{0,1\}$ and $\theta \in \R^{p+1}$, there exists  $v(x,y_0,\theta)\in\R$ such that if we let $\rho(s,x,d,y_0, \theta) := \exp(d \gamma + x_{s}'\beta - v(x,y_0, \theta))$ and  $\Omega_{x,y_0,\theta}(u) := \prod_{s=1}^T (1 + u [\rho(s,x,0,y_0, \theta) - 1]) (1 + u [\rho(s,x,1,y_0, \theta) - 1])$, the function $u\mapsto g(x,y_0,\Lambda^{-1}(u) - v(x,y_0,\theta),\theta)\times\Omega_{x,y_0,\theta}(u) $ defined on $(0,1)$ is a polynomial of degree at most $2T+1$. We let $\lambda_t(x,y_0,\theta)$ denote the coefficient of $u^t$ of this polynomial.
	\label{hyp:fct_obj_simple_dynamic}
\end{hyp}
Assumption \ref{hyp:fct_obj_simple_dynamic} holds for the aforementioned average effects. Under Assumption \ref{hyp:fct_obj_simple_dynamic},  for $U=\Lambda\left(\alpha+v(X,Y_0,\theta_0)\right)$ and $h_{x,y_0,\theta}(u)=g(x,y_0,\Lambda^{-1}(u)-v(x,y_0,\theta),\theta)$, we obtain
\begin{align}
	\delta_0(x, y_0)= & E\left[\frac{h_{x,y_0,\theta_0}(U)\Omega_{x,y_0,\theta_0}(U)}{\Omega_{x,y_0,\theta_0}(U)}|X=x, Y_0 = y_0\right]	 \notag \\
	= & \sum_{t=0}^{2T+1} \lambda_t(x,y_0,\theta_0) E\left[\frac{U^t}{\Omega_{x,y_0,\theta_0}(U)}|X=x, Y_0 = y_0\right]. \label{eq:Delta_x_dynamic}
\end{align}
Moreover, there exist identified variables $(Z_t)_{t=0,...,2T}$ such that $E[U^t/\Omega_{x,y_0,\theta_0}(U)|X=x,Y_0=y_0] = E[Z_t|X=x,Y_0=y_0]$. To see this, consider a vector of outcomes $y = (y_1, ... , y_T)'$. Then:
\begin{align*}
	& P(Y=y|X=x,Y_0=y_0) \\
	&= \E\left[\prod_{s=1}^T \Lambda\left(y_{s-1} \gamma_0 + x_{s}'\beta_0 + \alpha\right)^{y_s}   \left(1 - \Lambda\left(y_{s-1} \gamma_0 + x_{s}'\beta_0 + \alpha\right)\right)^{1-y_s} \bigg|X=x,Y_0=y_0 \right]  \\
	& =  \E\left[ \prod_{s=1}^T  \left[ \frac{  \rho_s(y_{s-1})U }{1 + U [\rho_s(y_{s-1})  - 1]}\right]^{y_s}  \left[ \frac{  1- U }{1 + U [\rho_s(y_{s-1})  - 1]}\right]^{1 - y_s} \bigg|X=x,Y_0=y_0 \right] \\
	& =  C(x,y,y_0, \theta_0) \, \E\left[ \frac{\left(\prod_{s=1}^T 1 + U [\rho_s(1-y_{s-1})  - 1] \right) U^{\sum y_s} (1 - U)^{T - \sum y_s}}{ \Omega_{x,y_0,\theta_0}(U ) } \bigg|X=x,Y_0=y_0 \right]
\end{align*}
where $\rho_s(y_{s-1})$ is a shorthand for $\rho(s,x,y_{s-1},y_0, \theta_0)$ and $C(x,y,y_0, \theta) := \prod_{s=1}^T \rho_s(y_{s-1})$. Denote by $\{y^{(1)}, ..., y^{(2^T)}\}$ all the elements of $\{0,1\}^T$ and let $b_t(x,y_0) = \E(U^t/\Omega_{x,y_0,\theta_0}(U)|X=x,Y_0=y_0)$ for $t = 0,..,2T$. The probabilities $\left(P(Y=y^{(k)}|X=x,Y_0=y_0)\right)_{k=1...2^T}$ are  linear combinations of $b_0(x,y_0), ..., b_{2T}(x,y_0)$. Therefore, there exists a matrix $Q(x,y_0,\theta)$ of size $2^T \times (2T+1)$ such that
\begin{equation}
	\begin{pmatrix}
		\frac{P(Y= y^{(1)}|X=x,Y_0=y_0)}{C(x,y,y_0, \theta_0)}\\
		\vdots\\
		\frac{P(Y=y^{(2^T)}|X=x,Y_0=y_0)}{C(x,y,y_0, \theta_0)}
	\end{pmatrix}
	= Q(x,y_0, \theta_0)
	\begin{pmatrix}
		b_0(x,y_0)\\
		\vdots\\
		b_{2T}(x,y_0)
	\end{pmatrix}.
\end{equation}
If $Q(x,y_0, \theta_0)$ has full column rank, then, for $t=0,...,2T$,
there exists known functions $(\mu_{k,t}(x,y_0))_{k=1,...,2^T}$ such that 
$$\E(U^t/\Omega_{x,y_0,\theta_0}(U)|X=x,Y_0=y_0) =  \sum_{k=1}^{2^T} \mu_{k,t}(x,y_0) P(Y= y^{(k)}|X=x,Y_0=y_0).$$
Letting  $Z_t=\sum_{k=1}^{2^T} \mu_{k,t}(x,y_0) \ind{Y=y^{(k)}}$, $t=0,...,2T$, by the law of iterated expectation $E[U^t/\Omega_{x,y_0,\theta_0}(U)|X=x,Y_0=y_0] = E[Z_t|X=x,Y_0=y_0]$.  This implies that only the last term of the sum in \eqref{eq:Delta_x_dynamic}  is partially identified in general. Define
$$m_t(x,y_0) := \frac{E[Z_t|X=x,Y_0=y_0]}{E[Z_0|X=x,Y_0=y_0]},$$
and let $m(x,y_0):=(m_0(x,y_0),...,m_{2T}(x,y_0))$. We obtain the following result on the closure of the sharp identified sets of $\delta_0(x,y_0)$ and $\delta_0$, denoted respectively by $\Delta(x,y_0)$ and $\Delta$.

\begin{thm}\label{thm:ident1_dynamic}
	If the distribution of $(X,Y,Y_0)$ and $\theta_0$ are identified, Assumptions \ref{hyp:model3}-\ref{hyp:fct_obj_simple_dynamic} hold and $Q(X,Y_0, \theta_0)$ is of full column rank almost surely, then $\Delta(x)=[\underline{\delta}(x),\, \overline{\delta}(x)]$ with
	\begin{equation*}
		\begin{array}{rcl}
			\underline{\delta}(x,y_0) & = & \sum_{t=0}^{2T} \E[Z_t|X=x,Y_0=y_0] \lambda_t(x,y_0,\theta_0) + \E[Z_0|X=x,Y_0=y_0] \lambda_{2T+1}(x,y_0,\theta_0)\\[2mm]
			& & \hspace{1.2cm} \big( \underline{q}_{2T}(m(x,y_0))  \ind{\lambda_{2T+1}(x,y_0,\theta_0)\geq 0}  +  \overline{q}_{2T}(m(x,y_0)) \ind{\lambda_{2T+1}(x,y_0,\theta_0)< 0} \big), \\[2mm]
			\overline{\delta}(x,y_0) & = & \sum_{t=0}^{2T} \E[Z_t|X=x,Y_0=y_0] \lambda_t(x,y_0,\theta_0) + \E[Z_0|X=x,Y_0=y_0] \lambda_{2T+1}(x,y_0,\theta_0) \\[2mm]
			& & \hspace{1.2cm} \big( \overline{q}_{2T}(m(x,y_0))\ind{\lambda_{2T+1}(x,y_0,\theta_0)\geq 0}  +  \underline{q}_{2T}(m(x,y_0)) \ind{\lambda_{2T+1}(x,y_0,\theta_0)< 0} \big).
		\end{array}
		\label{eq:bounds_Delta_dynamic}
	\end{equation*}
	Moreover,  $\Delta =[\underline{\delta},\, \overline{\delta}]$, with $\underline{\delta}=E(\underline{\delta}(X,Y_0))$ and $\overline{\delta}=E(\overline{\delta}(X,Y_0))$. $\delta_0$ is point identified if and only if
	$$P\left(\big\{\lambda_{2T+1}(X,Y_0,\theta_0)=0\big\} \cup \big\{|\Supp(\alpha|X,Y_0)|\le T \big\}\right)=1.$$
\end{thm}

The proof follows along the same lines as those of Theorems \ref{thm:ident1} and \ref{thm:ident1_ordered} and is thus omitted.

\subsection{Varying number of periods} 
\label{sub:varying_number_of_periods}

Missing data or attrition are common in panel data. A ``panel'' may also correspond to hierarchical data where $(i,t)$ corresponds to a unit $t$ belonging to a group $i$ (e.g. individuals within a household). In both cases, $T$ is a random variable varying from one individual (or group) to another. Our method still applies in this case, provided that $T$ is conditionally exogenous. Specifically, we assume that conditional on $(T,X,\alpha)$, $(\eps_1,...,\eps_T)$ are still i.i.d. with marginal logistic distribution. On the other hand, we remain agnostic on the dependence between $T$ and $(X,\alpha)$. 

\medskip

Under the independence condition above, the identification and estimation of $\beta_0$ remains unchanged. Moreover, Theorem \ref{thm:ident1} can be applied conditional on $T=\overline{t}$ to characterize the identified set of $\delta_0(\overline{t})$. The sharp identified set of $\delta_0$ is then obtained by integrating over $T$. 
Similarly, the first estimation method applies for each subpopulation satisfying $T=\overline{t}$, and then one can just sum over all $\overline{t}\in\Supp(T)$.

\medskip

Outer bounds can also be obtained easily. An inspection of $\tilde{\delta}$ reveals that the formula remains similar, with the Chebyshev polynomials used for the approximation $P^*_T(u,x)$ now varying with $T$. The estimator $\widehat{\tilde{\delta}}$ and the formulas of $\sigma^2$ and $\overline{b}$ should be adjusted in a similar way. These features are all included in the R package \texttt{MarginalFElogit} and the Stata command \texttt{mfelogit}.


\section{Further details on the simulations} 
	\label{sub:details_on_the_simulations}
	
	First, to estimate $\gamma_{0t}(x)$, we use a local linear estimator with a common bandwidth $h_t$ for the $T$ components of $X$. To choose $h_t$, we aim at reaching a certain ratio between the (integrated) bias and standard deviation of the estimator. Specifically, let $B_t(x,h)$ and $\sigma^2_t(x,h)$ denote respectively the asymptotic bias and variance of  $\widehat{\gamma}_t(x)$ with a bandwidth equal to $h$. Then, we have \citep[see, e.g.][]{ruppert1994}, 
	\begin{align*} B_t(x,h) & = h^2 \left(\int u^2 K(u)du\right) \sum_{j=1}^{pT} \Deriv{{}^2 \gamma_{0t}}{x_j^2}(x),\\
\sigma^2_t(x,h) & = \frac{1}{n h^{pT}} \frac{\left(\int K(u)^2du\right)^{pT} \gamma_{0t}(x) (1-\gamma_{0t}(x))}{f_X(x)}.
\end{align*}
Define $B^2_t(h):=E[B^2_t(X,h)]$ and $\sigma^2_t(h):=E[\sigma^2_t(X,h)]$. Assuming first that $B^2_t(h)$
and $\sigma^2_t(h)$ are known, we would choose $h_t$ so that $\sigma^2_t(h_t)=R_n\times B^2_t(h_t)$, where $R_n>0$ fixes 
the degree of undersmoothing. For instance, $R_n=1$ corresponds to the optimal bandwidth in terms of asymptotic mean
integrated squared error. We use $R_n=\log(n)$ in our simulations. Now, $B^2_t(h)$ and $\sigma^2_t(h)$ are actually unknown.
We estimate both assuming that $\alpha$ is constant. Then, we can estimate this constant by MLE (plugging the CMLE
$\widehat{\beta}$ in the log-likelihood) and then estimate $\gamma_{0t}(x)$ by plug-in, using \eqref{eq:constraints}.
	
\medskip
Finally, to obtain $\widehat{m}$, we must choose a threshold $c_n$. We actually slightly modify $\widehat{I}(x)$, by
letting
$$\widehat{I}(x):= \max\left\{t\in \{1,...,T\}: \underline{H}_t(\widetilde{m}_{\rightarrow t}(x)) \geq
\underline{c}_{nt}(x) \text{ and } \overline{H}_t(\widetilde{m}_{\rightarrow t}(x)) \geq \overline{c}_{nt}(x)\right\},$$
where $\underline{c}_{nt}(x):=\widehat{\underline{\sigma}}_t[2\ln\ln(n)/n]^{1/2}$, $\overline{c}_{nt}(x):=
\widehat{\overline{\sigma}}_t(x)[2\ln\ln(n)/n]^{1/2}$ and $\widehat{\underline{\sigma}}^2_t(x)$ (resp.
$\widehat{\overline{\sigma}}^2_t(x)$) is an estimator of the asymptotic variance of $\underline{H}_t(m_{\rightarrow t}(x))$
(resp. $\overline{H}_t(m_{\rightarrow t}(x))$).
	

\section{Additional proofs} 
\label{sec:proofs_of_asymptotic_results}
\subsection{Asymptotic of estimators of the sharp bounds}

To reduce the notational burden and save space, we prove Theorem \ref{thm:as_normality} and Proposition \ref{prop:CI1} in the case where $p_c=p$. The proofs rely on uniform consistency of $x\mapsto \widehat{\gamma}(x)$ and $x\mapsto \widehat{m}(x)$ on $\Supp(X)$ and boundedness of the derivatives of $x\mapsto \widehat{\gamma}(x)$. These conditions imply suitable approximations of $\sqrt{n}\left(\widehat{\underline{\delta}}-\underline{\delta}\right)$ and $\sqrt{n}\left(\widehat{\overline{\delta}}-\overline{\delta}\right)$ holds, see \eqref{eq:delta_l_equiv2} or \eqref{eq:approx_kink} below. In the case $1\le p_c < p$, the same reasoning can be used to show that $x^c\mapsto\widehat{\gamma}(x^c,x^d)$ and $x^c\mapsto\widehat{m}(x^c,x^d)$ are uniformly consistent, while the derivatives $x^c\mapsto \widehat{\gamma}(x^c,x^d)$ are bounded on $\Supp(X^c|X^d=x^d)$, for any value of $x^d\in\Supp(X^d)$. The approximations \eqref{eq:delta_l_equiv2} and \eqref{eq:approx_kink} follow. When $p_c=0$, the estimators of $\widehat{\gamma}(x^d)$ and $\widehat{m}(x^d)$ are root-$n$ consistent for any value of $x^d\in \Supp(X)$ and the asymptotic approximations \eqref{eq:delta_l_equiv2} and \eqref{eq:approx_kink} again hold.

\medskip
Before proving Theorem \ref{thm:as_normality} and Proposition \ref{prop:CI1}, we prove the consistency of $(\widehat{\underline{\delta}},\widehat{\overline{\delta}})$. To do so we introduce additional notation. For any vector of functions $\gamma=(\gamma_0,...,\gamma_T)$, let
$$	(c_0(\gamma,x,\beta),...,c_T(\gamma,x,\beta))' :=
\Gamma \left(\frac{\gamma_{0}(x) \exp(0\times x_T'\beta)}{C_0(x,\beta)}, \ ... \ , \frac{\gamma_{T}(x) \exp(T\times x_T'\beta)}{C_T(x,\beta)}\right)',$$
where $\Gamma$ is a square matrix of size $T+1$ with coefficients $\Gamma_{ij} = \binom{T-i+1}{j-i}\ind{i \leq j} $ for $i,j=1,...,T+1$.
Note that $\widehat{c}_t(x) = c_t(\widehat{\gamma}, x,\widehat{\beta})$.\\
For any function $f$ from a set $\mathcal{D}$ to $\R^q$, we let $\norminf{f}=\sup_{x\in\mathcal{D}}\norm{f(x)}$. For $a=(a_1,...,a_{pT})\in \mathbb{N}^{pT}$, let $|a|=\sum_{j=1}^{pT}a_j$ and $D^{(a)}=\frac{\partial^{|a|}}{\partial^{a_1}x_{11}...\partial^{a_{pT}}x_{pT}}$. Following Chen et al. (2003), let us define, for any function $\gamma$ from $\mathcal{X}:=\Supp(X)$ to $\mathbb{R}^{T+1}$ admitting continuous derivatives of order at least $\ell+1$,
$$\norm{\gamma}_{\mathcal{C}^{\ell+1}(\mathcal{X})} := \max_{|a|\leq \ell+1} \norm{D^{(a)}\gamma}_{\infty}.$$
For any $c>0$, we let $\mathcal{C}^{\ell+1}_{c}(\mathcal{X})$ denote the set of functions $\gamma$ such that $\norm{\gamma}_{\mathcal{C}^{\ell+1}(\mathcal{X})}\leq c$. By Assumptions \ref{hyp:iid_bdX}.\ref{assn:Xborne} and \ref{hyp:np_for_consistency}.\ref{assn:smoothgamma}, there exists $C$ such that $\gamma_0 \in \mathcal{C}^{\ell+1}_C(\mathcal{X})$. Because $\ell+1>pT/2$, $\mathcal{C}^{\ell+1}_C(\mathcal{X})$ is a Donsker class for any $C>0$ \citep[see for instance Corollary 2.7.2 and Section 2.5.2 in][]{VdV_Wellner}.

\medskip
In the following, we denote by $C$, $\underline{C}$ and $\overline{C}$ generic constants subject to changes from one line to the other.

\subsubsection{Consistency of $(\widehat{\underline{\delta}},\widehat{\overline{\delta}})$}\label{sec:app_consistency}  

\begin{thm}\label{thm:consistency}
	Suppose that Assumptions \ref{hyp:model}-\ref{hyp:np_for_consistency} hold, with Assumption \ref{hyp:np_for_consistency}.\ref{assn:hn_cn} replaced by the weaker conditions $h_n \to 0$ and $n h_n^{pT}/\ln n \to \infty$. Then
	$$(\widehat{\underline{\delta}},\widehat{\overline{\delta}}) \convP (\underline{\delta}, \overline{\delta}).$$
\end{thm}

\label{sub:proof_prop_consistency}

\textbf{Proof:}  We focus on $\widehat{\underline{\delta}}$ hereafter, as the proof for the upper bound is the same. The proof proceeds in three steps. First, we show the uniform consistency of $\widetilde{m}$ over $\Supp(X)$. Second, we prove that $\widehat{m}$ is also uniformly consistent. Finally, we show the consistency of $\widehat{\underline{\delta}}$.

\subsubsection*{Step 1: Uniform consistency of $\widetilde{m}$} 
\label{ssub:uniform_consistency_of_widetilde_m}

Remark that under Assumptions \ref{hyp:model}-\ref{hyp:iid_bdX}, $P\in \mathcal{P}'$ as defined in Lemma \ref{lem:conv_unif_P}, with some appropriate $\underline{A}$. Then, by Lemma \ref{lem:conv_unif_P}, $\widehat{\beta}\convP \beta_0$. Moreover, $\Supp(X)\times B$ is compact. Then, for all $(k, x, \beta)\in \{0,...,T\}\times\Supp(X)\times B$,
\begin{equation}
	\overline{C} > C_k(x,\beta) \geq \underline{C} > 0, \quad  \overline{C} >\exp(k x_T'\beta) \geq \underline{C}.	
	\label{eq:bound_elem}
\end{equation}
Moreover, by definition of $c_k(\gamma, x,\beta)$,
\begin{align*}
	& \norm{c(\widehat{\gamma},x, \widehat{\beta}) - c(\gamma_0,x, \beta_0)}   \\
	& \ \leq C \left| \left|   \left(\frac{\widehat{\gamma}_0(x) e^{0 \times x_T'\widehat{\beta}} }{C_0(x,\widehat{\beta})}, \ ... \ , \frac{\widehat{\gamma}_T(x)e^{T \times x_T'\widehat{\beta}} }{C_T(x,\widehat{\beta})}\right)' - \left(\frac{\gamma_{00}(x)e^{0 \times x_T'\beta_0} }{C_0(x,\beta_0)}, \ ... \ , \frac{\gamma_{0T}(x)e^{T \times x_T'\widehat{\beta}}}{C_T(x,\beta_0)}\right)'\right| \right|.
\end{align*}
Fix $0 \leq k \leq T$ and remember that $\gamma_{0k}(x)=P(S=k|X=x)\in [0,1]$. Then,
{\small $$\left|\frac{\widehat{\gamma}_k(x)e^{k \times x_T'\widehat{\beta}}}{C_k(x,\widehat{\beta})} - \frac{\gamma_{0k}(x) e^{k \times x_T'\beta_0} }{C_k(x,\beta_0)}	\right|  \leq
	\frac{|\widehat{\gamma}_k(x) - \gamma_{0k}(x)| e^{k \times x_T'\widehat{\beta}} }{C_k(x,\widehat{\beta})} + \gamma_{0k}(x) \left| \frac{e^{k \times x_T'\beta_0}}{C_k(x,\beta_0)} -\frac{e^{k \times x_T'\widehat{\beta}}}{C_k(x,\widehat{\beta})} \right|.$$}\\The derivatives of $\beta \mapsto  e^{k \times x_T'\beta} / C_k(x,\beta)$ are uniformly bounded over $(x,\beta) \in \Supp(X)\times B$. Combined with \eqref{eq:bound_elem}, this implies that,
$$	\left|\frac{\widehat{\gamma}_k(x)e^{k \times x_T'\widehat{\beta}}}{C_k(x,\widehat{\beta})} - \frac{\gamma_{0k}(x) e^{k \times x_T'\beta_0} }{C_k(x,\beta_0)}	\right| \leq C \left(|\widehat{\gamma}_k(x) - \gamma_{0k}(x)| + \| \beta_0 - \widehat{\beta}\| \right).$$
Therefore, recalling that $\widehat{c}=c(\widehat{\gamma}, x,\widehat{\beta})$,
\begin{equation}
	\norminf{\widehat{c}-c} \leq C \left(\norminf{\widehat{\gamma} - \gamma_0} + \norm{\beta_0 - \widehat{\beta}} \right).
	\label{eq:bound_c_hat}
\end{equation}

Next, by \eqref{eq:bound_elem} and $\sum_{j=0}^T \gamma_{0j}(x) = 1$, for all $(x,\beta)\in \Supp(X)\times \{\beta_0,\widehat{\beta}\}$, then
\begin{equation}
	c_0(\gamma_{0},x, \beta)=\sum_{j=0}^T\binom{T}{j}\frac{\gamma_{0j}(x)\exp(jx_T'\beta_0)}{C_0(x,\beta)} >  \sum_{j=0}^T \gamma_{0j}(x) \underline{C}/\overline{C} = \underline{C}/\overline{C}.
	\label{eq:lower_bound_c0}	
\end{equation}
Assumptions \ref{hyp:iid_bdX}.\ref{assn:Xborne} and \ref{hyp:np_for_consistency}.\ref{assn:densityX} ensure that Assumption X in \cite{Fan_Guerre_16} holds. Assumptions \ref{hyp:np_for_consistency}.\ref{assn:K} and  $h_n\rightarrow 0$ ensure Assumption K in \cite{Fan_Guerre_16} holds. Assumption \ref{hyp:np_for_consistency}.\ref{assn:smoothgamma} ensures Assumption S2 in  \cite{Fan_Guerre_16} holds. Assumption \ref{hyp:np_for_consistency}.\ref{assn:hn_cn} and \ref{hyp:np_for_consistency}.\ref{assn:densityX} ensure that $h_n^{-pT}=O\left(\frac{n^{(\nu-2)/\nu}}{\ln(n)}\right)$ for any $\nu>4$. Let $\varepsilon:=\mathds{1}\{S=k\}-\gamma_{0k}(x)$; because $|\varepsilon|\leq 1$, we have $\sup_{x\in\Supp(X)}E\left[|\varepsilon|^{2+\nu}|X=x\right]\leq 1$ for any $\nu>4$. The continuity (Assumption \ref{hyp:np_for_consistency}.\ref{assn:smoothgamma}) and positivity of $\gamma_{00}(x),...,\gamma_{0T}(x)$ ensure that $E(\varepsilon^2|X=x)=\gamma_{0k}(x)(1-\gamma_{0k}(x))$ is continuous and bounded away from 0 by Assumption \ref{hyp:iid_bdX}.\ref{assn:Xborne}.
This means that all the assumptions of Proposition 7 in \cite{Fan_Guerre_16} hold except that $\varepsilon$ does not admit a conditional density $f_{\varepsilon|X=x}(e)$ continuous in $(e,x)$. This assumption is necessary to define conditional quantiles of $Y|X$ and to estimate them \citep[see Propositions 8, 9 and 10 in][]{Fan_Guerre_16}. However, a careful inspection of the proof of the Proposition 7 in \cite{Fan_Guerre_16} that deals with conditional expectation ensures that its conclusion also holds also for the discrete variable $\ind{S=k}$. Thus, $\widehat{\gamma}$ is uniformly consistent and  $\norm{\widehat{\gamma}-\gamma_0}_{\infty}=O_P\left(\eta_n\right)$ for $\eta_n=\left(\frac{\ln(n)}{nh_n^{pT}}\right)^{1/2}+h_n^{\ell+1}$.

Given \eqref{eq:bound_c_hat} and \eqref{eq:lower_bound_c0}, we then have $c_0(\widehat{\gamma},x, \widehat{\beta})  > C$ with probability $1-O\left(\eta_n\right)$.

\medskip
By definition of $\widetilde{m}$, we have, for all $(k,x)\in \{0,...,T\}\times\Supp(X)$, with probability $1-O\left(\eta_n\right)$
\begin{align}
	|\widetilde{m}_k(x)-m_k(x) | \leq &  \frac{1}{c_0(\gamma_0,x, \beta_0)} |c_k(\widehat{\gamma},x, \widehat{\beta})-c_k(\gamma,x, \beta_0)| \notag \\
	&  + \frac{1}{\tilde{c}_0^2} |c_k(\widehat{\gamma},x, \widehat{\beta})|\times |c_0(\widehat{\gamma},x, \widehat{\beta})-c_0(\gamma_0,x, \beta_0) | \label{eq:bound_mtilde}
\end{align}
where $\tilde{c}_0^2 \geq \min(c_0(\gamma_0,x, \beta)^2, c_0(\widehat{\gamma},x, \widehat{\beta})^2) > C$ and $|c_k(\widehat{\gamma},x, \widehat{\beta})|$ is bounded in probability in view of \eqref{eq:bound_c_hat}. Therefore, by \eqref{eq:bound_mtilde} and, again, \eqref{eq:bound_c_hat},
\begin{align*}
	\norm{\widetilde{m}-m}_\infty  & \leq  C \left( \norminf{c - \widehat{c}} +  \norm{c_0-\widehat{c}_0}_\infty \right) \\
	& \leq C \left(\norminf{\widehat{\gamma} - \gamma_0} + \norm{\beta_0 - \widehat{\beta}} \right).
\end{align*}
The result follows by uniform consistency of $\widehat{\gamma}$ and consistency of $\widehat{\beta}$ and we have $\norm{\widetilde{m}-m}_\infty=O_P\left(\eta_n\right)$.


\subsubsection*{Step 2: Uniform consistency of $\widehat{m}$} 
\label{ssub:uniform_consistency_of_widehat_m}

We drop the dependence in $x$ and write $m$, $\widehat{m}$,... instead of $m(x)$, $\widehat{m}(x)$,... to simplify notation as all the statements to follow hold uniformly over $x \in \Supp(X)$. We start by showing that for all $\epsilon > 0$ and for $n$ large enough, if $\widehat{I} = t$ then $ |m_{t+1} - \widehat{m}_{t+1}| \leq 2 
\epsilon$. A first step is to notice that for all $\epsilon > 0$, there exists $N_0$ such that $n \geq N_0$, $m \in \M_T$ and $\underline{H}_{t+1}(m_1, \, ..., \, m_{t+1}) <  2 c_n^{1/2}$ implies $|m_{t+1} - \underline{q}_{t}(m_{\to t})| =  |m_{t+1} - \widehat{m}_{t+1}|\leq \epsilon$. To see this, suppose the contrary. Then there exists $\epsilon>0$ and a subsequence $(m^{\phi(n)}) \in \M_T^{\N}$ such that for all $n\in\N$,
$$0 <\underline{H}_{t+1}(m_1^{\phi(n)}, \, ..., \, m_{t+1}^{\phi(n)}) < 2c_{\phi(n)}^{1/2} \; \text{ and } |m_{t+1}^{\phi(n)} - \underline{q}_{t}(m_{\to t}^{\phi(n)})| > \epsilon.$$
The set $\M_{T}$ is compact, thus there exists a further subsequence $(m^{\phi'(n)})$ converging to some $m^0$. By continuity of the functions $\underline{q}_{t}$ and $\underline{H}_{t+1}$, we have $\underline{H}_{t+1}(m_1^0, \, ..., \, m_{t+1}^0)=0$ and $|m_{t+1}^0 - \underline{q}_{t}(m_{\to t}^0)| \geq \epsilon>0$. But this contradicts Proposition \ref{prop:extension_mom}. The same result holds for $\overline{H}_{t+1}$.

\medskip

Define $C'$ a Lipschitz constant valid for both $\overline{H}_t$ and $\underline{H}_t$ for all $t \leq T$. Take $\epsilon>0$, $N_1$ larger than the corresponding $N_0$ and such that $n > N_1$ implies
\begin{align*}
	&\forall \, t \leq T, \, \|m_{\to t} - m_{\to t}'\| \leq \eta_n \Rightarrow |\underline{q}_t(m_{\to t}) - \underline{q}_t(m_{\to t}')| \leq \epsilon,\\
	&\eta_n \leq \epsilon \text{ and } \eta_n \leq c_n^{1/2} / C'.
\end{align*}
Then for $n \geq N_1$, for all $t \leq T$, if  $\widetilde{m}_{\to t} \in \M_t$ and $0 <\underline{H}_{t+1}(\widetilde{m}_1,...,\widetilde{m}_t,\widetilde{m}_{t+1}) < c_n^{1/2}$ then
$0 \leq \underline{H}_{t+1}(m_1, \, ..., \, m_{t+1}) \leq  c_n^{1/2} + C' \times \eta_n \leq 2 c_n^{1/2} $. Thus if $\widehat{I} = t$ and we are in the case $0 <\underline{H}_{t+1}(\widetilde{m}_1,...,\widetilde{m}_t,\widetilde{m}_{t+1}) < c_n^{1/2}$  then
\begin{align*}
	|m_{t+1} - \widehat{m}_{t+1}| & =	|m_{t+1} - \underline{q}_{t}(\widetilde{m}_{\to t})| \leq | m_{t+1} - \underline{q}_{t}(m_{\to t}) | + | \underline{q}_{t}(m_{\to t}) - \underline{q}_{t}(\widetilde{m}_{\to t}) | \\
	& \leq 2 \epsilon.
\end{align*}
The same result holds for $\overline{H}_{t+1}$.
We can then proceed by induction, as
\begin{align*}
	|m_{t+2} - \widehat{m}_{t+2}| & =	|m_{t+2} - \underline{q}_{t+1}(\widetilde{m}_{\to t},  \widehat{m}_{t+1})| \\
	& \leq |\overline{q}_{t+1}(m_{\to t+1}) -\underline{q}_{t+1}(m_{\to t+1}) | + |\underline{q}_{t+1}(m_{\to t+1}) - \underline{q}_{t+1}(\widetilde{m}_{\to t},  \widehat{m}_{t+1}) |\\
	& \leq  |\overline{q}_{t+1}(m_{\to t+1}) -\overline{q}_{t+1}(\widetilde{m}_{\to t},  \widehat{m}_{t+1})| + |\underline{q}_{t+1}(\widetilde{m}_{\to t},  \widehat{m}_{t+1}) -\underline{q}_{t+1}(m_{\to t+1}) | \\
	& \quad + |\underline{q}_{t+1}(m_{\to t+1}) - \underline{q}_{t+1}(\widetilde{m}_{\to t},  \widehat{m}_{t+1}) |
\end{align*}
where the last inequality follows from
$\underline{q}_{t+1}(\widetilde{m}_{\to t},  \widehat{m}_{t+1}) = \overline{q}_{t+1}(\widetilde{m}_{\to t},  \widehat{m}_{t+1})$. Using recursively the uniform continuity of $\overline{q}_{t'}$ and $\underline{q}_{t'}$  as functions of $m_{\to t}$ over $\M_t$ and properly adjusting recursive choices of the $\epsilon$'s, we then obtain the uniform convergence of $\widehat{m} - m$  to $0$.


\subsubsection*{Step 3: Consistency of the  bounds} 
\label{ssub:consistency_of_the_lower_bound}

We just consider estimation of the lower bound. Similar reasoning applies to the upper bound.
Let $\widehat{A}(x):= \widehat{c}_0(x)  \lambda_{T+1}(x, \widehat{\beta}) {\underline{q}}_T(\widehat{m}(x))$ and $\widehat{B}(x):= \widehat{c}_0(x)  \lambda_{T+1}(x, \widehat{\beta}) {\overline{q}}_T(\widehat{m}(x))$. By Equation \eqref{eq:def_estim_bds}, $\widehat{\underline{\delta}}$ satisfies
\begin{equation}
	\widehat{\underline{\delta}}  = \frac{1}{n} \sum_{i=1}^n r(X_i,S_i,\widehat{\beta}) + \frac{1}{n} \sum_{i=1}^n \min\left(\widehat{A}(X_i),\widehat{B}(X_i)\right).
	\label{eq:decomp_borne}
\end{equation}
Since $\lambda_t$ is infinitely differentiable for all $t \leq T$, $\Supp(X)$ is compact and $\widehat{\beta}$ is consistent, $x \mapsto \lambda_t(x,\widehat{\beta})$ converges uniformly in probability to $x \mapsto \lambda_t(x,\beta_0)$. The same holds for $(x,s) \mapsto C_s(x,\widehat{\beta})$ and $(x,s) \mapsto \exp(sX_T'\widehat{\beta}$). Because $ C_s(x,\beta) > \underline{C}$ for all $(x,\beta) \in \Supp(X)\times B$ and $s \leq T$,
$(x,s) \mapsto r(x,s,\widehat{\beta})$ converges uniformly in probability to $(x,s) \mapsto r(x,s,\beta_0)$. Then, by the triangle inequality and the law of large numbers (LLN),
$$ \frac{1}{n} \sum_{i=1}^n r(X_i,S_i,\widehat{\beta})  \convP \E\left(r(X,S,\beta_0)\right).$$
Let us show the convergence in probability of the second term in \eqref{eq:decomp_borne}. The functions $\overline{q}_T$ and $\underline{q}_T$ are continuous and thus uniformly continuous over the compact set $\M_T$. Then, by Step 2 and since by construction $(m(x),\widehat{m}(x))\in\M_T^2$, $x \mapsto \underline{q}_T(\widehat{m}(x))$ and   $x \mapsto \overline{q}_T(\widehat{m}(x))$  converge uniformly in probability to $x \mapsto \underline{q}_T(m(x))$ and   $x \mapsto \overline{q}_T(m(x))$ respectively. $x\mapsto \lambda_{T+1}(x,\widehat{\beta})$ converges uniformly to $x\mapsto \lambda_{T+1}(x,\beta_0)$ for the same reasons than $\lambda_t$ for $t\leq T$. Uniform consistency of $\widehat{\gamma}$ and \eqref{eq:bound_c_hat} ensure uniform consistency of $\widehat{c}_0(x)$ to $c_0(x)$. Because $\sup_{x\in \Supp (X)}|c_0(x)|<\infty$, $\sup_{(x,b)\in \Supp (X)\times B} |\lambda_{T+1}(x,b)|<0$, $\sup_{m\in \mathcal{M}_T}|\overline{q}_T(m)|\leq 1$ and $\sup_{m\in \mathcal{M}_T}|\underline{q}_T(m)|\leq 1$, the functions $\widehat{A}$ and $\widehat{B}$ converge uniformly in probability to their corresponding limits $A(.)=c_0(.)\lambda_{T+1}(.,\beta_0)\underline{q}_{T}(m(.))$ and $B(.)=c_0(.)\lambda_{T+1}(.,\beta_0)\overline{q}_{T}(m(.))$. Since $\min(A,B) = (A + B - |A - B|)/2$, the function $x \mapsto \min(\widehat{A}(x),\widehat{B}(x))$ also converges uniformly in probability to $x\mapsto \min(A(x),B(x))$. Then, by the triangle inequality and the LLN,
$$ \frac{1}{n} \sum_{i=1}^n \min\left(\widehat{A}(X_i),\widehat{B}(X_i)\right)\convP E\left(\min\left(A(X),B(X)\right)\right).$$
The result follows.



\subsubsection{Theorem \ref{thm:as_normality}} 
\label{sub:proof_prop_asnormality_old}

Theorem 10.7 in \cite{schmudgen2017moment} combined with Lemma \ref{lem:cushion_bis} ensure that $I(x)=\max\{t\in\{1,...,T\}: \forall j\leq t, \underline{H}_{j}(m_{\rightarrow j}(x))\overline{H}_{j}(m_{\rightarrow j}(x))>0\}$ does not depend on $x$ and $I$ denotes this common value.

Then, let
\begin{align}m(\gamma,x,\beta)&:=\left(1,\frac{c_1(\gamma,x,\beta)}{c_0(\gamma,x,\beta)},...,\frac{c_I(\gamma,x,\beta)}{c_0(\gamma,x,\beta)}\right),\label{eq:defin_m}\end{align}
so that $m(\gamma_0,x,\beta_0)=m_{\rightarrow I}(x)$ and $m(\widehat{\gamma},x,\widehat{\beta})=\widetilde{m}_{\rightarrow I}(x)$. Now, if $I=T$, we let, with a slight abuse of notation, $\underline{q}_T(\gamma,x,\beta)=\underline{q}_T(m(\gamma,x,\beta))$. If $I<T$, by Assumption \ref{hyp:cushion_m} and Proposition \ref{prop:extension_mom},
$m_{I+1}(x)=\underline{q}_I(m_{\rightarrow I}(x))$ or $m_{I+1}(x)=\overline{q}_I(m_{\rightarrow I}(x))$. Then, by Proposition \ref{prop:extension_mom} again and a straightforward induction, we can define $m_t(x)$ for $t\in\{I+1,...,T\}$ as a function of $m_{\rightarrow I}(x)$. We let $\text{Ext}(.)$ denote the corresponding extension function. Then $m(x)=\text{Ext}(m_{\rightarrow I}(x))$. Finally, we let (with again a slight abuse of notation)
\begin{align}\underline{q}_T(\gamma,x,\beta):=\underline{q}_T(\text{Ext}(m(\gamma,x,\beta))).\label{eq:ext}\end{align}
We define similarly $\overline{q}_T(\gamma, x,\beta)$. Note that $\underline{q}_T(\cdot,\cdot,\cdot)$ and $\overline{q}_T(\cdot,\cdot,\cdot)$ depend on the unknown $I$ and on the true unknown functions $m$ and not on $\widehat{m}$ or $\widehat{I}$. Conversely, $\underline{q}_T(\widehat{m}(x))$ and $\overline{q}_T(\widehat{m}(x))$ depend on $\widehat{I}(x)$ and $\widehat{m}(x)$. However, we show in the proof of Theorem \ref{thm:as_normality} below that with probability approaching one, $\underline{q}_T(\widehat{m}(x)) = \underline{q}_T(\widehat{\gamma}, x,\widehat{\beta})$.

\medskip
Then, we also define
\begin{align}
	\underline{h}( x, s, \gamma, \beta,\widetilde{\beta}  )  = & r(x,s,\beta) + c_0(\gamma, x, \beta)  \lambda_{T+1}(x, \beta)  \big[\underline{q}_T(\gamma, x, \beta) \ind{\lambda_{T+1}(x, \widetilde{\beta}) > 0} \notag\\
	& + \overline{q}_T(\gamma, x, \beta) \ind{ \lambda_{T+1}(x, \widetilde{\beta}) < 0}\big], \label{eq:defin_underline_h_long}\\
	\overline{h}( x, s, \gamma, \beta ,\widetilde{\beta} )  = & r(x,s,\beta) +  c_0(\gamma, x, \beta) \lambda_{T+1}(x, \beta)  \big[\overline{q}_T(\gamma, x, \beta) \ind{\lambda_{T+1}(x, \widetilde{\beta}) > 0} \notag\\
	& + \underline{q}_T(\gamma, x, \beta) \ind{ \lambda_{T+1}(x, \widetilde{\beta}) < 0}\big].\label{eq:defin_overline_h_long}
\end{align}

Note that $\underline{h}( x, s, \gamma, \beta, \widetilde{\beta} )$ (and similarly $\overline{h}( x, s, \gamma, \beta,\widetilde{\beta})$) depends on $\gamma$ only through $\gamma(x)$. Also, $\underline{h}$ is differentiable with respect to the vectors $\beta$ and $\gamma(x)$. Let $\underline{v}_\beta$ denote $E\left(\deriv{\underline{h}(X,S,\gamma,\beta,\widetilde{\beta})}{\beta}\right)$ for $(\gamma, \beta, \widetilde{\beta})=(\gamma_0,\beta_0,\beta_0)$. Note that we consider here the gradient with respect to $\beta$ (but not with respect to $\widetilde{\beta}$). Let $\underline{v}_\gamma(x,s)$ the gradient of $\underline{h}(x, s, \gamma_0, \beta_0,\beta_0)$ with respect to $\gamma(x)$. If $\beta=\widetilde{\beta}$, we denote (with a slight abuse of notation) $\underline{h}( x, s, \gamma, \beta)$ for $\underline{h}( x, s, \gamma, \beta, \beta )$ and $\underline{h}(x,s)=\underline{h}(x,s,\gamma_0,\beta_0)$. We define similarly $\overline{v}_{\beta}$, $\overline{v}_{\gamma}(x,s)$, $\overline{h}(x,s,\gamma, \beta)$ and $\overline{h}(x,s)$.

\medskip
When $a(\beta_{0k})\neq 0$, the influence functions of $\widehat{\underline{\delta}}$ and $\widehat{\overline{\delta}}$ are:
\begin{align}
	\underline{\psi}_i= & \underline{h}( X_i, S_i) -  \E[\underline{h}(X, S)] + \underline{v}_{\beta}' \phi_i + \underline{v}_\gamma(X_i, S_i)' [\Gamma_i - \gamma_0(X_i)], 	\label{eq:def_psi_inf}
	\\
	\overline{\psi}_i= & \overline{h}(X_i, S_i) -  \E[\overline{h}(X,S)] + \overline{v}_\beta' \phi_i  + \overline{v}_\gamma(X_i, S_i)' [\Gamma_i - \gamma_0(X_i)],
	\label{eq:def_psi_sup}
\end{align}
where $\Gamma_i = (\ind{S_i = 0}, \ ... ,  \ind{S_i = T})'$ and
$\phi_i = \mathcal{I}_0^{-1}\deriv{\ell_c}{\beta}(Y_i|X_i;\beta_0)$ is the influence function of $\widehat{\beta}$.
We let $\Sigma$ denote the variance-covariance matrix of $(\underline{\psi},\overline{\psi})$. \medskip

When $a(\beta_{0k})=0$, the estimator is no more asymptotically linear (and next the influence functions are no more defined). For any $h\in \mathbb{R}^p$, we define
\begin{align*}
	\underline{k}(x,s,h)=&E\left(\Deriv{r}{\beta}(X,S,\beta_0)\right)'h\\&+c_0(\gamma_0,x,\beta_0)\min\left(\underline{q}_T(m(x))\Deriv{\lambda_{T+1}}{\beta}(x,\beta_0)'h, \overline{q}_T(m(x))\Deriv{\lambda_{T+1}}{\beta}(x,\beta_0)'h\right), \\
	\overline{k}(x,s,h)=&E\left(\Deriv{r}{\beta}(X,S,\beta_0)\right)'h\\&+c_0(\gamma_0,x,\beta_0)\max\left(\underline{q}_T(m(x))\Deriv{\lambda_{T+1}}{\beta}(x,\beta_0)'h, \overline{q}_T(m(x))\Deriv{\lambda_{T+1}}{\beta}(x,\beta_0)'h\right),
\end{align*}
and $\underline{K}(h)=E\left(\underline{k}(X,S,h)\right)$, $\overline{K}(h)=E\left(\overline{k}(X,S,h)\right)$. Even if $\underline{\delta}$ does not have an influence function, we have \begin{align}\sqrt{n}\left(\widehat{\underline{\delta}}-\underline{\delta}\right)&=\left[\frac{1}{\sqrt{n}}\sum_{i=1}^nr(X_i,S_i,\beta_0)-E\left(r(X,S,\beta_0)\right)\right]+\underline{K}\left(\frac{1}{\sqrt{n}}\sum_{i=1}^n\phi_i\right)+o_P(1),\label{eq:approx_kink}\end{align}
and similar approximation holds for $\sqrt{n}\left(\widehat{\overline{\delta}}-\overline{\delta}\right)$. Let $\Omega$ the covariance matrix of $(r(X,S,\beta_0), \phi)$.

\bigskip

\textbf{Proof of Theorem \ref{thm:as_normality}:}

First, we show that with probability approaching one (wpao), $\widehat{I}(X_i)=I$ for all $i$. Next, we obtain some stochastic approximations. When $a(\beta_{0k})\neq 0$, the approximation is linear and the asymptotic normality follows directly the central limit theorem. When $a(\beta_{0k})= 0$, the approximation is not  linear anymore and the limit distribution follows from the continuous mapping theorem.

\subsubsection*{Step 1: with probability approaching one, $\widehat{I}(X_i)=I$ for all $i$.} 

First, let $G_t(m) := \underline{H}_t(m) \overline{H}_t(m)$. By definition of $\widehat{I} (x)$ and $I$,
$$
\widehat{I} (x) > I  \ \Rightarrow \ G_{I + 1}(m(x)) = 0 \text{ and }  G_{I + 1}(\widetilde{m}(x))  > c_n.
$$
Moreover,
$$
G_{I + 1}(\widetilde{m}(x))  > c_n \Rightarrow \underline{H}_{I+1}(\widetilde{m}(x)) \overline{H}_{I+1}(\widetilde{m}(x)) -  \underline{H}_{I+1}(m(x)) \overline{H}_{I+1}(m(x)) > c_n,
$$
The functions $\underline{H}_{I+1}$ and $\overline{H}_{I+1}$ are infinitely differentiable on the compact set $\M_{I+1}$. As the product of these functions, $G_{I+1}$ is thus Lipschitz on this set. By induction, $(\widetilde{m}_0(x), ..., \widetilde{m}_{I+1}(x))$ lies in $\M_{I+1}$. Indeed, otherwise we would not have $\widehat{I}(x)\geq I+1$.
This implies that  for any given value $x  \in \Supp(X)$, 
$$
G_{I + 1}(\widetilde{m}(x))  > c_n \Rightarrow \|\widetilde{m}(x) - m(x)\| > C c_n.
$$
Because $\eta_n/c_n \to 0$, this cannot occur for any $x \in \Supp(X)$, with probability approaching one. Hence, with probability approaching one, $\widehat{I}(X_i) \leq  I  $ for all $i$.

Now, assume that $\widehat{I} (  x) < I$ for some $x\in \Supp(X)$. Then,
\begin{equation*}
	\exists k = \widehat{I}(x) + 1 \leq I, \ G_k(\widetilde{m}(x))\leq c_n  \text{ and }   G(m(x))  > 0 \text{ for any }j\leq k.
\end{equation*}
We know $m_{\rightarrow k}(x) \in \text{int} \left(\mathcal{M}_k\right)$ for any $k \leq I$. $G_k$ and $m_{\rightarrow k}$ are  continuous functions and $\Supp(X)$ is a compact set, thus as in Lemma \ref{lem:cushion_bis}, $\inf_{x \in \Supp(X)} G_k(m(x)) >0$.
So for sufficiently large $n$, $\inf_{x \in \Supp(X)} G_k(m(x))/2>c_n>0$.
By triangle inequality and Lipschitz property of $G_k$ we deduce:
\begin{align*}
	c_n\geq G_k(\widetilde{m}(x)) & \geq  \inf_{x \in \Supp(X)} G_k(m(x)) \ -  |G_k(\widetilde{m}(x)) - G_k(m(x)) | \\
	& \geq 2 c_n - C \|m - \widetilde{m}\|_{\infty}>2c_n-O_P(\eta_n),
\end{align*}
By $\eta_n=o(c_n)$ we conclude that with probability approaching one $\widehat{I} (x) \geq I$ for any $x\in \Supp(X)$.

\medskip
In conclusion, with probability approaching one, we have $\widehat{I} (X_i) = I$ for all $i\in\{1,...,n\}$.

\subsubsection*{Step 2 : asymptotic approximation when $a(\beta_{0k})\neq 0$} 
\label{ssub:part_1}
We focus on $\widehat{\underline{\delta}}$: the proof for $\widehat{\overline{\delta}}$ is similar. This second step is divided into two sub-steps.  In the first sub-step, we prove that for an event $A_n$ with probability approaching zero and a random variable $R_n$, we have:
\begin{align}
	\widehat{\underline{\delta}} =& \frac{1}{n} \sum_{i=1}^n \underline{h}(X_i,S_i,\widehat{\gamma},\widehat{\beta},\beta_0) + R_n \ind{A_n} + o_P(n^{-1/2}).
	\label{eq:delta_l_equiv}
\end{align}
Next, in a second sub-step we deduce that:
\begin{align}\sqrt{n}\left(\widehat{\underline{\delta}}-\underline{\delta}\right)=\frac{1}{\sqrt{n}}\sum_{i=1}^n\underline{\psi}_i+o_P(1)\label{eq:delta_l_equiv2}\end{align} for $\underline{\psi}_i$ defined in \eqref{eq:def_psi_inf}.

In this step of the proof, $R_n$ and $A_n$ denote a generic random variable and a generic event with probability approaching to zero which may differ in different places.

\subsubsection*{Step 2, sub-step 1: proof of \eqref{eq:delta_l_equiv}}
Let $A_n=\{\forall \, i\in\{1,...,n\}, \ (\underline{q}_T(\widehat{m}(X_i)),\overline{q}_T(\widehat{m}(X_i)) )= (\underline{q}_T(\widehat{\gamma}, X_i,\widehat{\beta}),\overline{q}_T(\widehat{\gamma}, X_i,\widehat{\beta}))\}^c$, where recall that $B^c$ denotes the complement of a set $B$, and $R_n=\widehat{\underline{\delta}} - \left[\frac{1}{n}\sum_{i=1}^n   \underline{h}(X_i,S_i,\widehat{\gamma},\widehat{\beta})\right]$. From Step 1, $A_n$ has a probability approaching zero. This in turn implies
\begin{align}
	\widehat{\underline{\delta}}  = \left[\frac{1}{n}\sum_{i=1}^n   \underline{h}(X_i,S_i,\widehat{\gamma},\widehat{\beta})\right]+R_n\ind{A_n} . \label{eq:equiv_step2}
\end{align}
To obtain  (\ref{eq:delta_l_equiv}), we define
\begin{align*}
	J_n = \ & \frac{1}{n}  \sum_{i=1}^n\underline{h}(X_i,S_i,\widehat{\gamma},\widehat{\beta})-\underline{h}(X_i,S_i,\widehat{\gamma},\widehat{\beta},\beta_0)\\=\ & \frac{1}{n}  \sum_{i=1}^n \widehat{c}_0(X_i) \lambda_{T+1}(X_i, \widehat{\beta}) \left[ \overline{q}_T(\widehat{\gamma}, X_i,\widehat{\beta})  \left(  \ind{\lambda_{T+1}(X_i,\widehat{\beta})\geq 0} - \ind{\lambda_{T+1}(X_i,\beta_0) \geq 0 } \right) \right.\\
	& \qquad \qquad \qquad \qquad \qquad \ \  \left. + \  \underline{q}_T(\widehat{\gamma}, X_i,\widehat{\beta})  \left( \ind{\lambda_{T+1}(X_i,\widehat{\beta}) < 0} -   \ind{\lambda_{T+1}(X_i,\beta_0) < 0 } \right) \right] \\	
	= \ &  \frac{1}{n}  \sum_{i=1}^n \widehat{c}_0(X_i) \lambda_{T+1}(X_i, \widehat{\beta})  \left[ \underline{q}_T(\widehat{\gamma}, X_i,\widehat{\beta}) - \overline{q}_T(\widehat{\gamma}, X_i,\widehat{\beta})  \right]\\
	& \qquad \qquad \qquad \qquad \qquad \qquad \
	\left[ \ind{\lambda_{T+1}(X_i,\widehat{\beta}) < 0} -   \ind{\lambda_{T+1}(X_i,\beta_0) < 0} \right].
\end{align*}
We prove now that $J_n=o_P(n^{-1/2})$ which will guarantee that (\ref{eq:delta_l_equiv}) holds. By definition of $c(\gamma,x,\beta)$, continuity of $C_k(x,\beta)$ and $\exp(kx'\beta)$ ensures that $c_0(X_i)<C$ and by Equation \eqref{eq:bound_c_hat} and because $|\widehat{\overline{q}}_T(m)-\widehat{\underline{q}}_T(m)|\leq 1$ for any $m\in \mathcal{M}_T$, we have
\begin{align*}
	J_n	\leq \  \left(C+O_p(\eta_n)\right) \frac{1}{n}  \sum_{i=1}^n \left| \lambda_{T+1}(X_i,\widehat{\beta}) \right|  \left| \ind{\lambda_{T+1}(X_i,\widehat{\beta}) < 0} -   \ind{\lambda_{T+1}(X_i,\beta_0) < 0} \right|.
\end{align*}

Because $\left|\lambda_{T+1}(X_i,\widehat{\beta})-\lambda_{T+1}(X_i,\beta_0)\right|\leq C ||\widehat{\beta}-\beta_0||$, we have:
\begin{align}
	&\sqrt{n}|J_n|\nonumber\\&\leq (C+O_p(\eta_n))\frac{1}{\sqrt{n}}\sum_{i=1}^n|\lambda_{T+1}(X_i,\beta_0)|\left|\ind{  \lambda_{T+1}(X_i,\widehat{\beta}) < 0  }  -   \ind{\lambda_{T+1}(X_i,\beta_0) < 0}\right|\nonumber\\
	&+(C'+O_p(\eta_n))\left(\sqrt{n}||\widehat{\beta}-\beta_0||\right)\frac{1}{n}\sum_{i=1}^n\left|\ind{  \lambda_{T+1}(X_i,\widehat{\beta}) < 0  }  -  \ind{\lambda_{T+1}(X_i,\beta_0) < 0}\right|, \label{eq:Jn}
	\end{align}
and  $\left|\mathds{1}\{\lambda_{T+1}(X_i,\widehat{\beta})<0\}-\ind{\lambda_{T+1}(X_i,\beta_0) < 0}\right|\leq\ind{|\lambda_{T+1}(X_i,\beta_0)|\leq C||\widehat{\beta}-\beta_0||}$.\\ Let $F_{|\lambda_{T+1}|}$ the cumulative distribution function of $|\lambda_{T+1}(X_1,\beta_0)|$. The Glivenko-Cantelli theorem ensures that \begin{align*}\frac{1}{n}\sum_{i=1}^n\left|\mathds{1}\{\lambda_{T+1}(X_i,\widehat{\beta})<0\}-\ind{\lambda_{T+1}(X_i,\beta_0) < 0}\right|&\leq F_{|\lambda_{T+1}|}(C||\widehat{\beta}-\beta_0||)+o_P(1).\end{align*}

Because $P\left(\lambda_{T+1}(X_1,\beta_0)=0\right)=0$ and $||\widehat{\beta}-\beta_0||=o_P(1)$, we have $F_{|\lambda_{T+1}|}(C||\widehat{\beta}-\beta_0||)=o_P(1)$ ensuring that the second term in the right hand side of \eqref{eq:Jn} is $o_P(1)$.\\ Let $G(c)=E\left(\left|\lambda_{T+1}(X_1,\beta_0)\right|\ind{\left|\lambda_{T+1}(X_1,\beta_0)\right|\leq c}\right)$,\\ $\mathbb{G}_n(c)=\frac{1}{\sqrt{n}}\sum_{i=1}^n\left[\left|\lambda_{T+1}(X_i,\beta_0)\right|\ind{  \left|\lambda_{T+1}(X_i,\beta_0)\right| \leq c  }-G(c)\right]$ and $\mathbb{G}(c)$ the centered Gaussian process of covariance: $$\rho(c_1,c_2)=E\left(\lambda_{T+1}^2(X_1,\beta_0)\ind{|\lambda_{T+1}(X_1,\beta_0)|\leq c_1\wedge c_2}\right)-G(c_1) G(c_2).$$  We have:
\begin{align*}
	&\frac{1}{\sqrt{n}}\sum_{i=1}^n\left|\lambda_{T+1}(X_i,\beta_0)\right|\left|\ind{  \lambda_{T+1}(X_i,\widehat{\beta}) < 0  }  -   \ind{  \lambda_{T+1}(X_i,\beta_0) < 0  }\right|\\&\leq 	\mathbb{G}_n(C||\widehat{\beta}-\beta_0||)+\sqrt{n}G(C||\widehat{\beta}-\beta_0||)
	\end{align*}
Note that $G(c)=\int_0^{\infty}\int_0^{\infty}\ind{t\leq u\leq c}dt dF_{|\lambda_{T+1}|}(u)=\int_0^{c}F_{|\lambda_{T+1}|}(c)-F_{|\lambda_{T+1}|}(t)dt$ and next $|G(c)|\leq c (F_{|\lambda_{T+1}|}(c)-F_{|\lambda_{T+1}|}(0))$. Because $F_{|\lambda_{T+1}|}$ is right continuous, we have $G(c)=o(c)$ for $c\downarrow 0$ and next $\sqrt{n}G(C||\widehat{\beta}-\beta_0||)= o_P(\sqrt{n}||\widehat{\beta}-\beta_0||)=o_P(1)$. $\phi(u,v)=u\times v$ is Lipschitz on $[0;\sup_{x}|\lambda_{T+1}(x,\beta_0)|]\times [0,1]$. Stability of Donsker properties by Lipschitz transformation ensures that $\{f(x,c)=\left|\lambda_{T+1}(x,\beta_0)\right|\ind{\left|\lambda_{T+1}(x,\beta_0)\right|\leq c};c\in \mathbb{R}\}$ is a Donsker class.  Then we have $\mathbb{G}_n(C||\widehat{\beta}-\beta_0||)=\mathbb{G}(C||\widehat{\beta}-\beta_0||)+o_P(1)$. Moreover, $\mathbb{G}$ admits almost-surely uniformly continuous sample paths. Next $||\widehat{\beta}-\beta_0||=o_P(1)$ implies $\mathbb{G}(C||\widehat{\beta}-\beta_0||)=o_P(1)$ because $\mathbb{G}(0)=0$. This ensures that the first term in the right hand side of \eqref{eq:Jn} is also $o_P(1)$. It follows that $\sqrt{n}|J_n|=o_P(1)$ and \eqref{eq:delta_l_equiv} holds.

\subsubsection*{Step 2, sub-step 2: proof of \eqref{eq:delta_l_equiv2}}

Define $H_n(\gamma, \beta) := \frac{1}{n} \sum_{i=1}^n \underline{h}( X_i, S_i, \gamma, \beta,\beta_0) $ and $H(\gamma, \beta) := \E( \underline{h}( X, S, \gamma, \beta,\beta_0))$, so that $H(\gamma_0, \beta_0) = \underline{\delta}$. Let $\mathbb{H}_n(\gamma, \beta)=\sqrt{n}\left(H_n(\gamma, \beta)-H(\gamma, \beta)\right)$. Then, by \eqref{eq:delta_l_equiv},
$$\widehat{\underline{\delta}} = H_n(\widehat{\gamma}, \widehat{\beta}) + R_n \ind{A_n} + o_P(n^{-1/2}).$$
Next, for $R_n$ a generic sequence of random variable and $A_n$ an event tending to zero in probability:
\begin{align*}
	\sqrt{n}\left(\widehat{\underline{\delta}}-\underline{\delta}\right)=&\left[\mathbb{H}_n(\widehat{\gamma}, \widehat{\beta})+\sqrt{n}H(\widehat{\gamma}, \widehat{\beta})-\sqrt{n}H(\gamma_0, \beta_0)\right]+ R_n \ind{A_n} + o_P(1)\\
	=&\left[\mathbb{H}_n(\gamma_0, \beta_0)+\sqrt{n}\left(H(\widehat{\gamma}, \widehat{\beta})-H(\gamma_0, \beta_0)\right)\right]\\
	&+\left[\mathbb{H}_n(\widehat{\gamma}, \widehat{\beta})-\mathbb{H}_n(\gamma_0, \beta_0)\right]+ o_P(1)
\end{align*}

To prove $\sqrt{n}\left(\widehat{\underline{\delta}}-\underline{\delta}\right)=\frac{1}{\sqrt{n}}\sum_{i=1}^n\underline{\psi}_i+o_P(1)$ for $\underline{\psi}_i$ defined in \eqref{eq:def_psi_inf}, we will prove the two following properties: \begin{align}\mathbb{H}_n(\widehat{\gamma}, \widehat{\beta})-\mathbb{H}_n(\gamma_0, \beta_0)=&o_P(1)\label{eq:cont_emp_proc}\\\sqrt{n}\left(H(\widehat{\gamma}, \widehat{\beta})-H(\gamma_0, \beta_0)\right)=&\frac{1}{\sqrt{n}}\sum_{i=1}^n\underline{v}_{\gamma}(X_i,S_i)'\left(\Gamma_i-\gamma_{0}(X_i)\right)+\frac{1}{\sqrt{n}}\sum_{i=1}^n\underline{v}_{\beta}'\phi_i+o_P(1).\label{eq:deriv_first_step}
\end{align}

In Lemma \ref{lem:cond_as_nor} in the supplementary material, we show the three following conditions \citep[see][for a discussion of these conditions]{newey94}:
\begin{enumerate}
	\item \label{cond.lin} Linearization: There exists $b(\cdot)$ such that $\E(b(X_i)) < \infty$ and some $\eta>0$ such that for any $\gamma,\beta$ such that $\|\gamma - \gamma_0\|_{\infty} + \|\beta - \beta_0\|\leq \eta$, we have
	\begin{align*}
		& \big|\underline{h}(X_i, S_i, \gamma, \beta,\beta_0) - \underline{h}( X_i, S_i, \gamma_0, \beta_0) \\
		& - \underline{v}_\gamma (X_i, S_i)' [\gamma(X_i) - \gamma_0(X_i)] - \Deriv{\underline{h}}{\beta} ( X_i, S_i, \gamma_0, \beta_0, \beta_0)' [\beta - \beta_0] \big| \\
		\leq & \; b(X_i) \left(\|\gamma - \gamma_0\|_{\infty}^2 + \|\beta - \beta_0\|^2\right),
	\end{align*}		
	and $\sqrt{n}\|\widehat{\gamma} - \gamma_0\|_{\infty}^2 =o_P(1)$.
	\item \label{cond.sto_equi} Stochastic equicontinuity: we have
	\begin{align*}
		\frac{1}{\sqrt{n}}\sum_{i=1}^n\left( \underline{v}_{\gamma}(X_i,S_i)'(\widehat{\gamma}(X_i)-\gamma_0(X_i))- \int \underline{v}_{\gamma}(x,s)'(\widehat{\gamma}(x)-\gamma_0(x))dP(x,s)\right)
		&=o_P(1)
		\end{align*}
	\item \label{cond.mean_sq_cont} Mean-square continuity: we have $\E\left[\left(\underline{v}_{\gamma}(X_1,S_1)'(\Gamma_1-\gamma_0(X_1))\right)^2\right]<\infty$ and
	\begin{align*}
		\sqrt{n}\left(\int \underline{v}_{\gamma}(x,s)'(\widehat{\gamma}(x)-\gamma_0(x))dP(x,s)-\frac{1}{n}\sum_{i=1}^n\underline{v}_{\gamma}(X_i,S_i)'(\Gamma_i-\gamma_0(X_i))\right)
		&=o_P(1)
		\end{align*}
	\end{enumerate}

We also need to take into account the first step estimation of the nuisance parameter $\beta_0$. Because $\E\left[\left|\left|\deriv{\underline{h}}{\beta}(X_1,S_1,\gamma_0,\beta_0,\beta_0)\right|\right|\right]<\infty$ (by similar arguments as those used to prove Conditions \ref{cond.lin}, \ref{cond.sto_equi} and \ref{cond.mean_sq_cont}), the LLN and the asymptotic properties of the CMLE imply:
\begin{align}
	\frac{1}{\sqrt{n}}\sum_{i=1}^n \Deriv{\underline{h}}{\beta}(X_i,S_i,\gamma_0,\beta_0,\beta_0)'(\widehat{\beta}-\beta_0)
	&=\sqrt{n}\underline{v}_{\beta}'(\widehat{\beta}-\beta_0)+o_P(1)\label{cond.sto_equi_beta}\\
	\sqrt{n}\underline{v}_{\beta}'(\widehat{\beta}-\beta_0)&=\frac{1}{\sqrt{n}}\sum_{i=1}^n\underline{v}_{\beta}'\phi_i+o_P(1),\label{cond.mean_sq_cont_beta}
	\end{align}
for $\phi_i$ the influence function of the CMLE $\widehat{\beta}$.

Next, \eqref{eq:cont_emp_proc} follows from the triangular inequality, Conditions \ref{cond.lin}, \ref{cond.sto_equi} and Equation  \eqref{cond.sto_equi_beta}:
\begin{align*}\left|\mathbb{H}_n(\widehat{\gamma},\widehat{\beta})-\mathbb{H}_n(\gamma_0,\beta_0)\right|&\leq \left(\frac{1}{n}\sum_{i=1}^n b(X_i)+E\left(b(X_1)\right)\right)\sqrt{n}\left(\|\widehat{\gamma} - \gamma_0\|_{\infty}^2 + \|\widehat{\beta} - \beta_0\|^2\right)+o_P(1)\\
&=o_P(1),\end{align*}
and \eqref{eq:deriv_first_step} follows from triangular inequality, Conditions \ref{cond.lin}, \ref{cond.mean_sq_cont} and \eqref{cond.mean_sq_cont_beta}. Next, \eqref{eq:delta_l_equiv2} follows. And similarly, we have $\sqrt{n}\left(\widehat{\overline{\delta}}-\overline{\delta}\right)=\frac{1}{\sqrt{n}}\sum_{i=1}^n\overline{\psi}_i+o_P(1)$ for $\overline{\psi}_i$ defined in \eqref{eq:def_psi_sup}. The Central Limit Theorem ensures asymptotic normality.


\subsubsection*{Step 3: asymptotic approximation  when $a(\beta_{0k})=0$.} 
\label{sub:part_2}

Stability properties of Donsker classes ensure that  $\underline{\mathcal{K}}:=\{(x,s)\mapsto \underline{h}(x,s,\gamma,\beta): \beta\in B, \gamma\in \mathcal{C}^{\ell+1}_{C}(\Supp(X))\}$ and $\overline{\mathcal{K}}:=\{(x,s)\mapsto \overline{h}(x,s,\gamma,\beta): \beta\in B, \gamma\in \mathcal{C}^{\ell+1}_{C}(\Supp(X))\}$ are Donsker classes. Let $\underline{\kappa}_n(\gamma,\beta):=\frac{1}{n}\sum_{i=1}^n\underline{h}(X_i,S_i,\gamma,\beta)$ and
$\underline{\kappa}(\gamma,\beta):=\int \underline{h}(x,s,\gamma,\beta)dP(x,s)$; $\underline{h}$ and $\underline{\kappa}$ are not differentiable at $(\gamma_0,\beta_0)$ with respect to $\beta$ but directionally Hadamard differentiable \citep[see Definition 2.1 in][]{Fang_Santos_18}. Their directional derivatives with respect to $\beta$ are $\underline{k}(x,s,h)$ and $\underline{K}(h)$, see the expressions  before the Proof of Theorem \ref{thm:as_normality}, and their directional derivatives with respect to $\gamma$ are $0$. For $\alpha>0$ we have $\alpha \underline{K}(h)=\underline{K}(\alpha h)$ but we do not have $\underline{K}(h_1+h_2)=\underline{K}(h_1)+\underline{K}(h_2)$. For $g_t\rightarrow g$ in $\mathcal{C}^{\ell+1}_C(\Supp(X))$ and $h_t\rightarrow h$ in $\mathbb{R}^p$ when $t\downarrow 0$, we have  by arguments similar to those proving   the linearization condition in the proof of Lemma \ref{lem:cond_as_nor},
\begin{align*}
	\left|\left(\underline{\kappa}(\gamma_0+tg_t,\beta_0+th_t)-\underline{\kappa}(\gamma_0,\beta_0)\right)-t\underline{K}(h)\right|&\leq C t^2 (||h_t||^2+||g_t||_{\infty}^2)
	\end{align*}
Let $\underline{\mathbb{K}}_n(\gamma,\beta)=\sqrt{n}\left(\underline{\kappa}_n(\gamma,\beta)-\underline{\kappa}(\gamma,\beta)\right)$. We have $\int \left(\underline{h}(x,s,\widehat{\gamma},\widehat{\beta})-\underline{h}(x,s,\gamma_0,\beta_0)\right)^2 dP(x,s)=o_P(1)$ by arguments similar to those of the proof of Lemma \ref{lem:cond_as_nor}. Thus,
\citep[cf. Lemma 19.24 in][]{vandervaart_2000} $\underline{\mathbb{K}}_n(\widehat{\gamma},\widehat{\beta})=\underline{\mathbb{K}}_n(\gamma_0,\beta_0)+o_P(1)$ and \begin{align*}\sqrt{n}\left(\underline{\kappa}(\widehat{\gamma},\widehat{\beta})-\underline{\kappa}(\gamma_0,\beta_0)\right)&=\sqrt{n}~\underline{K}(\widehat{\beta}-\beta_0)+\sqrt{n}O_p\left(\left|\left|\widehat{\gamma}-\gamma_0\right|\right|^2_{\infty}+\norm{\widehat{\beta}-\beta_0}^2\right)\\&=\underline{K}\left(\sqrt{n}(\widehat{\beta}-\beta_0)\right)+o_P(1)
\end{align*}
By Equation \eqref{eq:equiv_step2}, we have $\sqrt{n}\left(\widehat{\underline{\delta}}-\underline{\delta}\right)=\underline{\mathbb{K}}_n(\gamma_0,\beta_0)+\underline{K}\left(\sqrt{n}\left(\widehat{\beta}-\beta_0\right)\right)+o_p(1)$, where $\underline{\mathbb{K}}_n(\gamma_0,\beta_0)=\frac{1}{\sqrt{n}}\sum_{i=1}^n \left(r(X_i,S_i,\beta_0)-\E\left(r(X_1,S_1,\beta_0)\right)\right)$. Similar reasoning for $\widehat{\overline{\delta}}$ ensures $$\left(\begin{array}{c}\sqrt{n}\left(\widehat{\underline{\delta}}-\underline{\delta}\right)\\\sqrt{n}\left(\widehat{\overline{\delta}}-\overline{\delta}\right)\end{array}\right)=\left(\begin{array}{c}\frac{1}{\sqrt{n}}\sum_{i=1}^n \left(r(X_i,S_i,\beta_0)-\E\left(r(X_1,S_1,\beta_0)\right)\right)+\underline{K}\left(\sqrt{n}\left(\widehat{\beta}-\beta_0\right)\right)\\\frac{1}{\sqrt{n}}\sum_{i=1}^n \left(r(X_i,S_i,\beta_0)-\E\left(r(X_1,S_1,\beta_0)\right)\right)+\overline{K}\left(\sqrt{n}\left(\widehat{\beta}-\beta_0\right)\right)\end{array}\right)+o_p(1),$$
with \begin{align*}
	\left(\frac{1}{\sqrt{n}}\sum_{i=1}^n \left(r(X_i,S_i,\beta_0)-\E\left(r(X_1,S_1,\beta_0)\right)\right),
		\sqrt{n}\left(\widehat{\beta}-\beta_0\right)\right)\convD(N_1,N_2)\sim \mathcal{N}\left(0,\Omega\right).
	\end{align*}
Because $\overline{K}$ and $\underline{K}$ are continuous functions on $\mathbb{R}^{p}$, the continuous mapping theorem ensures:
$$\sqrt{n}\left(\begin{array}{c}\widehat{\overline{\delta}}-\overline{\delta}\\\widehat{\underline{\delta}}-\underline{\delta}\end{array}\right)\convD \left(\begin{array}{c}N_1+\overline{K}(N_2)\\N_1+\underline{K}(N_2)\end{array}\right).$$




\subsection{Proposition \ref{prop:CI1}} 
\label{sub:proposition_ref_prop_ci1}

We introduce $\widehat{\phi}_i=-\left[\frac{1}{n}\sum_{j=1}^n \deriv{{}^2 \ell_c}{\beta \partial \beta'}(Y_j|X_j;\widehat{\beta})\right]^{-1}\deriv{\ell_c}{\beta}(Y_i|X_i;\widehat{\beta})$ as the sample analog of  $\phi_i$. 
To construct sample analogs of $ \underline{v}_{\gamma}$ and $ \underline{v}_{\beta}$, we need to take into account that $\underline{h}(x,s,\gamma,\beta,\tilde{\beta})$ defined in \eqref{eq:defin_underline_h_long} depends on $I$ which is unknown. For $\tilde{I}\in\{1,...,T\}$, let
$\underline{h}(x,s,\gamma,\beta,\tilde{\beta},\tilde{I})$ defined as  \eqref{eq:defin_underline_h_long} with $\widetilde{I}$ replacing $I$ in \eqref{eq:defin_m} and \eqref{eq:ext}. 
We also let $\widehat{\underline{v}}_{\beta, i}=\deriv{\underline{h}}{\beta}(x,s, \gamma, \beta,\widetilde{\beta},\tilde{I})$  for $(x,s, \gamma, \beta,\widetilde{\beta},\tilde{I})=(X_i,S_i,\widehat{\gamma}, \widehat{\beta}, \widehat{\beta},\widehat{I}(X_i))$, $\widehat{\underline{v}}_{\beta}=\frac{1}{n}\sum_{i=1}^n\widehat{\underline{v}}_{\beta, i}$
and 
$\widehat{\underline{v}}_{\gamma}(X_i,S_i)=\deriv{\underline{h}}{\gamma(x)}(x,s,\gamma,\beta, \widetilde{\beta},\tilde{I})$ for $(x,s, \gamma, \beta,\widetilde{\beta},\tilde{I})=(X_i,S_i,\widehat{\gamma}, \widehat{\beta}, \widehat{\beta},\widehat{I}(X_i))$. This approach is valid because we proved that  $\widehat{I}(X_i)=I$ for all $i$ wpao 1 in Step 1 of the proof of Theorem \ref{thm:as_normality}. We define similarly $\widehat{\overline{v}}_{\beta}$ and $\widehat{\overline{v}}_{\gamma}(X_i,S_i)$. The sample analogs of $\underline{\psi}_i$ and $\overline{\psi}_i$ are
\begin{align}
	\widehat{\underline{\psi}}_i=    & \ \widehat{\underline{h}}( X_i, S_i)  -  \frac{1}{n} \sum_{j=1}^n \widehat{\underline{h}}( X_j, S_j) + \widehat{\underline{v}}_{\beta}' \widehat{\phi}_i + \widehat{\underline{v}}_\gamma(X_i, S_i)' [\Gamma_i - \widehat{\gamma}(X_i)],	\label{eq:def_psihat_inf} \\
	\widehat{\overline{\psi}}_i = & \ \widehat{\overline{h}}( X_i, S_i)  -  \frac{1}{n} \sum_{j=1}^n \widehat{\overline{h}}( X_j, S_j) + \widehat{\overline{v}}_{\beta}' \widehat{\phi}_i + \widehat{\overline{v}}_\gamma(X_i, S_i)' [\Gamma_i - \widehat{\gamma}(X_i)].
	\label{eq:def_psihat_sup}
\end{align}
We finally estimate  $\Sigma$ by $\widehat{\Sigma} = \frac{1}{n} \sum_{i=1}^n (\widehat{\underline{\psi}}_i , \, \widehat{\overline{\psi}}_i)' (\widehat{\underline{\psi}}_i , \, \widehat{\overline{\psi}}_i)$.

\medskip

First assume that $\delta_0$ is the AME or ATE and $a(\beta_{0k})=0$. This implies that $\beta_{0k}=0$, thus $\delta_0=0$ and 
$$P\left(\delta_0 \in \CI{1}\right)\geq P(\varphi_{\alpha}=0)\to 1-\alpha,$$
where the latter follows since $\varphi_{\alpha}$ has asymptotic level $\alpha$.

\medskip

Now, assume $a(\beta_{0k})\neq 0$. We first prove that $\widehat{\Sigma}$ is consistent. 
Let  
$\widetilde{\underline{v}}_{\beta, i}^0 = \deriv{\underline{h}}{\beta}(X_i,S_i, \widehat{\gamma}, \widehat{\beta}, \beta_0)$, $\widetilde{\underline{v}}_{\beta}^0=\frac{1}{n}\sum_{i=1}^n\widetilde{\underline{v}}_{\beta, i}^0$,  $\widetilde{\underline{v}}^0_{\gamma}(X_i,S_i) := \deriv{\underline{h}}{\gamma(x)}(X_i,S_i,\widehat{\gamma}, \widehat{\beta}, \beta_0)$ 
and $\widetilde{\underline{\psi}}_i$   be defined as in \eqref{eq:def_psihat_inf} replacing $\widehat{\underline{v}}_{\beta}$ with $\widetilde{\underline{v}}_{\beta}^0$ and $ \widehat{\underline{v}}_\gamma(X_i, S_i)$ with $\widetilde{\underline{v}}^0_{\gamma}(X_i,S_i)$. Defining the corresponding quantities for $\widetilde{\overline{\psi}}_i$, we have wpoa 1 for all $i$,
\begin{align*}
	\underline{\nu}_i &:= \widehat{\underline{\psi}}_i - \widetilde{\underline{\psi}}_i = \left[\widehat{\underline{v}}_{\beta}-\widetilde{\underline{v}}_{\beta}^0\right]' \widehat{\phi}_i + \left[\widehat{\underline{v}}_\gamma(X_i, S_i) - \widetilde{\underline{v}}^0_{\gamma}(X_i,S_i) \right]' [\Gamma_i - \widehat{\gamma}(X_i)] \\
	\overline{\nu}_i &:= \widehat{\overline{\psi}}_i - \widetilde{\overline{\psi}}_i = \left[\widehat{\overline{v}}_{\beta}-\widetilde{\overline{v}}_{\beta}^0\right]' \widehat{\phi}_i + \left[\widehat{\overline{v}}_\gamma(X_i, S_i) - \widetilde{\overline{v}}^0_{\gamma}(X_i,S_i) \right]' [\Gamma_i - \widehat{\gamma}(X_i)].
\end{align*}
Let $\widetilde{\Sigma} = \frac{1}{n} \sum_{i=1}^n (\widetilde{\underline{\psi}}_i , \, \widetilde{\overline{\psi}}_i)' (\widetilde{\underline{\psi}}_i , \, \widetilde{\overline{\psi}}_i)$, then $\widehat{\Sigma} = \widetilde{\Sigma} + \frac{1}{n} \sum_{i=1}^n (\underline{\nu}_i , \, \overline{\nu}_i)' (\widetilde{\underline{\psi}}_i , \, \widetilde{\overline{\psi}}_i) +  \frac{1}{n} \sum_{i=1}^n (\widetilde{\underline{\psi}}_i , \, \widetilde{\overline{\psi}}_i)' (\underline{\nu}_i , \, \overline{\nu}_i) + \frac{1}{n} \sum_{i=1}^n (\underline{\nu}_i , \, \overline{\nu}_i)' (\underline{\nu}_i , \, \overline{\nu}_i)$. By the smoothness and the compactness conditions imposed throughout, the fact that $\beta_0 \in B$ and the proof of Condition 2 in Lemma \ref{lem:cond_as_nor}, there exists a constant $C$ such that wpoa 1, for all $i$, $|\underline{\nu}_i|, |\overline{\nu}_i|, |\widetilde{\underline{\psi}}_i|, |\widetilde{\overline{\psi}}_i|, ||\widehat{\phi}_i||  \leq C$. Thus for $(a,b) \in \{1,2\}^2$,
\begin{align*}
	&|\widehat{\Sigma}_{ab} - \widetilde{\Sigma}_{ab}|  \leq \frac{3C}{n} \sum_{i=1}^n \max(|\underline{\nu}_i| , |\overline{\nu}_i|)\\
	& \leq C \max(||\widehat{\underline{v}}_{\beta}-\widetilde{\underline{v}}_{\beta}^0 ||, || \widehat{\overline{v}}_{\beta}-\widetilde{\overline{v}}_{\beta}^0||) + \frac{C}{n} \sum_{i=1}^n ||\widehat{\overline{v}}_\gamma(X_i, S_i) - \widetilde{\overline{v}}^0_{\gamma}(X_i,S_i)|| +|| \widehat{\underline{v}}_\gamma(X_i, S_i) - \widetilde{\underline{v}}^0_{\gamma}(X_i,S_i) ||,
\end{align*}
where we allow the constant $C$ to change value from one line to another. Each term in the above inequality is $o_P(1)$. We show it for the first term using arguments  from Step 2, sub step 1 of the proof of Theorem \ref{thm:as_normality}, but the reasoning applies to all terms.
\begin{align*}
	||\widehat{\underline{v}}_{\beta}-\widetilde{\underline{v}}_{\beta}^0 || &\leq \frac{C}{n} \sum_{i=1}^n \left|\ind{  \lambda_{T+1}(X_i,\widehat{\beta}) < 0  }  -  \ind{\lambda_{T+1}(X_i,\beta_0) < 0}\right|\\
		&\leq  \frac{C}{n} \sum_{i=1}^n \ind{|\lambda_{T+1}(X_i,\beta_0)|\leq C||\widehat{\beta}-\beta_0||}\leq F_{|\lambda_{T+1}|}(C||\widehat{\beta}-\beta_0||)+o_P(1) = o_P(1).
\end{align*}
Thus we obtain $\widehat{\Sigma} = \widetilde{\Sigma} + o_P(1)$. We now show that the functions of $(X_i,S_i)$ appearing in $\widetilde{\underline{\psi}}_i$ and $\widetilde{\overline{\psi}}_i$ converge uniformly to their pointwise limits.
Similarly to what we argued in the proof of Theorem \ref{thm:consistency}, 
$$
(x,s) \mapsto \widehat{\underline{h}}(  x,s)  -  \frac{1}{n} \sum_{j=1}^n \widehat{\underline{h}}( x,s)
$$ 
converges uniformly in probability to 
$$
(x,s) \mapsto \underline{h}( x,s, \gamma_0, \beta_0) - \E(\underline{h}( X, S, \gamma_0, \beta_0)).
$$ 
The smoothness arguments given in Condition 1 above implies in particular that the derivatives of $\underline{h}$ with respect to both the vector $\gamma(x)$ and $\beta$ are Lipschitz continuous on $\mathcal{C}_C^\alpha$ and $B$, with Lipschitz constant uniform over $x \in \Supp(X)$. This implies that $(x,s,\Gamma) \mapsto  \widetilde{\underline{v}}^0_\gamma( x, s)' [\Gamma - \widehat{\gamma}(x)]$ converges uniformly in probability to $(x,s,\Gamma) \mapsto  \underline{v}_{\gamma}( x, s)' [\Gamma - \gamma_0(x)]$. The same results follow for $\overline{h}$. By $\mathcal{I}_0$ nonsingular,  $\beta_0 \in B$ and $C_s(x,\beta)$ bounded away from $0$ uniformly over $(s,x,\beta)$, the derivatives of $\beta \in B \mapsto \left[ \frac{1}{n}\sum_{j=1}^n  \deriv{{}^2 \ell_c}{\beta \partial \beta'}(Y_j|X_j;\beta) \right]^{-1}    \deriv{\ell_c}{\beta}(y|x;\beta)$ are uniformly bounded over $(y,x,\beta)$ wpao. Thus $\widehat{\phi}_i$ converges uniformly in probability to $\phi_i$. In conclusion,  the functions of $(X_i,S_i)$ appearing in $\widetilde{\underline{\psi}}_i$ and $\widetilde{\overline{\psi}}_i$ converge uniformly to their pointwise limits. This implies that $(\widetilde{\underline{\psi}}_i , \, \widetilde{\overline{\psi}}_i) (\widetilde{\underline{\psi}}_i , \, \widetilde{\overline{\psi}}_i)'$ converges uniformly to $(\underline{\psi}_i,\overline{\psi}_i)(\underline{\psi}_i,\overline{\psi}_i)'$. As in Theorem \ref{thm:consistency}, we obtain using the LLN that $\widehat{\Sigma}\convP \Sigma$.

\medskip

Now, because $\varphi_\alpha$ is consistent, $\varphi_\alpha\convP 1$, so that $\CI{1}$ takes the first form wpao. Suppose first that $\underline{\delta}<\overline{\delta}$. By consistency of the bounds, consistency of $\widehat{\Sigma}$ and $\min(\Sigma_{11},\Sigma_{22})>0$, we have
$$\frac{n^{1/2}\left(\widehat{\overline{\delta}}-\widehat{\underline{\delta}}\right)}{\max\left(\widehat{\Sigma}^{1/2}_{11},\widehat{\Sigma}^{1/2}_{22}\right)}\convP \infty.$$
Then, by Lemma 5.10 of \cite{vandervaart_2000}, $c_\alpha\convP c_\alpha^0 := \Phi^{-1}(1-\alpha)$. Using arguments similar to Lemma 3 of \cite{imbens2004confidence} and taking the class of DGPs to be a singleton, we obtain that  
$
 \liminf_n \inf_{\delta_0\in[\underline{\delta},\overline{\delta}]} P(\delta_0\in\CI{1})= 1-\alpha.$

\medskip

Next, assume $\underline{\delta}=\overline{\delta}$. Then, because $\widehat{\overline{\delta}}\geq \widehat{\underline{\delta}}$ a.s., $\overline{N}-\underline{N}$ must be degenerate, implying in turn  $\overline{N}=\underline{N}$ a.s. Hence,
$$\frac{n^{1/2}\left(\widehat{\overline{\delta}}-\widehat{\underline{\delta}}\right)}{\max\left(\widehat{\Sigma}^{1/2}_{11},\widehat{\Sigma}^{1/2}_{22}\right)}=o_P(1).$$
Then, by Lemma 5.10 of \cite{vandervaart_2000}, $c_\alpha\convP c_{\alpha/2}^0 = \Phi^{-1}(1-\alpha/2)$. Note that  $P(\underline{\delta} = \overline{\delta}\in\CI{1}) = 1 -  P\left(\underline{\delta} \leq \widehat{\underline{\delta}} - c_\alpha (\widehat{\Sigma}_{11}/n)^{1/2}\right) - P\left( \overline{\delta} \geq \widehat{\overline{\delta}} + c_\alpha (\widehat{\Sigma}_{22}/n)^{1/2}  \right)$. By Theorem \ref{thm:as_normality} and consistency of $\widehat{\Sigma}_{11}$, 
$$P\left(\underline{\delta} \leq \widehat{\underline{\delta}} - c_\alpha (\widehat{\Sigma}_{11}/n)^{1/2}\right) \to P(\underline{N} \ge \Phi^{-1}(1-\alpha/2) \Sigma_{11}) = \frac{\alpha}{2}.$$
Similarly, $P\left( \overline{\delta} \geq \widehat{\overline{\delta}} + c_\alpha (\widehat{\Sigma}_{22}/n)^{1/2}  \right)\to  \alpha/2$. The result follows.


\subsection{Theorem \ref{thm:CI_triv}} 
\label{sub:theorem_ref_prop_ci_triv}

First assume that $\overline{b}=0$. Then $\tilde{\delta}=\delta_0$ and $\widehat{\overline{b}}\ge 0$. By Lemma \ref{lem:q_alpha} (\ref{lem:q_alpha_increasing}), $b\mapsto q_\alpha(b)$ is increasing on $\R^+$ thus we obtain
\begin{align*}
	P\left(n^{1/2}\left|\frac{\widehat{\tilde{\delta}}- \delta_0}{\widehat{\sigma}}\right| \le q_\alpha\left(n^{1/2} \frac{\widehat{\overline{b}}}{\widehat{\sigma}}\right) \right) & \ge P\left(n^{1/2}\left|\frac{\widehat{\tilde{\delta}}- \tilde{\delta}}{\widehat{\sigma}}\right| \le q_{\alpha}(0) \right) \\
	& \to 1-\alpha,
\end{align*}
where the convergence follows by Lemma \ref{lem:as_nor_triv}.

\medskip
Next, consider the case $\abs{\tilde{\delta}-\delta_0}<\overline{b}$. Let us define the event
$$E_n:=\left\{\widehat{\sigma} q_\alpha\left(n^{1/2} \frac{\widehat{\overline{b}}}{\widehat{\sigma}}\right)  \ge \sigma q_\alpha\left(n^{1/2} \frac{\tilde{\delta}-\delta_0}{\sigma}\right)\right\}.$$
By Lemma \ref{lem:as_nor_triv},
$\widehat{\sigma} \convP \sigma$.
Moreover under Assumption \ref{hyp:lambda}, we have  by Lemma \ref{lem:conv_unif_R}, $\widehat{\overline{b}}\convP \overline{b}>\abs{\tilde{\delta}-\delta_0}$.
Thus, with probability approaching one, $\widehat{\sigma} > \sigma -\eta$ and $\widehat{\overline{b}}>\abs{\tilde{\delta}-\delta_0}+ n^{-1/2} \left[\eta z_{1-\alpha} +  \sigma (z_{1-\alpha/2} - z_{1-\alpha}) \right]$.
If so,
\begin{align*}
	\widehat{\sigma}  q_\alpha\left(n^{1/2} \frac{\widehat{\overline{b}}}{\widehat{\sigma}}\right)
	& \ge n^{1/2} \widehat{\overline{b}}+ (\sigma -\eta) z_{1-\alpha}  \\
	& \ge  n^{1/2} \abs{\tilde{\delta}-\delta_0}+ \eta z_{1-\alpha} +  \sigma (z_{1-\alpha/2} - z_{1-\alpha})  + (\sigma -\eta) z_{1-\alpha}\\
	& =  \sigma \left(n^{1/2} \frac{\abs{\tilde{\delta}-\delta_0}}{\sigma} + z_{1-\alpha/2} \right) \\
	& \ge \sigma q_\alpha\left(n^{1/2} \frac{\tilde{\delta}-\delta_0}{\sigma}\right).
\end{align*}
where the first  and third inequalities hold by Lemma \ref{lem:q_alpha}.\ref{lem:q_alpha_ineq}.
As a result, $P(E_n)\to 1$. Then, using $P(A\cap B)\ge P(A)+P(B)-1$,
\begin{align*}
	P\left(n^{1/2}\left|\widehat{\tilde{\delta}}- \delta_0\right| \le \widehat{\sigma} q_\alpha\left(n^{1/2} \frac{\widehat{\overline{b}}}{\widehat{\sigma}}\right) \right) & \ge P\left(n^{1/2}\left|\widehat{\tilde{\delta}}- \delta_0 \right| \le \sigma q_\alpha\left(n^{1/2} \frac{\tilde{\delta}-\delta_0}{\sigma}\right)\right)  \\
	& \; + P(E_n) - 1 \\
	&  \ge P\left(\left|\tilde{Z}_n + n^{1/2} \frac{\tilde{\delta}-\delta_0}{\sigma} \right| \le q_\alpha\left(n^{1/2} \frac{\tilde{\delta}-\delta_0}{\sigma}\right)\right)  \\
	& \; + P(E_n) - 1,
\end{align*}
where $\tilde{Z}_n:=n^{1/2}(\widehat{\tilde{\delta}} - \widetilde{\delta})/\sigma$. Hence, by what precedes,
$$\liminf_{n\to\infty} P\left(n^{1/2}\left|\widehat{\tilde{\delta}}- \delta_0\right| \le \widehat{\sigma} q_\alpha\left(n^{1/2} \frac{\widehat{\overline{b}}}{\widehat{\sigma}}\right) \right) \ge \liminf_{n\to\infty} P\left(\left|\tilde{Z}_n + n^{1/2} \frac{\tilde{\delta}-\delta_0}{\sigma} \right| \le q_\alpha\left(n^{1/2} \frac{\tilde{\delta}-\delta_0}{\sigma}\right)\right).$$
Now, let $F_n$ denote the cdf of $\tilde{Z}_n$ and let $Z\sim\mathcal{N}(0,1)$. We have:
\begin{align*}
	& \left|P\left(\left|\tilde{Z}_n + n^{1/2} \frac{\tilde{\delta}-\delta_0}{\sigma} \right| \le q_\alpha\left(n^{1/2} \frac{\tilde{\delta}-\delta_0}{\sigma} \right)\right) - (1-\alpha)\right| \\
	= &  \left|P\left(\left|\tilde{Z}_n + n^{1/2} \frac{\tilde{\delta}-\delta_0}{\sigma} \right| \le q_\alpha\left(n^{1/2} \frac{\tilde{\delta}-\delta_0}{\sigma} \right)\right) - P\left(\left|Z + n^{1/2} \frac{\tilde{\delta}-\delta_0}{\sigma} \right| \le q_\alpha\left(n^{1/2} \frac{\tilde{\delta}-\delta_0}{\sigma} \right)\right)\right| \\
	\le & \sup_{x\in\R} \left|P\left(\left|\tilde{Z}_n + x \right| \le q_\alpha\left(x\right)\right) - P\left(\left|Z + x \right| \le q_\alpha\left(x\right)\right)\right| \\
	= & \sup_{x\in\R} \left|F_n(x+q_\alpha(x)) - \Phi(x+q_\alpha(x)) - F_n(x-q_\alpha(x)) + \Phi(x-q_\alpha(x))\right| \\
	\le & 2 \sup_{x\in\R} \left|F_n(x) - \Phi(x)\right|.
\end{align*}
Finally, Lemma \ref{lem:as_nor_triv} implies that for all $x$, $F_n(x) \to \Phi(x)$ with $\Phi$ continuous. By Lemma
2.11 in \cite{vandervaart_2000}, the convergence is uniform. The result follows.

	
\subsection{Lemma \ref{lem:cond_CI2}} 
\label{sub:lem_cond_CI2}

$P(\lambda_{T+1}(X,\beta_0)\ne 0)>0$ is implied by $\overline{b}>0$. Now, assume that $\delta_0-\tilde{\delta}=\overline{b}$. By Lemma \ref{lem:optim}, this implies that for any set of measures $\left(\mu_x\right)_{x \in \Supp(X)}$ such that $\mu_x\in \mathcal{D}(m(x))$ for all $x \in \Supp(X)$,
\begin{equation}
	\E\left[ \lambda_{T+1}(X, \beta)  \E[Z_0 |X] \int_0^1 \mathbb{T}_{T+1}(u) d\mu_X(u)  \right] = \frac{1}{2\times 4^T} E\left[ \E[Z_0|X]|\lambda_{T+1}(X,\beta_0)|\right].	
	\label{eq:reaching_bsup}
\end{equation}
By the proof of Proposition \ref{prop:outer_ident} and since $Z_0> 0$ almost surely, \eqref{eq:reaching_bsup} implies that for almost all $x$ such that $\lambda_{T+1}(x,\beta_0) > 0$, $\Supp(\mu_x)\subseteq \arg\max_{u\in[0,1]} \mathbb{T}_{T+1}(u)$. Similarly,  for almost all $x$ such that $\lambda_{T+1}(x,\beta_0) < 0$, $\Supp(\mu_x)\subseteq \arg\min_{u\in[0,1]} \mathbb{T}_{T+1}(u)$. Since $\Supp(U|X=x)=\Supp(\mu_X|X=x)$ by the proof of Lemma \ref{lem:optim}, we obtain $$P\left(\Supp(U|X)\subseteq \mathcal{R}_{T,X} | \lambda_{T+1}(X,\beta_0)\ne 0\right)=1.$$ 
Similarly, $\delta_0-\tilde{\delta}=- \overline{b}$ implies $P\left(\Supp(U|X)\subseteq \mathcal{R}'_{T,X} | \lambda_{T+1}(X,\beta_0)\ne 0\right)=1$. 


	\subsection{Theorem \ref{thm:CI_triv2}}
	

	The asymptotic variance of $a(\widehat{\beta}_{0k})$ and its estimators are
		$\sigma_a^2  = a'(\beta_{0k})^2  \tau_k$ and $\widehat{\sigma}_a^2 = a'(\widehat{\beta}_k)^2 \widehat{\tau}_k$,
	where $\widehat{\tau}_k$ (respectively $\tau_k$) is the $(k,k)$ element of $\widehat{\tau}=\frac{1}{n}\sum_{i=1}^n\widehat{\phi}_i\widehat{\phi}'_i$ (respectively $\tau=E(\phi\phi')$) for $\widehat{\phi}$ the plug-in estimator of the influence function $\phi$ of $\widehat{\beta}$ (cf. Section \ref{sub:asymptotic_properties}). Lemma \ref{lem:conv_unif_P} ensures uniform consistency of the estimator $\widehat{\tau}_k$ to $\tau_k$. By the smoothness conditions on $a$ and compactness of $B$, $\widehat{\sigma}_a^2$ converges uniformly to $\sigma_a^2$.
	
	To show the result, it suffices to show that
	\begin{equation}
		\liminf_n P_n\left(\delta_0\in \CI{3}\right)\ge 1-\alpha,
		\label{eq:conv_unif_simple}
	\end{equation}
	for any sequence of probability distributions $(P_n)_{n\geq 1}$ in $\mathcal{P}$. Note that to simplify notation, we do not index parameters by $P_n$ (nor by $n$). We proceed in three sub-steps. We first show that
	\begin{equation}
		\liminf_n P_n\left(|a\left(\widehat{\beta}_k\right)|+ z_{1-\alpha_1} n^{-1/2}\widehat{\sigma}_a\ge | a\left(\beta_{0k}\right) |\right)\geq 1-\alpha_1. \label{eq:Eprime}
	\end{equation}
	Let $b=\tilde{\delta}-\delta_0$ and $\alpha_2 = \alpha - \alpha_1$. We then prove that for any $\eta$ small enough,
	\begin{equation}
		\lim \inf_{n}P_n\left(\widehat{\sigma}	q_{\alpha_2}\left( \frac{n^{1/2}\widehat{\overline{b}}_{\alpha_1}}{\widehat{\sigma}} \right) \ge\sigma q_{\alpha_2}\left(\frac{n^{1/2}b}{\sigma} \right)-\eta\right) \geq 1-\alpha_1 \label{eq:CI3_step2}.
	\end{equation}
	We finally establish \eqref{eq:conv_unif_simple} in the third step.
	
	\subsubsection*{Step 1: (\ref{eq:Eprime}) holds.}
	We consider the two cases given in Assumption \ref{hyp:lambda2}. If $a'(\beta_{k})= 0$ for all $\beta \in B$, then $\widehat{\sigma}_a = 0$. Moreover by Assumption \ref{hyp:iid_bdX}, $a$ is constant and (\ref{eq:Eprime}) holds trivially. If $|a'(\beta_{k})|> 0$ for all $\beta \in B$, define $Z_n=n^{1/2}\left[a\left(\widehat{\beta}_k\right)-a\left(\beta_{0k}\right)\right]/\sigma_a$. Let $F_n$ be its cdf.
	If  $A'\cap B\subseteq A\cap B$, then $P_n(A)\ge P_n(A')-P_n(A'\cap B^c)+P_n(A\cap B^c)\ge P_n(A')-P_n(B^{c})$. Thus for any $\eta>0$ and for $z_{1-\alpha_1}$ the $1-\alpha_1$-quantile of the standard normal distribution, 
	\begin{align*}
		P_n\left(|a\right. & \left.\left(\widehat{\beta}_k\right)|+ z_{1-\alpha_1} n^{-1/2}\widehat{\sigma}_a   \ge |a\left(\beta_{0k}\right)| \right)\\
		&=P_n\left(|Z_n + n^{1/2}a\left(\beta_{0k}\right)/\sigma_a| \ge |n^{1/2}a\left(\beta_{0k}\right)/\sigma_a|-z_{1-\alpha_1}  \widehat{\sigma}_a/\sigma_a  \right) \notag\\
		& \ge P_n\left(|Z_n + n^{1/2}a\left(\beta_{0k}\right)/\sigma_a| \ge |n^{1/2}a\left(\beta_{0k}\right)/\sigma_a|-z_{1-\alpha_1} (1- \eta)\right) -P_n\left( \widehat{\sigma}_a/\sigma_a < 1- \eta\right) \notag \\
		& \ge \left[\inf_{|x|\ge z_{1-\alpha_1} (1- \eta)} P_n\left(|Z_n + x| \ge |x| -z_{1-\alpha_1} (1- \eta)\right)\right] - o(1) \notag \\	
		& \ge \min \left[F_n\left(z_{1-\alpha_1} (1- \eta)\right), 1 -  F_n\left(-z_{1-\alpha_1} (1- \eta)\right)\right]-o(1) \notag\\
		& \to \min \left[ \Phi(z_{1-\alpha_1}(1-\eta)), 1 -  \Phi(-z_{1-\alpha_1}(1-\eta))\right]= \Phi(z_{1-\alpha_1}(1-\eta))
	\end{align*}
	The second inequality holds because $P_n(|Z_n-x|\geq |x|-b)=1$ for $|x|<b$ and by uniform convergence of $ \widehat{\sigma}_a/\sigma_a$ to $1$, which is guaranteed by uniform convergence of $ \widehat{\sigma}_a$ to $\sigma_a$ and $\sigma_a > \varepsilon$ for some $\varepsilon>0$ by Assumptions \ref{hyp:iid_bdX} and \ref{hyp:lambda2} and the definition of $\mathcal{P}$ given in \eqref{eq:def_P}. The third inequality holds because for any $a>0$ and $b$, we have $$\inf_{|x|\geq a}P(|Z-x|\geq |x|+b)\geq \min \left(\inf_{x\geq a}P(x-Z\geq x+b), \inf_{x\leq-a}P(Z-x\geq -x+b)\right).$$
	To obtain the limit above, we use Lemma \ref{lem:conv_unif_P} and the delta method, see, e.g., Theorem B.3 in  \cite{belloni2015program}.
	Because $\eta$ is arbitrarily small and $\Phi$ is continuous everywhere, we conclude that
	\begin{align*}
		\liminf_n P_n\left(|a\left(\widehat{\beta}_k\right)|+ z_{1-\alpha_1} n^{-1/2}\widehat{\sigma}_a   \ge |a\left(\beta_{0k}\right)| \right)\geq 1-\alpha_1.
	\end{align*}
	
	\subsubsection*{Step 2: (\ref{eq:CI3_step2}) holds.}

		Let $E_{n,\eta}=\left\{\widehat{\sigma}	q_{\alpha_2}\left( \frac{n^{1/2}\widehat{\overline{b}}_{\alpha_1}}{\widehat{\sigma}} \right) \ge\sigma q_{\alpha_2}\left(\frac{n^{1/2}(\tilde{\delta}-\delta_0)}{\sigma} \right)-\eta\right\}$. Let us first suppose that $P_n$ is a sequence such that for each $n$, $P_n\in \mathcal{P}_{1n}$ for $\mathcal{P}_{1n}=\mathcal{P}\cap \left\{P: n^{1/2}|\tilde{\delta}-\delta_0|/\sigma>2 z_{1-\alpha_2/2}/\zeta\right\}$ (recall that $\zeta$ appears in \eqref{eq:def_P}). We have
	\begin{align}
		q_{\alpha_2}\left(\frac{n^{1/2}b}{\sigma}\right) & \le \frac{n^{1/2}\abs{b}}{\sigma} + z_{1-\alpha_2/2} \notag\\
		& \le n^{1/2} \frac{|a\left(\beta_{0k}\right)|R(1+\zeta/2)}{\sigma} \label{eq:ineq1_unif}
	\end{align}
	where the first inequality uses the second statement of Lemma \ref{lem:q_alpha}. From the second statement of Lemma \ref{lem:q_alpha}, we also have:
	\begin{align}
		\widehat{\sigma}  q_{\alpha_2}\left(n^{1/2} \frac{\widehat{\overline{b}}_{\alpha_1}}{\widehat{\sigma}}\right)
		& \ge \widehat{\sigma} \left(n^{1/2} \frac{\widehat{\overline{b}}_{\alpha_1}}{\widehat{\sigma}}+z_{1-\alpha_2}  \right) \notag \\
		& = n^{1/2} \left(|a\left(\widehat{\beta}_k\right)|+n^{-1/2} \widehat{\sigma}_a z_{1-\alpha_1}\right)\widehat{\overline{R}} + \widehat{\sigma}z_{1-\alpha_2} . \label{eq:ineq2_unif}
	\end{align}
	Moreover, 
	\begin{align*}
		&\lim\inf_{n} P_n\bigg[n^{1/2} \left(|a\left(\widehat{\beta}_k\right)|+n^{-1/2} \widehat{\sigma}_a z_{1-\alpha_1}\right)\widehat{\overline{R}} +\widehat{\sigma} z_{1-\alpha_2} \ge n^{1/2} |a\left(\beta_{0k}\right)|R(1+\zeta/2)\bigg]\\
		& \ge \lim\inf_{n} P_n\left(\left\{ |a\left(\widehat{\beta}_k\right)|+ z_{1-\alpha_1} n^{-1/2}\widehat{\sigma}_a\ge |a\left(\beta_{0k}\right)|\right\}\cap \left\{\widehat{\overline{R}}>R(1+\zeta/2)\right\} \right)\\
		&\ge 1-\alpha_1,
	\end{align*}
	where the second inequality follows from \eqref{eq:Eprime}, $P_n(A\cap B)\ge P_n(A)+P_n(B) -1$ and  $\lim_n P_n(\widehat{\overline{R}}>R(1+\zeta/2))=1$. This last limit equality holds because $\widehat{\overline{R}}$ converges uniformly to $\overline{R}$ and $\overline{R} \geq \omega / (2 \times 4^T\times \sup_{u,x}\Omega_{x,\beta_0}(u))$ is bounded away from $0$ by Assumptions \ref{hyp:iid_bdX} and \ref{hyp:lambda2}, which implies that $\widehat{\overline{R}} / \overline{R}$ converges uniformly to $1$. Combined with \eqref{eq:ineq1_unif}-\eqref{eq:ineq2_unif}, this yields for any $\eta>0$ and any sequence $P_n$ such that $P_n\in \mathcal{P}_{1n}$:
	\begin{align}
		\lim\inf_{n} P_n\left(E_{n,\eta}\right) &\geq 1-\alpha_1. \label{eq:case1}
	\end{align}

\medskip
Next, let us suppose that $P_n$ is a sequence such that for any $n$, $P_n\in \mathcal{P}_{2n}$ for $\mathcal{P}_{2n}=\mathcal{P}\cap \left\{P: n^{1/2}\abs{\tilde{\delta}-\delta_0}/\sigma\le 2 z_{1-\alpha_2/2}/\zeta\right\}$. Take $\eta >0$.
	Define
	\begin{align*}
		\mathcal{F}_n := & \left\{|a\left(\widehat{\beta}_k\right)|+ z_{1-\alpha_1} n^{-1/2}\widehat{\sigma}_a \ge |a\left(\beta_{0k}\right)|\right\} \cap \left\{\widehat{\overline{R}}\ge  \overline{R} (1 + \zeta)^{-1/2} \right\} \\
		& \ \cap \left\{ \widehat{\sigma} \leq  \sigma (1 + \zeta)^{1/2} \right\} \cap \left\{ \widehat{\sigma} \geq \sigma - \eta / q_{\alpha_2}(2 z_{1-\alpha_2/2}/\zeta)\right\}.
	\end{align*}
	On $\mathcal{F}_n$, because $x\mapsto q_{\alpha_2}(x)$ is increasing,
	\begin{align*}
		\widehat{\sigma}  q_{\alpha_2}\bigg(n^{1/2} \frac{\widehat{\overline{b}}_{\alpha_1}}{\widehat{\sigma}}\bigg) \ge \ & \widehat{\sigma} q_{\alpha_2}\bigg(n^{1/2} \frac{|a\left(\beta_{0k}\right)| \overline{R} (1 + \zeta)^{-1/2}}{\sigma (1 + \zeta)^{1/2}}\bigg) = \widehat{\sigma} q_{\alpha_2}\bigg(n^{1/2} \frac{|a\left(\beta_{0k}\right)| \overline{R} }{\sigma (1 + \zeta)}\bigg)\\
		\ge \ & \left(\sigma - \frac{\eta} {q_{\alpha_2}(2 z_{1-\alpha_2/2}/\zeta)}\right) q_{\alpha_2}\bigg(n^{1/2} \frac{ \tilde{\delta}-\delta_0 }{\sigma }\bigg)\\
		\ge \ & \sigma  q_{\alpha_2}\bigg(n^{1/2} \frac{\tilde{\delta}-\delta_0}{\sigma }\bigg) - \eta
	\end{align*}
	where the second inequality holds by $\overline{R} > (1+\zeta)R$ and the last inequality holds by $n^{1/2}\abs{b}/\sigma\le 2 z_{1-\alpha_2/2}/\zeta$.
	Equation \eqref{eq:Eprime}, $\zeta > 0 $, uniform convergence of $\widehat{\sigma}/\sigma$ to $1$ (in view of $\sigma\ge \underline{\sigma}>0$ and Lemma \ref{lem:as_nor_triv}) and previous arguments ensure that $\lim_n P_n\left(\mathcal{F}_n\right) \ge 1 - \alpha_1$. Thus for any $\eta>0$ and any sequence $P_n$ such that $P_n\in \mathcal{P}_{2n}$ we have:
	\begin{align}
		\lim \inf_{n}P_n\left(E_{n,\eta}\right) &\geq 1-\alpha_1 \label{eq:case2}.
	\end{align}
Now consider $P_n$ is a sequence in $\mathcal{P}$. If $P_n\in \mathcal{P}_{1n}$ for any $n$ sufficiently large, then \eqref{eq:case1} ensures \eqref{eq:CI3_step2}. Else if $P_n\in \mathcal{P}_{2n}$) for any $n$ sufficiently large, then \eqref{eq:case2} ensures \eqref{eq:CI3_step2}. Else there exist $\sigma_1,\sigma_2$ strictly increasing functions from $\mathbb{N}$ to $\mathbb{N}$ such that $\sigma_1(\N)\cap \sigma_2(\N)=\emptyset$, $\sigma_1(\N)\cup \sigma_2(\N)=\mathbb{N}$ and for any $n$, $P_{\sigma_1(n)}\in\mathcal{P}_{1 \sigma_1(n)}$ and $P_{\sigma_2(n)}\in\mathcal{P}_{2 \sigma_2(n)}$. Then, we have :
\begin{align*}
	\lim \inf_n P_n\left(E_{n,\eta}\right)
	&=\lim_{k\rightarrow \infty} \inf_{n \geq k} P_n\left(E_{n,\eta}\right)\\
	&=\lim_{k\rightarrow \infty} \inf\left(\inf_{n\in \sigma_1(\N),n\geq k}P_n(E_{n,\eta}),\inf_{n\in \sigma_2(\N),n\geq k}P_n(E_{n,\eta})\right)\\
	&=\lim_{k\rightarrow \infty} \inf\left(\inf_{n\geq \sigma_1^{-1}(k)}P_{\sigma_1(n)}(E_{\sigma_1(n),\eta}),\inf_{n\geq \sigma_2^{-1}(k)}P_{\sigma_2(n)}(E_{\sigma_2(n),\eta})\right)
	\end{align*}
For $i=1,2$,  $\inf_{n\geq \sigma_i^{-1}(k)}P_{\sigma_i(n)}(E_{\sigma_i(n),\eta})$ is increasing with $k$ then admits a limit when $k$ tends to $\infty$ and $(a,b)\mapsto \inf(a,b)$ is continuous. It follows that:
 \begin{align*}
 	\lim \inf_n P_n\left(E_{n,\eta}\right)
 	&=\inf\left(\lim_{k\rightarrow \infty}\inf_{n\geq \sigma_1^{-1}(k)}P_{\sigma_1(n)}(E_{\sigma_1(n),\eta}),\lim_{k\rightarrow \infty}\inf_{n\geq \sigma_2^{-1}(k)}P_{\sigma_2(n)}(E_{\sigma_2(n),\eta})\right)\\
 	&\geq 1-\alpha_1
 	\end{align*}
	where the last inequality holds by monotonicity of $(a,b)\mapsto \inf(a,b)$ and because \eqref{eq:case1} and \eqref{eq:case2} hold for sub-sequences. We finally obtain \eqref{eq:CI3_step2}.

	\subsubsection*{Step 3: conclusion.}
	
	\medskip
	Following the same line as in the proof of simple convergence, we have:
	\begin{align*}
		P_n\left(\delta_0 \in \CI{3}\right) &= P_n\left(n^{1/2}\left|\widehat{\tilde{\delta}}- \delta_0\right| \le \widehat{\sigma} q_{\alpha_2}\left(n^{1/2} \frac{\widehat{\overline{b}}_{\alpha_1}}{\widehat{\sigma}}\right) \right) \\
		& \ge P_n\left(n^{1/2}\left|\widehat{\tilde{\delta}}-\delta_0 \right| \le \sigma q_{\alpha_2}\left(n^{1/2} \frac{\tilde{\delta}-\delta_0}{\sigma}\right)-\sigma\eta\right)  \\
		& \; + P_n(E_{n,\sigma\eta}) - 1 \\
		&  \ge P_n\left(\left|\tilde{Z}_n + n^{1/2} \frac{\tilde{\delta}-\delta_0}{\sigma} \right| \le q_{\alpha_2}\left(n^{1/2} \frac{\tilde{\delta}-\delta_0}{\sigma}\right)-\eta\right)  \\
		& \; + P_n(E_{n,\underline{\sigma}\eta}) - 1,
	\end{align*}
	where $\tilde{Z}_n:=n^{1/2}(\widehat{\tilde{\delta}}- \tilde{\delta})/\sigma$. Hence, by \eqref{eq:CI3_step2} and for sufficiently small $\eta$:
	\begin{equation*}
		\liminf_{n\to\infty} P_n\left(\delta_0 \in \CI{3}\right)  \ge \liminf_{n\to\infty} P_n \left(\left|\tilde{Z}_n + n^{1/2} \frac{\tilde{\delta}-\delta_0}{\sigma} \right| \le q_{\alpha_2}\left(n^{1/2} \frac{\tilde{\delta}-\delta_0}{\sigma}\right)-\eta\right)-\alpha_1.
	\end{equation*}
	Now, let $F_n$ denote the cdf of $\tilde{Z}_n$ under $P_n$ and let $Z\sim\mathcal{N}(0,1)$. We have,
	\begin{align*}
		& \left|P_n\left(\left|\tilde{Z}_n + n^{1/2} \frac{\tilde{\delta}-\delta_0}{\sigma} \right| \le q_{\alpha_2}\left(n^{1/2} \frac{\tilde{\delta}-\delta_0}{\sigma} \right)-\eta\right) - (1-\alpha_2)\right| \\
		\le &  \left|P_n\left(\left|\tilde{Z}_n + n^{1/2} \frac{\tilde{\delta}-\delta_0}{\sigma} \right| \le q_{\alpha_2}\left(n^{1/2} \frac{\tilde{\delta}-\delta_0}{\sigma} \right)-\eta\right) - P_n\left(\left|Z + n^{1/2} \frac{\tilde{\delta}-\delta_0}{\sigma} \right| \le q_{\alpha_2}\left(n^{1/2} \frac{\tilde{\delta}-\delta_0}{\sigma} \right)-\eta\right)\right| \\
		& + \left|P_n\left(\left|Z + n^{1/2} \frac{\tilde{\delta}-\delta_0}{\sigma} \right| \le q_{\alpha_2}\left(n^{1/2} \frac{\tilde{\delta}-\delta_0}{\sigma}\right)-\eta \right) - P_n\left(\left|Z + n^{1/2} \frac{\tilde{\delta}-\delta_0}{\sigma} \right| \le q_{\alpha_2}\left(n^{1/2} \frac{\tilde{\delta}-\delta_0}{\sigma}\right)\right)\right| \\
		\le & \sup_{x\in\R} \left|P_n\left(\left|\tilde{Z}_n + x \right| \le q_{\alpha_2}\left(x\right)-\eta\right) - P_n\left(\left|Z + x \right| \le q_{\alpha_2}\left(x\right)-\eta\right)\right| + \sqrt{\frac{2}{\pi}}\eta\\
		= & \sup_{x\in\R} \left|F_n(x+q_{\alpha_2}(x)-\eta) - \Phi(x+q_{\alpha_2}(x)-\eta) - F_n(x-q_{\alpha_2}(x)+\eta) + \Phi(x-q_{\alpha_2}(x)+\eta)\right| + \sqrt{\frac{2}{\pi}}\eta\\
		\le & 2 \sup_{x\in\R} \left|F_n(x) - \Phi(x)\right| + \sqrt{\frac{2}{\pi}}\eta.
	\end{align*}
	Finally, Lemma \ref{lem:conv_unif_P} implies that for all $x$, $F_n(x) \to \Phi(x)$ with $\Phi$ continuous. By Lemma
	2.11 in van der Vaart (2000), the convergence is uniform. Since $\eta$ was arbitrary small we obtain
	$$\liminf_{n\to\infty} P_n\left(\delta_0 \in \CI{3}\right)  \ge 1 - \alpha_2 - \alpha_1.$$
	The result follows because $\alpha=\alpha_1+\alpha_2$.
	
	\subsection{Theorem \ref{thm:ident1_ordered}} 
	\label{sub:lemma_ref_lem_optim4}
	
	We prove below that there exist identified variables $(Z_t)_{t=0,...,(J-1)T}$ such that $E[U^t/\Omega_{x,\theta_0}(U)|X=x] = E[Z_t|X=x]$. Then, the proof of the theorem follows along the lines of the proofs of Lemma \ref{lem:optim} and Theorem \ref{thm:ident1}. To show the existence of such $(Z_t)_{t=0,...,(J-1)T}$, we first prove that
	\begin{align}
		& \text{span}\left\{u\mapsto P(Y=y|X=x,U=u), \; y\in \{1,...,J\}^{T}\right\}  \nonumber \\
		=&   \text{span}\left\{u\mapsto \frac{u^t}{\Omega_{x,\theta_0}(u)}, \; t\in\{0,...,(J-1)T\}\right\}. \label{eq:from_alpha_to_U_ordered}
	\end{align}
Let $\lpi$ be the set of functions from $\{1,...,T\}$ into $\{0,1,...,J-1\}$. First, we show that the set of conditional probabilities $\left(P(S^{\pi}=s|X,U)\right)_{s=0,...,T, \pi\in\lpi}$ is in one-to-one linear mapping with $P(Y=y|X,U)_{y\in\{0,1,...,J-1\}^T}$. First, $$P(Y=(J-1,...,J-1)|X,U)=P(S^{\overline{\pi}}=T|X,U)$$ with $\overline{\pi}$ such that $\overline{\pi}(t)=J-1$ for all $t$. Next, for any $y=(y_1,...,y_T)\in \{0,1,...,J-1\}^T$, let $\pi \in \lpi$ be such that $\pi(t)=y_t$.  Then:
	$$P(Y=y|X,U)=P(S^{\pi}=T|X,U)-\sum_{\substack{y': y'\neq y \\ \forall t, y'_t\geq y_t}}P(Y= y'|X,U).$$
	Hence, by a decreasing induction on $y$, using the lexicographic order, $P(Y=y|X,U)$ is a linear combination of the $\left(P(S^{\pi}=T|X,U)\right)_{\pi\in \lpi}$. Conversely, $P(S^{\pi}=s|X,U)=\sum_{y\in \mathcal{Y}^{\pi}_s}P(Y=y|X,U)$ with $\mathcal{Y}^{\pi}_s=\left\{y\in\{0,1,...,J-1\}^T :\sum_{t}\mathds{1}\{y_t\geq \pi(t)\}=s\right\}$.
	This ensures that $\left(S^{\pi}\right)_{\pi\in \lpi}$ is exhaustive for $U$ and
	\begin{align*}
		&\text{span}\left\{u\mapsto P(Y=y|X,U=u), \, y\in\{0,1,...,J-1\}^T\right\} \\
		=&\text{span}\left\{u\mapsto P(S^{\pi}=s|X,U=u), \, s=0,...,T, \pi\in\lpi\right\}.
	\end{align*}
	To conclude the proof of \eqref{eq:from_alpha_to_U_ordered}, it suffices to show that
	\begin{align}
		& \text{ span}\left\{u\mapsto P(S^{\pi}=s|X=x,U=u),\, \pi \in \lpi, s=0,...,T\right\}\notag \\
		=&\text{ span}\left\{u\mapsto \frac{u^t}{\Omega_{x,\theta}(u)},\, t=0,...,(J-1)T\right\}.\label{eq:span2}
	\end{align}
	
	For $\pi \in \lpi$, let $\mathcal{T}^{\pi}_+=\{t:\pi(t)>0\}$ and for $k\leq |\mathcal{T}^{\pi}_+|$, let $\mathcal{D}^{\pi}_k=\{d\in \{0,1\}^{\mathcal{T}^{\pi}_+}:\sum_{t\in \mathcal{T}^{\pi}_+} d_t=k\}$ and
	\begin{align*}
		C^{\pi}_k(x, \beta,\gamma):= & \sum_{d\in \mathcal{D}^{\pi}_k} \exp\left(\sum_{t\in \mathcal{T}^{\pi}_+} d_t (x_t'\beta -\gamma_{\pi(t)})\right).
	\end{align*}
	For any $\pi \in \lpi$, let $s_0^{\pi}=T-|\mathcal{T}^{\pi}_+|$. We have
	\begin{align*}
		&P\left(S^{\pi}=s|X=x,U=u\right)\\
		=&\frac{C^{\pi}_{s-s^{\pi}_0}(x,\beta_0,\gamma)\exp(-(s-s^{\pi}_0)v(x,\theta))u^{s-s^{\pi}_0}(1-u)^{T-s}}{\prod_{t\in \mathcal{T}^{\pi}_+}\left[1+u\rho(\pi(t),t,x)\right]}\ind{s_0^{\pi}\leq s\leq T}.
	\end{align*}
	
	The Bernstein polynomials $\{u\mapsto u^{s-s^{\pi}_0}(1-u)^{T-s}, \,s=s_0^{\pi},...,T\}$ are a basis of polynomials of degree lower than $|\mathcal{T}^{\pi}_+|$. Thus,
	\begin{align}
		& \text{ span}\left\{u\mapsto P(S^{\pi}=s|X=x,U=u), \, \pi \in \lpi, \, s=0,...,T\right\} \notag \\
		=& \text{ span}\left\{u\mapsto \frac{u^t}{\prod_{t\in \mathcal{T}^{\pi}_+}\left[1+u\rho(\pi(t),t,x)\right]}, \, \pi \in \lpi, \, t=0,...,|\mathcal{T}^{\pi}_+|\right\} \label{eq:span} \\
		\subseteq & \text{ span}\left\{u\mapsto   \frac{u^t}{\Omega_{x,\theta}(u)},\, t=0,...,(J-1)T\right\}.\notag
	\end{align}
	Conversely, let $\sim$ be the equivalence relation on $\{1,...,J-1\}\times \{1,...,T\}$ defined by:
	$$(j,t)\sim(j',t')\Leftrightarrow \rho(j,t,x)=\rho(j',t',x).$$
	Then let $[(j,t)]$ denote the equivalence class of $(j,t)$. Denote with $[]_0$ the equivalence class such that if $(j,t) \in []_0$, then $\rho(j,t,x) = 0$. Let $\mathcal{E}$ be a set of representatives of all the equivalence classes, except $[]_0$. Let also $n(j,t)=|[(j,t)]|$ and $n_0 = |[]_0|$. Using partial fraction decompositions, we obtain
\begin{align}
		& \text{ span}\left\{u\mapsto u^t \Omega_{x,\theta}(u),\,t=0,...,(J-1)T\right\} \notag \\
		& \subseteq  \text{ span} \bigg\{u\mapsto u^d,\, d=0,...,n_0, \; u\mapsto \left(1+u\rho(j,t',x)\right)^{-d}, (j,t')\in \mathcal{E}, \, d=1,...,n(j,t')\bigg\}. \label{eq:inclusion}
	\end{align}
	Fix $(j,t')\in \mathcal{E}$, $d\in \{1,...,n(j,t')\}$ and let $(j_1,t_1),...,(j_d,t_d)$ denote $d$ distinct elements of $[(j,t')]$. By definition of $\rho$, $t_1,...,t_d$ are all distinct. Then, define $\pi\in\lpi$ as $\pi(t_i)=j_i$ for $i=1,...,d$ and $\pi(t)= 0$ for $t\notin\{t_1,...,t_d\}$. Then:
	\begin{equation}
		\frac{1}{\left(1+u\rho(j,t',x)\right)^d} = \frac{1}{\prod_{t\in \mathcal{T}^{\pi}_+}\left[1+u\rho(\pi(t),t,x)\right]}.
		\label{eq:constr_pi1}
	\end{equation}
	Next, fix $d\in\{0,...,n_0\}$, and let $(j_1,t_1),...,(j_d,t_d)$ denote $d$ distinct elements of $[]_0$. Define $\pi\in\lpi$ exactly as above if $d>0$ and $\pi(t)=0$ for all $t$ if $d=0$. Using $\rho(j_i,t_i,x)=0$ for $i=1,...,d$ and the definition of $\mathcal{T}^{\pi}_+$, we obtain $d=|\mathcal{T}^{\pi}_+|$ and
	\begin{equation}
		u^d=\frac{u^d}{\prod_{t\in \mathcal{T}^{\pi}_+}\left[1+u\rho(\pi(t),t,x)\right]}.
		\label{eq:constr_pi2}
	\end{equation}
	Using \eqref{eq:inclusion} \eqref{eq:constr_pi1} and \eqref{eq:constr_pi2} and then \eqref{eq:span}, we finally obtain
	\begin{align*}
		& \text{ span}\left\{u\mapsto \frac{u^t}{\Omega_{x,\theta}(u)},\,t=0,...,(J-1)T\right\} \\
		\subseteq & \text{ span}\left\{u\mapsto \frac{u^t}{\prod_{t\in \mathcal{T}^{\pi}_+}\left[1+u\rho(\pi(t),t,x)\right]}\, \pi \in \lpi, \, t=0,...,|\mathcal{T}^{\pi}_+|\right\} \\
		= &  \text{ span}\left\{u\mapsto P(S^{\pi}=s|X=x,U=u),\, \pi \in \lpi, s=0,...,T\right\}.
	\end{align*}
	Equation \eqref{eq:span2} follows, and thus \eqref{eq:from_alpha_to_U_ordered} holds as well. This implies that for all $t=0,...,(J-1)T$, there exists functions $(\mu_{k,t}(x))_{k=1,...,J^T}$ such that
\begin{align*}
	\frac{u^t}{\Omega_{x,\theta_0}(u)} = & \sum_{k=1}^{J^T} \mu_{k,t}(x) P(Y=y_k|X=x,U=u) \\
	= & E\left[\sum_{k=1}^{J^T} \mu_{k,t}(x) \ind{Y=y_k} |X=x,U=u\right],
\end{align*}
where $y_1,...,y_{J^T}$ is an enumeration of $\{1,...,J\}^T$. Moreover, the functions $\mu_{k,t}(x)$ are known since $P(Y=y_k|X=x,U=u)$ is a known function of $(u,x)$. Let $Z_t=\sum_{k=1}^{J^T} \mu_{k,t}(x) \ind{Y=y_k}$, $t=0,...,(J-1)T$. By the law of iterated expectation, $E[U^t/\Omega_{x,\theta_0}(U)|X=x]=E[Z_t|X=x]$. The result follows.
		
	
	
	\newpage
	\pagenumbering{arabic}
	\begin{center}
		{\Large Identification and Estimation of Average Causal Effects in Fixed Effects Logit Models \\[2mm] Supplementary material } \\ {\large (not for publication) \\[2mm] Laurent Davezies \quad Xavier D’Haultf{\oe}uille\quad Louise Laage}	
	\end{center}
	
This supplementary material gathers technical lemmas used in some of our proofs, and the proofs of these lemmas. %
Below, we let $\mathcal{B}(u, \epsilon)$ denote the closed ball centered at $u\in \R^d$ and with radius $\epsilon>0$. 
	
	\begin{lem}\label{lem:cushion_bis}
		Suppose that Assumptions \ref{hyp:model}-\ref{hyp:cushion_m} hold. Then, there exists $\epsilon >0$ and $I \in \{1,...,T\}$ such that for all $x \in \Supp(X)$; (i) $\mathcal{B}(m_{\rightarrow I}(x), \epsilon)\subseteq \interior_I$; (ii) if $I<T$, $m_{\rightarrow I+1}(x)\in\frontier_{I+1}$.
	\end{lem}
	
	\textbf{Proof:} First, by \eqref{eq:mom_Z}, we have for all $x\in\Supp(X)$,
	$$m_1(x)= \frac{E[U/\Omega_{x,\beta_0}(U)|X=x]}{E[1/\Omega_{x,\beta_0}(U)|X=x]},$$
	with $U:=\Lambda(v(x,\beta_0)+\alpha)$ and $\Omega_{x,\beta_0}(U)>0$ a.s.. Because $0 < U/\Omega_{x,\beta_0}(U)< 1/\Omega_{x,\beta_0}(U)$ a.s., we have for all $x\in\Supp(X)$, $m_1(x)\in (0,1)$. In other words, $m_{\to 1}(x)\in \interior_1$. This implies that $I(x):=\max\{t\in\{1,...,T\}:m_{\to t}(x)\in\interior_t\}$ is well-defined. We first prove that $I(x)$ does not depend on $x$.
	
	\medskip
	If $m(x)\in\interior_T$ for all $x\in\Supp(X)$, $I(x)=T$ for all $x$ and we have nothing to prove. Otherwise, there exists $x\in\Supp(X)$ such that $m(x)\in\frontier_T$. Then $I(x)<T$. For any $\mu\in \mathcal{D}$, let
	$$\text{ind}(\mu)=2|\Supp(\mu)| - \ind{0\in\Supp(\mu)}-\ind{1\in\Supp(\mu)}.$$
	Then, for any $m\in\M_t$, let ind$(m)=\inf_{\mu\in\mathcal{D}(m)}$ind$(\mu)$. By Theorem 10.7 in \cite{schmudgen2017moment},
	$$I(x)<\text{ind}(m_{\to I(x)}(x))\le \text{ind}(m_{\to I(x)+1}(x)) \le I(x)+1.$$
	As a result, $\text{ind}(m_{\to I(x)}(x))=\text{ind}(m_{\to I(x)+1}(x))=I(x)+1$. Theorem 10.7 in \cite{schmudgen2017moment} also indicates that $\mathcal{D}(m_{\to I(x)+1}(x))$ is a singleton. By $U:=\Lambda(v(x,\beta_0)+\alpha)$, the corresponding distribution  has the same number of support points as $F_{\alpha|X=x}$. As a result, $\text{ind}(m_{\to I(x)+1}(x))=2|\Supp(\alpha|X=x)|$. Hence,
	$$I(x)+1 = 2|\Supp(\alpha|X=x)|.$$
	Since $I(x)<T$, this implies that $2|\Supp(\alpha|X=x)|\le T$. Now, take $x'\ne x$ and assume that $m(x')\in\interior_T$. Then
	$$T<\text{ind}(m(x'))\le 2|\Supp(\alpha|X=x')|,$$
	a contradiction since Assumption \ref{hyp:cushion_m} implies $2|\Supp(\alpha|X=x')|=2|\Supp(\alpha|X=x)|\le T$. Thus, for all $x'\ne x$, we also have $m(x')\in\frontier_T$. Reasoning as above but now for $x'$, we obtain
	$$I(x')+1 = 2|\Supp(\alpha|X=x')|= 2|\Supp(\alpha|X=x)|=I(x)+1.$$
	Hence, $I(\cdot)$ is constant.
	
	\medskip
	Now, let $I=I(x)$, independent of $x$ by what precedes. Point (ii) holds by definition of $I(x)$, so we just need to prove (i). Let us define $M_I^\epsilon:=\cup_{x\in\Supp(X)} \mathcal{B}(m_{\rightarrow I}(x),\epsilon)$ for $\epsilon\ge 0$ and
	$$f(\epsilon):=\inf_{m\in M^\epsilon_I}\underline{H}_I(m)\overline{H}_I(m).$$
	By Theorem 10.8 in \cite{schmudgen2017moment}, it suffices to show that there exists $\epsilon>0$ such that $f(\epsilon)>0$. To this end, first note that  $\underline{H}_I$ and $\overline{H}_I$ are continuous. Moreover, by continuity of $\gamma_0$ (Assumption \ref{hyp:np_for_consistency}.\ref{assn:smoothgamma}) and the fact that $m_t(x)=c_t(x)/c_0(x)$ for $t\in\{0,...,T\}$ and \eqref{eq:otherdef_m}, $x\mapsto m_{\rightarrow I}(x)$ is continuous. $M_I^0$ is thus compact, as the image of the compact set $\Supp(X)$ (in view of Assumption \ref{hyp:iid_bdX}.\ref{assn:Xborne}) by a continuous function. Hence,
	$$f(0)=\min_{m\in M_I}\underline{H}_I(m)\overline{H}_t(m)>0.$$
	The result thus follows if $f$ is continuous at 0. To this end, we apply Berge maximum theorem \citep[see, e.g.][Theorem 2.3]{carter2001foundations}. First, $M_I^\epsilon$ is bounded, since $M_I^0$ itself is bounded. Also, by Berge maximum theorem, the function $d(\cdot, M_I^0)$ defined by $d(x, M_I^0)=\inf_{m\in M_I^0} \|x-m\|$ is continuous. $M_I^\epsilon$ is thus closed as the preimage of $[0,\epsilon]$ by $d(\cdot, M_I^0)$. Hence, $M_I^\epsilon$ is compact. Next, let us prove that the correspondence $\epsilon \mapsto M_I^\epsilon$ is continuous. Let  $\epsilon_n\to \epsilon$, $m_n\in M_I^{\epsilon_n}$ tending to $m$. Then, by continuity,
	$$d(m, M_I^0) = \lim_{n\to\infty} d(m_n, M^0_I) \le \lim_{n\to\infty} \epsilon_n = \epsilon.$$
	Hence, $m\in M_I^\epsilon$ and $\epsilon \mapsto M_I^\epsilon$ is upper hemicontinuous. Finally, fix $(\eps_n) \to \eps$ and $m\in M_I^\eps$. Then there exists $m_0\in M_I^0$ such that $\|m-m_0\|\le \eps$. Let $t_n=\min(1,\eps_n/\eps)$ and $m_n=t_n m +(1-t_n)m_0$. Then, we have $m_n\in M_I^{\eps_n}$ and $\|m-m_n\|\le |\eps-\eps_n|$. As a result, $m_n\to m$, and $\epsilon \mapsto M_I^\epsilon$ is lower hemicontinuous. This proves that $\epsilon \mapsto M_I^\epsilon$ is continuous. By Berge maximum theorem, $f$ is continuous at 0, and the result follows.
	
	\medskip
	\begin{lem}\label{lem:cond_as_nor}
		Suppose that Assumptions \ref{hyp:model}-\ref{hyp:cushion_m} hold. Then, the three conditions in Step 2, sub-step 2, of the proof of Theorem \ref{thm:as_normality} hold.
	\end{lem}
	
	\textbf{Proof:} We use the same notation as that introduced in the proof of Theorem \ref{thm:as_normality}.
	
	\medskip
	\textit{Condition \ref{cond.lin} (linearization):} 
	Note that even if our estimator of $\underline{\delta}$ is:
	$$\widehat{\underline{\delta}}=\frac{1}{n}\sum_{i=1}^n\underline{\widehat{h}}(X_i,S_i),$$
	where, according to \eqref{eq:def_estim_bds}, $\underline{\widehat{h}}(x,s)$ depends on $\widehat{I}(x)$, \eqref{eq:defin_underline_h_long} ensures that
	$\underline{h}(x,s,\gamma,\beta,\tilde{\beta})$ depends only on $I$ defined in Lemma \ref{lem:cushion_bis} (but not on $\widehat{I}(.)$).\\
Arguments leading to \eqref{eq:bound_c_hat} and \eqref{eq:lower_bound_c0} ensure that there exist $\eta>0$ such that:
\begin{align*}
	\norm{c(\gamma,x,\beta)-c(\gamma_0,x,\beta_0)}_{\infty}&\leq C\left(\norm{\gamma-\gamma_0}_{\infty}+\norm{\beta-\beta_0}\right)\\
	c_{0}\left(\gamma,x,\beta\right)&>C
\end{align*}
for any $(\gamma,\beta)$ such that $\norm{\gamma-\gamma_0}_{\infty}+\norm{\beta-\beta_0}\leq \eta$. It follows that
	\begin{align*}
		\norm{m_{\rightarrow I}(\gamma,x,\beta)-m_{\rightarrow I}(\gamma_0,x,\beta_0)}_{\infty}&\leq C \left(\norm{\gamma-\gamma_0}_{\infty}+\norm{\beta-\beta_0}\right)
	\end{align*}
	for any $(\gamma,\beta)$ such that $\norm{\gamma-\gamma_0}_{\infty}+\norm{\beta-\beta_0}\leq \eta$.
And for  sufficiently small $\eta$, Lemma \ref{lem:cushion_bis} ensures that  ${m}(\gamma,x,\beta):=\left(c_t(\gamma,x,\beta)/c_0(\gamma,x,\beta)\right)_{t=0,...,I}$ lies in $\M_I^{\epsilon/2}$ for all $x\in\Supp(X)$ where
	$$\M_I^{\omega}:=\text{cl}\{ m \in \M_I \, ; \, \mathcal{B}(m, \omega) \subset \interior_I \}.$$
	The function $(\gamma,x,\beta) \mapsto m(\gamma,x,\beta)$  depends on $\gamma$ only through $\gamma_{\rightarrow I}(x):=(\gamma_{0}(x),...,\gamma_{I}(x))$, and is infinitely differentiable with respect to the vector $\gamma(x)$ and with respect to $\beta$ for $(\gamma,\beta)$ in a neighborhood of $(\gamma_0,\beta_0)$. This is also the case for $c_0(\gamma,x,\beta)$, $\lambda_{T+1}(x,\beta)$ and $r(x,s,\beta)$ (the derivatives with respect to $\gamma(x)$ are null for $\lambda_{T+1}$ and $r$).
	Lastly, it is known that $\underline{q}_I$ and $\overline{q}_I$ are infinitely differentiable on $\M_I^{\omega}$ for any $\omega>0$. Then $(x,s,\gamma,\beta)\mapsto \underline{h}(x,s,\gamma,\beta,\beta_0)$ is therefore infinitely differentiable in the vector $\gamma(x)$ and in $\beta$ for $\gamma$ and $\beta$ sufficiently close to $\gamma_0$ and $\beta_0$. Recall that $\underline{v}_{\gamma}(X_i, S_i)$ is the gradient of $\underline{h}(x,s,\gamma,\beta,\beta_0)$ with respect to $\gamma(x)$, evaluated at $(X_i, S_i, \gamma_0, \beta_0,\beta_0)$. Next for any $x,s$ and any $\gamma,\beta\in \mathcal{C}_1^0\times B$, $g_{x,s}(t):=\underline{h}(x,s,t\gamma+(1-t)\gamma_0,t\beta+(1-t)\beta_0,\beta_0)$ admits a Taylor expansion at $t=0$, i.e., $g_{x,s}(1)=g_{x,s}(0)+g_{x,s}'(0)+\frac{1}{2}g_{x,s}''(\xi)$ for some $\xi_{x,s}\in [0,1]$. This means that for any $i=1,...,n$ and any $(\gamma,\beta)$ such that $\norm{\gamma-\gamma_0}_{\infty}+\norm{\beta-\beta_0}\leq \eta$ for sufficiently small $\eta>0$ there exists $t_i\in [0;1]$ such that
	\begin{align*}
		& \underline{h}( X_i, S_i, \gamma, \beta,\beta_0) - \underline{h}( X_i, S_i, \gamma_0, \beta_0,\beta_0)- \Deriv{\underline{h}}{\beta}( X_i, S_i, \gamma_0, \beta_0,\beta_0)' \left(\beta - \beta_0\right)\\
		&- \underline{v}_{\gamma}(X_i, S_i)' \left(\gamma(X_i) - \gamma_0(X_i)\right)-g''_{X_i,S_i}(t_i)=0.
	\end{align*}
 For any $s=0,...,T$, $\underline{h}_I( x, s, \gamma, \beta,\beta_0)$ and its derivatives with respect to $(\gamma(x),\beta)$ are continuous in $x,\gamma(x),\beta$. Because $\Supp{(X)}\times [0,1]^{T}\times B$ is compact we have $\sup_{x,s,t}||g_{x,s}''(t)||<C_0 \left[\norm{\gamma-\gamma_0}_{\infty}^2+\norm{\beta-\beta_0}^2\right]$. We conclude that:
	\begin{align*}
	& \biggl[\underline{h}_I( X_i, S_i, \gamma, \beta,\beta_0) - \underline{h}_I( X_i, S_i, \gamma_0, \beta_0,\beta_0)- \Deriv{\underline{h}_I}{\beta}(X_i, S_i, \gamma_0, \beta_0,\beta_0)' \left(\beta - \beta_0\right)\biggr.\\
	&\biggl.- \underline{v}_{\gamma}(X_i, S_i)' \left(\gamma(X_i) - \gamma_0(X_i)\right) \biggr] \leq C_0 \left[\norm{\gamma-\gamma_0}_{\infty}^2+\norm{\beta-\beta_0}^2\right],
\end{align*}
	for $\gamma,\beta$ such that $\norm{\gamma-\gamma_0}_{\infty}+\norm{\beta-\beta_0}\leq \eta$ for some $\eta>0$. Remember that $\norm{\widehat{\gamma}-\gamma_0}_{\infty}=O_P\left(\eta_n\right)$ for $\eta_n=\left(\frac{\ln(n)}{nh_n^{pT}}\right)^{1/2}+h_n^{\ell+1}$ (cf. Section \ref{ssub:uniform_consistency_of_widetilde_m}). Assumption \ref{hyp:np_for_consistency}.\ref{assn:hn_cn} ensures that $\sqrt{n}\eta_n^2\rightarrow 0$ and next $\sqrt{n}\norm{\widehat{\gamma}-\gamma_0}^2_{\infty}=0$.\\
\textit{Condition \ref{cond.sto_equi} (Stochastic equicontinuity):}	 
Let $$J_n=\left|\frac{1}{\sqrt{n}}\sum_{i=1}^n\left( \underline{v}_{\gamma}(X_i,S_i)'(\widehat{\gamma}(X_i)-\gamma_0(X_i))- \int \underline{v}_{\gamma}(x,s)'(\widehat{\gamma}(x)-\gamma_0(x))dP(x,s)\right)\right|.$$
First assume that the following condition holds:
\begin{align}
	\forall \eta>0, \exists C>0 \text{ such that } \limsup_n P\left(\widehat{\gamma}\notin \mathcal{C}^{\ell+1}_{C}(\mathcal{X})\right)<\eta. \label{eq:cond_Donsker_asymp_equi}
	\end{align}
Fix any $\eta>0$, and consider $C$ such that $\limsup_n P\left(\widehat{\gamma}\notin \mathcal{C}^{\ell+1}_{C}(\mathcal{X})\right)<\eta$.\\ $\{x,s\mapsto \underline{v}_{\gamma}(x,s)'\left(\gamma(x)-\gamma_0(x)\right): \gamma \in \mathcal{C}^{\ell+1}_{C}(\mathcal{X})\}$ is a Donsker class \citep[see Example 19.20 in][]{vandervaart_2000}. Because $\norm{\underline{v}_{\gamma}}_{\infty}<\infty$ and $\norm{\widehat{\gamma}-\gamma_0}_{\infty}=O_p(\eta_n)$, we have $$\int \left[\underline{v}_{\gamma}(x,s)'\left(\widehat{\gamma}(x)-\gamma_0(x)\right)\right]^2dP(x,s)\leq O_p(\eta_n^2)=o_p(1)$$ and next, Lemma \ref{lem:random_func} ensures
for any $\varepsilon>0$, we have $\limsup_n P\left(J_n>\varepsilon\right)<\eta$ and next Condition \ref{cond.sto_equi} holds because $\eta$ is arbitrary small.\\
So we only have to show that \eqref{eq:cond_Donsker_asymp_equi} holds.
Let $\mathbb{N}^{pT}_{\leq \ell}$ the set of vectors $(k_1,...,k_{pT})\in \mathbb{N}^{pT}$ such that $\sum_{j=1}^{pT}k_{j}\leq \ell$. For $(k)\in \mathbb{N}^{pT}_{\leq \ell}$, $|k|$ denotes the degree of $(k)$ that is $|k|=\sum_{j=1}^{pT}k_{j}$ and $(k)!=\prod_{j=1}^{pT}k_j!$. We consider that $\mathbb{N}^{pT}_{\leq \ell}$ is ordered by increasing degree and lexicographic order and $rk(k)$ denotes the rank of $(k)$ with respect to this order. For any $(k)\in\mathbb{N}^{pT}_{\leq \ell}$, $e_{(k)}$ denotes the vector of $\mathbb{R}^{\left|\mathbb{N}^{pT}_{\leq \ell}\right|}$ with all components equal to 0 except the $rk(k)$-component equal to 1. For $x=(x_{11},...,x_{pT})\in \mathbb{R}^{pT}$ and $(k)\in \mathbb{N}^{pT}_{\leq \ell}$, let $x^{(k)}=(x_{11}^{k_1},...,x_{pT}^{k_{pT}})$ and $w(x)=(x^{(k)})_{(k)\in \mathbb{N}^{pT}_{\leq \ell}}$. For $j=0,...,T$, $x\in \mathbb{R}^{pT}$, $h>0$, let $\widehat{\rho}_j(x,h)\in \mathbb{R}^{\left|\mathbb{N}^{pT}_{\leq \ell}\right|}$ the weighted least square estimator regressing $\ind{S_i=j}$ on $w(X_i-x)$ weighted by $K\left(\frac{X_i-x}{h}\right)$. Let the diagonal matrix $D(h)=\text{Diag}(h^{|k|})_{(k)\in \mathbb{N}^{pT}_{\leq \ell}}$. For any $x\in \mathbb{R}^{pT}$ and any $h>0$, we have $D(h)w(x/h)=w(x)$. Let
\begin{align*}	
	\widehat{\Omega}_1(x,h_n)&=\frac{1}{nh_n^{pT}}\sum_{i=1}^nw\left(\frac{X_i-x}{h_n}\right)w\left(\frac{X_i-x}{h_n}\right)'K\left(\frac{X_i-x}{h_n}\right).
\end{align*}
From Lemma 5 in \cite{Fan_Guerre_16}, $\underline{\lambda}_{\widehat{\Omega}_1(x,h_n)}$ the smallest eigenvalue  of $\widehat{\Omega}_1(x,h_n)$ is such that $\inf_{x\in \Supp(X)}\underline{\lambda}_{\widehat{\Omega}_1(x,h_n)}$ tends in probability to a positive constant as soon as $\ln(n)/(nh_n^{pT})=o(1)$ and $h_n=o(1)$. This means that $\widehat{\Omega}_1(x,h_n)$ is nonsingular for any $x\in \Supp(X)$ with probability tending to one and:
\begin{align}
	\forall \eta>0, \exists C>0\text{ such that } P\left(\sup_{x\in \mathcal{X}}\left|\left|\widehat{\Omega}_1^{-1}(x,h_n)\right|\right|>C\right)\leq \eta,\label{eq:cond_Donsker_asymp_equi2}
	\end{align}
where $||A||$ refers here to the spectral norm of the matrix $A$. For any $j=0,1,...,T$, we have, when $\widehat{\Omega}_1(x,h_n)$ is nonsingular for all $x\in \mathcal{X}$ (this holds with probability tending to one),
\begin{align*}
	\widehat{\rho}_j(x,h_n)&=D(h_n)^{-1}\widehat{\Omega}_1^{-1}(x,h_n)\widehat{V}_{j}(x,h),
	\end{align*}
with $\widehat{V}_{j}(x,h_n)=\frac{1}{nh_n^{pT}}\sum_{i=1}^n\ind{S_i=j}w\left(\frac{X_i-x}{h_n}\right) K\left(\frac{X_i-x}{h_n}\right)$. And next, with probability approaching one, we have for any $j\in \{0,...,T\}$, $x\in \mathcal{X}$:
\begin{align*}
	\widehat{\gamma}_j(x)&=e'_{(0)}D(h_n)^{-1}\widehat{\Omega}_1^{-1}(x,h_n)\widehat{V}_{j}(x,h)\\
	&=e'_{(0)}\widehat{\Omega}_1^{-1}(x,h_n)\widehat{V}_{j}(x,h_n).
\end{align*}
Because $K$ admit continuous derivatives up to order $\ell+1$, this is also the case for each component of $\widehat{\Omega}_1(x,h_n)$ and $\widehat{V}_j(x,h_n)$ for any $j$. Because for any nonsingular matrix $M$, we have $ (M+H)^{-1}-M^{-1}=-M^{-1}HM^{-1}+o\left(||H||\right)$, each component of $\widehat{\gamma}(x)$ admits continuous derivatives up to order $\ell+1$. We denote $\widehat{\gamma}_j^{(k)}(x)$ the derivative $\frac{\partial^{|k|}\widehat{\gamma}}{\partial x_{11}^{k_1}...\partial x_{pT}^{k_{pT}}}$ for $(k)\in \mathbb{N}^{pT}_{\leq \ell+1}$. Similarly, $\widehat{\Omega}_1^{(k)}(x,h_n)$, $\left[\widehat{\Omega}_1^{-1}\right]^{(k)}(x,h_n)$ and $\widehat{V}_j^{(k)}(x,h_n)$ denote the derivatives of order $(k)$ of $\widehat{\Omega}_1(x,h_n)$, $\widehat{\Omega}_1^{-1}(x,h_n)$ and $\widehat{V}_j(x,h_n)$. To prove \eqref{eq:cond_Donsker_asymp_equi}, it only remains to show that $\sup_{x\in \mathcal{X}}\left|\widehat{\gamma}_j^{(k)}(x)\right|$ are bounded in probability, i.e.:
\begin{align}
	\forall (k)\in \mathbb{N}^{pT}_{\leq \ell+1}, \forall \eta>0, \exists C>0\text{ s.t. }   P\left(\sup_{x\in \mathcal{X}}\left|\widehat{\gamma}_j^{(k)}(x)\right|>C\right)<\eta. \label{eq:cond_Donsker_asymp_equi3}
	\end{align}
We have:
\begin{align*}
	\widehat{\gamma}_j^{(k)}(x)&=e'_{(0)}\sum_{(0)\leq (k')\leq (k)}\left[ \prod_{s=1}^{pT}\binom{k_s}{k'_s}\right]\left[\widehat{\Omega}_1^{-1}\right]^{ (k-k')}(x,h_n)\widehat{V}_j^{(k')}(x,h_n)
	\end{align*}

For $q\in\{1,...,p\}$, $t\in \{1,...,T\}$, we have $\frac{\partial\widehat{\Omega}_1^{-1}}{\partial x_{q t}}(x,h_n)=-\widehat{\Omega}_1^{-1}(x,h_n)\frac{\partial\widehat{\Omega}_1}{\partial x_{q t}}(x,h_n)\widehat{\Omega}^{-1}_1(x,h_n)$. 

\medskip
Next, because \eqref{eq:cond_Donsker_asymp_equi2} holds, to show \eqref{eq:cond_Donsker_asymp_equi3}, it is sufficient to show for any $(k)\in \mathbb{N}^{pT}_{\ell+1}$:
\begin{align}
	\forall \eta>0, \exists C>0\text{ s.t. } P\left(\sup_{x\in \mathcal{X}}\left|\left|\widehat{\Omega}^{(k)}_1(x,h_n)\right|\right|>C\right)<\eta,\label{eq:cond_Donsker_asymp_equi4}\\
	\forall \eta>0, \exists C>0\text{ s.t. }  P\left(\sup_{x\in \mathcal{X}}\left|\left|\widehat{V}^{(k)}_j(x,h_n)\right|\right|>C\right)<\eta.\label{eq:cond_Donsker_asymp_equi5}
\end{align}

For $z\in \mathbb{R}^{pT}$, let $\omega_1(z)=w(z)w'(z)K(z)$ and $q(z)=w(z)K(z)$ and $r_j(z)=\gamma_{0j}(z)f_{X}(z)$. Let $\Omega_1^{(k)}(x,h):=E(\widehat{\Omega}^{(k)}_1(x,h))$ and $V_1^{(k)}(x,h):=E\left(\widehat{V}_1^{(k)}(x,h)\right)$. We have by dominated convergence: \begin{align*}\Omega_1^{(k)}(x,h)&=\frac{(-1)^{|k|}}{h^{pT+|k|}}E\left[\omega_1^{(k)}\left(\frac{X-x}{h}\right)\right]\\&=\left[E\left[\frac{1}{h^{pT}}\omega_1\left(\frac{X-x}{h}\right)\right]\right]^{(k)}\\&=\left[\int\omega_1(z)f_X(x+hz)dz\right]^{(k)}\\&=\int\omega_1(z)f^{(k)}_X(x+hz)dz.
\end{align*}
Similarly, $V_j^{(k)}(x,h)=\int q(z)r_j^{(k)}(x+hz)dz$. This ensures that $\sup_{x\in \mathcal{X},h\in(0;1]}||\Omega_1^{(k)}(x,h)||<\infty$ and $\sup_{x\in \mathcal{X},h\in(0;1]}||V_1^{(k)}(x,h)||<\infty$. For the component $(a,b)$ of $\widehat{\Omega}_1^{(k)}(x,h_n)$, we have
\begin{align*}
	\text{Var}\left[\left(\widehat{\Omega}_1^{(k)}(x,h_n)\right)_{a,b}\right]&\leq \frac{1}{nh_n^{2pT+2|k|}}E\left[\left({\omega}_1^{(k)}\left(\frac{X-x}{h_n}\right)\right)_{a,b}^2\right]\\&\leq\frac{1}{nh_n^{pT+2|k|}}\int \left(\omega_{1ab}^{(k)}(z)\right)^2f_{X}(x+zh_n)dz\\&\leq \frac{1}{nh_n^{pT+2|k|}}||f_{X}||_{\infty}\int \left(\omega_{1ab}^{(k)}(z)\right)^2dz,
	\end{align*}
and a similar equality holds for each component of $\widehat{V}_j^{(k)}(x,h_n)$. 

Let $\varepsilon>0$ and $(x_1,...,x_{N_{\varepsilon}})\in \mathcal{X}^{N_{\varepsilon}}$ such that $\mathcal{X}\subseteq \bigcup_{j=1}^{N_{\varepsilon}}B(x_i,\varepsilon)$ and $N_{\varepsilon}\leq C_{\mathcal{X}}\varepsilon^{-pT}$ for some $C_{\mathcal{X}}>0$ that does not depend on $\varepsilon>0$. By the triangle inequality, we have
\begin{align}
	&\sup_{x\in \mathcal{X}}\left|\left|\widehat{\Omega}_1^{(k)}(x,h_n)-\Omega_1^{(k)}(x,h_n)\right|\right|\nonumber\\&\leq \max_{i=1,...,N_{\varepsilon}}\left|\left|\widehat{\Omega}_1^{(k)}(x_i,h_n)-\Omega_1^{(k)}(x_i,h_n)\right|\right|
	+\max_{i=1,...,N_{\varepsilon}}\sup_{x\in B(x_i,\varepsilon)\cap\mathcal{X}}\left|\left|\Omega_1^{(k)}(x,h_n)-\Omega_1^{(k)}(x_i,h_n)\right|\right|\nonumber\\
	&+\max_{i=1,...,N_{\varepsilon}}\sup_{x\in B(x_i,\varepsilon)\cap\mathcal{X}}\left|\left|\widehat{\Omega}_1^{(k)}(x,h_n)-\widehat{\Omega}_1^{(k)}(x_i,h_n)\right|\right|\nonumber\\
	\leq& \max_{i=1,...,N_{\varepsilon}}\left|\left|\widehat{\Omega}_1^{(k)}(x_i,h_n)-\Omega_1^{(k)}(x_i,h_n)\right|\right|+\overline{C}(1+h_n^{-pT-|k|-1})\varepsilon,\label{eq:control_hat_Omega^k}
	\end{align}
where the last inequality holds for some $\overline{C}>0$ because $\left|\left|\omega_1^{(k)}(x)\right|\right|$ and $|f_X^{(k)}(x)|$ are Lipschitz on $\mathcal{X}$ for any $(k)$ such that $|k|\leq \ell+1$. Bernstein inequality ensures for any $i=1,...,N_{\varepsilon}$ and for any component $j,j'$ of $\widehat{\Omega}^{(k)}_1(x_i,h_n)-{\Omega}^{(k)}_1(x_i,h_n)$ and any $t>0$:
\begin{align*}
	P\left(\left|\left[\widehat{\Omega}^{(k)}_1(x_i,h_n)-\Omega_1^{(k)}(x_i,h_n)\right]_{j,j'}\right|>t\right)&\leq 2\exp\left(-\frac{t^2/2}{\frac{||f_X||_{\infty}\int (\omega_{1jj'}^{(k)}(z))^2dz}{nh_n^{pT+2|k|}}+\frac{||\omega_{1jj'}||_{\infty}t}{3nh_n^{pT}}}\right).
	\end{align*}
Next, there exists $\overline{C}'>0$ such that for any $t>0$:
\begin{align*}
	P\left(\left|\left|\widehat{\Omega}^{(k)}_1(x_i,h_n)-\Omega_1^{(k)}(x_i,h_n)\right|\right|>t\right)&\leq \overline{C}'\exp\left(-\frac{nh_n^{pT+2|k|}t^2}{\overline{C}'(1+t)}\right).
\end{align*}
Now choosing a sequence $\varepsilon_n=h_n^{pT+|k|+1}$, using the union bound we deduce:
\begin{align*}
	P\left(\max_{i=1,...,N_{\varepsilon_n}}\left|\left|\widehat{\Omega}^{(k)}_1(x_i,h_n)-\Omega_1^{(k)}(x_i,h_n)\right|\right|>t\right)&\leq \overline{C}'N_{\varepsilon_n}\exp\left(-\frac{t^2nh_n^{pT+2|k|}}{\overline{C}'(1+t)}\right)\\
	&\leq \overline{C}' C_{\mathcal{X}}\exp\left(-\frac{t^2nh_n^{pT+2|k|}}{\overline{C}'(1+t)}-pT\ln(\varepsilon_n)\right),
	\end{align*}
with $\ln(\varepsilon_n)=O(\ln(h_n))=o(nh_n^{pT+2(\ell+1)})=o(nh_n^{pT+2|k|})$.
And next, for any $t>0$, $\lim_n P\left(\max_{i=1,...,N_{\varepsilon_n}}\left|\left|\widehat{\Omega}^{(k)}_1(x_i,h_n)-\Omega_1^{(k)}(x_i,h_n)\right|\right|>t\right)=0$. Thus for any $t>\overline{C}=\lim_n \overline{C}(1+h_n^{-pT-|k|-1})\varepsilon_n$, \eqref{eq:control_hat_Omega^k} ensures $\lim_n P\left(\sup_{x\in \mathcal{X}}\left|\left|\widehat{\Omega}^{(k)}_1(x_i,h_n)-\Omega_1^{(k)}(x_i,h_n)\right|\right|>t\right)=0$. Because $\sup_{x,h}||\Omega^{(k)}_1(x,h)||<\infty$, this ensures \eqref{eq:cond_Donsker_asymp_equi4}. \eqref{eq:cond_Donsker_asymp_equi5} follows from a similar reasoning.

\textit{Condition \ref{cond.mean_sq_cont} (Mean square continuity):}
Let $\mathbb{N}^{pT}_{\leq \ell}$ the set of vectors $(k_1,...,k_{pT})\in \mathbb{N}^{pT}$ such that $\sum_{j=1}^{pT}k_{j}\leq \ell$. For $(k)\in \mathbb{N}^{pT}_{\leq \ell}$, $|k|$ denotes the degree of $(k)$ that is $|k|=\sum_{j=1}^{pT}k_{j}$ and $(k)!=\prod_{j=1}^{pT}k_j!$. We consider that $\mathbb{N}^{pT}_{\leq \ell}$ is ordered by increasing degree and lexicographic order and $rk(k)$ denotes the rank of $(k)$ with respect to this order. For any $(k)\in\mathbb{N}^{pT}_{\leq \ell}$, $e_{(k)}$ denotes the vector of $\mathbb{R}^{\left|\mathbb{N}^{pT}_{\leq \ell}\right|}$ with all components equal to 0 except the $rk(k)$-component equal to 1. For $x=(x_{11},...,x_{pT})\in \mathbb{R}^{pT}$ and $(k)\in \mathbb{N}^{pT}_{\leq \ell}$, let $x^{(k)}=(x_{11}^{k_1},...,x_{pT}^{k_{pT}})$ and $w(x)=(x^{(k)})_{(k)\in \mathbb{N}^{pT}_{\leq \ell}}$. Last, for $h>0$ let the diagonal matrix $D(h)=\text{Diag}(h^{|k|})_{(k)\in \mathbb{N}^{pT}_{\leq \ell}}$. For any $x\in \mathbb{R}^{pT}$ and any $h>0$, we have $D(h)w(x/h)=w(x)$.\\ For $j=0,...,T$, let $\widehat{\rho}_j(x,h)\in \mathbb{R}^{\left|\mathbb{N}^{pT}_{\leq \ell}\right|}$ the weighted least square estimator regressing $\ind{S_i=j}$ on $w(X_i-x)$ weighted by $K\left(\frac{X_i-x}{h}\right)$ and let $\overline{\rho}_j(x,h)$ its limit when $h$ is fixed and $n\rightarrow \infty$. The derivative $\gamma^{(k)}_j(x)$ of order $(k)$ of $\gamma_j(x)$ is estimated by $\widehat{\gamma^{(k)}_j}(x,h_n):=\frac{|k|!}{(k)!}e'_{(k)}\widehat{\rho}_j(x,h_n)$. Similarly, let $\overline{\gamma}_j^{(k)}(x,h)=\frac{|k|!}{(k)!}e'_{(k)}\overline{\rho}_j(x,h)$. Last, $\rho_j(x)$ is the vector such that $\gamma^{(k)}(x)=\frac{|k|!}{(k)!}e'_{(k)}\rho_j(x)$.

For any $k\in \mathbb{N}^{pT}_{\leq \ell}$ and any $j=0,1,...,T$, we have:
\begin{align*}
	\widehat{\rho}_j(x,h_n)&=D(h_n)^{-1}\widehat{\Omega}_1^{-1}(x,h_n)\frac{1}{nh_n^{pT}}\sum_{i=1}^n\ind{S_i=j}w\left(\frac{X_i-x}{h_n}\right)K\left(\frac{X_i-x}{h_n}\right),\\
	\overline{\rho}_j(x,h_n)&=D(h_n)^{-1}\Omega^{-1}_1(x,h_n)\frac{1}{h_n^{pT}}E\left[\ind{S_i=j}w\left(\frac{X_i-x}{h_n}\right)K\left(\frac{X_i-x}{h_n}\right)\right],\end{align*}
with
\begin{align*}	
	\widehat{\Omega}_1(x,h_n)&=\frac{1}{nh_n^{pT}}\sum_{i=1}^nw\left(\frac{X_i-x}{h_n}\right)w\left(\frac{X_i-x}{h_n}\right)'K\left(\frac{X_i-x}{h_n}\right),\\
	{\Omega}_1(x,h_n)&=\frac{1}{h_n^{pT}}\E\left(w\left(\frac{X_1-x}{h_n}\right)w\left(\frac{X_1-x}{h_n}\right)'K\left(\frac{X_1-x}{h_n}\right)\right)\\
	&=\int w(z)w'(z)K(z)f_X(x+h_nz)dz.
	\end{align*}	
Note that if $x$ is an inner point of the support of $X$, we have 
$$\lim_{h\downarrow 0} \Omega_1(x,h)=f_X(x)\int w(z)w'(z)K(z) dz,$$
but if $x$ belong to the boundary of the support of $X$, the limit will depend on the local shape of $\Supp(X)$. However, assumptions on the support of $X$ and on $f_X$ ensures that $||\Omega_1(x,h)||$ is bounded away from 0 and infinity when $h$ tends to 0 for any $x$.

Theorem 2 in \cite{Fan_Guerre_16} ensures that:
$$\sup_{x\in \Supp{(X)}}\left|\left|\gamma_j(x)-\overline{\gamma}^{(0)}_j(x,h_n)\right|\right|=O(h_n^{\ell+1}).$$

Let $\xi_{ji}(x,h)=\gamma_j(X_i)-w(X_i-x)'\overline{\rho}_j(x,h_n)$ and $\epsilon_{ji}=\ind{S_i=j}-\gamma_j(X_i)$. Note that $\frac{1}{nh_n^{pT}}\sum_{i=1}^nw\left(\frac{X_i-x}{h_n}\right)w\left(X_i-x\right)'\overline{\rho}_j(x,h_n)K\left(\frac{X_i-x}{h_n}\right)=\widehat{\Omega}_1(x,h_n)D(h_n)\overline{\rho}_j(x,h_n)$. It follows:
\begin{align}
	D(h_n)\left(\widehat{\rho}_j(x,h_n)-\overline{\rho}_j(x,h_n)\right)=&\widehat{\Omega}_1^{-1}(x,h_n)\frac{1}{nh_n^{pT}}\sum_{i=1}^nw\left(\frac{X_i-x}{h_n}\right)\xi_{ji}(x,h_n)K\left(\frac{X_i-x}{h_n}\right)\nonumber\\
	&+\widehat{\Omega}_1^{-1}(x,h_n)\frac{1}{nh_n^{pT}}\sum_{i=1}^nw\left(\frac{X_i-x}{h_n}\right)\epsilon_{ji}K\left(\frac{X_i-x}{h_n}\right)\label{eq:mean_sq_cont_1}
	\end{align}
Lemma 5 in \cite{Fan_Guerre_16} ensures:
\begin{align}\sup_{x}\norm{\widehat{\Omega}_1(x,h_n)-\Omega_1(x,h_n)}&=O_p\left(\left(\frac{\ln(n)}{nh_n^{pT}}\right)^{1/2}\right),\label{eq:mean_sq_cont_2}\end{align}
and  \begin{align}\inf_{h}\inf_{x\in \Supp{(X)}}\norm{\Omega_1(x,h)}>0.\label{eq:mean_sq_cont_3}
\end{align}
Independence and identical distribution across $i$ and $E(\epsilon_{ji}|X_i)=0$ ensure:
\begin{align}&E\left[\norm{\int\frac{1}{\sqrt{n}h_n^{pT}}\sum_{i=1}^nw\left(\frac{X_i-x}{h_n}\right)\epsilon_{ji}K\left(\frac{X_i-x}{h_n}\right)dP_{X,S}(x,s)}^2\right]\nonumber\\&= E\left[\norm{\frac{1}{\sqrt{n }}\sum_{i=1}^n\epsilon_{ji}\int w\left(z\right)K\left(z\right)f_X(X_i-zh_n)d(z)}^2\right]\nonumber\\
	&= \E\left(\epsilon_{j1}^2\norm{\int w\left(z\right)K\left(z\right)f_X(X_i-hz)dz}^2\right)\nonumber\\ &\leq\norm{f_X}_{\infty}^2\left(\int \norm{w(z)}K(z)dz\right)^2<\infty.\label{eq:mean_sq_cont_4}\end{align}
Because  \begin{align*}\overline{\rho}_j(x,h_n)&=\arg\min_{b}E\left(\left(\ind{S_1=j}-w(X_1-x)'b\right)^2K\left(\frac{X_i-x}{h_n},
	\right)\right)\end{align*}
we have $\E\left(w\left(\frac{X_1-x}{h}\right)\xi_{j1}(x,h)K\left(\frac{X_1-x}{h}\right)\right)=0$ for any $x$ and the Fubini Theorem ensures that $\E\left(\int w\left(\frac{X_1-x}{h}\right)\xi_{j1}(x,h)K\left(\frac{X_1-x}{h}\right)f_X(x)dx\right)=0$. Independence across $i$ implies: 
\begin{align*}
	&\E\left[\norm{\int \frac{1}{\sqrt{n}h_n^{pT}}\sum_{i=1}^nw\left(\frac{X_i-x}{h_n}\right) \xi_{ji}(x,h_n)K\left(\frac{X_i-x}{h_n}\right)f_X(x)dx}^{2}\right] \\
&=\E\left[\norm{\int \frac{1}{h_n^{pT}}w\left(\frac{X_i-x}{h_n}\right)\xi_{ji}(x,h_n)K\left(\frac{X_i-x}{h_n}\right)f_X(x)dx}^{2}\right]\\
	&=\int \norm{\int  w\left(u\right)\left(\gamma(y)-w\left(u\right)'D(h_n)\overline{\rho}_j(y-uh_n,h_n)\right)K\left(u\right)f_X(y-uh_n)du}^2f_X(y)dy.
\end{align*}
Arguments used to prove Theorem 2 in \cite{Fan_Guerre_16} ensure that there exists $C>0$ such that for any $h>0$ \begin{align*}\sup_{y,u\in \Supp{(X)}\times \Supp{(K)},y-uh\in \Supp(X)}\left|\gamma(y)-w\left(u\right)'D(h)\overline{\rho}_j(y-uh,h)\right|<Ch^{\ell+1}.
\end{align*}
This ensures
\begin{align}
	\E\left[\norm{\int \frac{1}{\sqrt{n}h_n^{pT}}\sum_{i=1}^nw\left(\frac{X_i-x}{h_n}\right)\xi_{ji}(x,h_n) K\left(\frac{X_i-x}{h_n}\right)f_X(x)dx}^{2}\right]&=O(h_n^{2(\ell+1)}).\label{eq:mean_sq_cont_5}
\end{align}

Together, Equations \eqref{eq:mean_sq_cont_1}-\eqref{eq:mean_sq_cont_5}  ensure that for any $j=0,...,T$:
\begin{align}
	&\int \norm{\sqrt{n }\left(\widehat{\gamma}_j(x)-\gamma_{0j}(x)\right)-e'_{(0)}\Omega^{-1}(x,h_n)\frac{1}{\sqrt{n}h_n^{pT}}\sum_{i=1}^nw\left(\frac{X_i-x}{h_n}\right)\epsilon_{ji}K\left(\frac{X_i-x}{h_n}\right)}^2dP_{X,S}(x,s)\nonumber\\&=O_p\left(h_n^{2(\ell+1)}+\frac{\ln(n)}{nh_n^{pT}}\right)=o_p(1).\label{eq:mean_sq_cont_6}
	\end{align}
Let $\underline{v}_{\gamma j}(x,s)$ the component $j$ of $\underline{v}_{\gamma}(x,s)$ and let $\widetilde{\lambda}_{j}(x)=\E\left(\underline{v}_{\gamma j}(X,S)|X=x\right)$. Because $\E \left(\underline{v}_{\gamma j}^2(x,s)\right)<\infty$, Cauchy-Schwarz inequality and \eqref{eq:mean_sq_cont_6} implies:
\begin{align}
	&\sqrt{n}\int \underline{v}_{\gamma j}(x,s)(\widehat{\gamma}_j(x)-\gamma_{0j}(x))dP_{X,S}(x,s)\nonumber\\
	&=\frac{1}{\sqrt{n}}\sum_{i=1}^n\epsilon_{ji}e'_{(0)}\int \Omega^{-1}(x,h_n)\frac{1}{h_n^{pT}}w\left(\frac{X_i-x}{h_n}\right)\widetilde{\lambda}_{j}(x)K\left(\frac{X_i-x}{h_n}\right)f_{X}(x)dx+o_p(1)\nonumber\\
	&=\frac{1}{\sqrt{n}}\sum_{i=1}^n\epsilon_{ji}e'_{(0)}\int \Omega^{-1}(X_i-zh_n,h_n)w\left(z\right)\widetilde{\lambda}_{j}(X_i-zh_n)K\left(z\right)f_{X}(X_i-zh_n)dz\nonumber\\&~~~~+o_p(1).\label{eq:mean_sq_cont_7}
	\end{align}
Because $f_X$ is Lipschitz on $\Supp (X)$, we have:
\begin{align*}h_n^{-1}\sup_{x\in \Supp{X}}\norm{\Omega(x,h_n)-f_X(x)\int w(z)w'(z)K(z)dz}<\infty,
\end{align*}
we also have $\inf_{x\in \Supp{X}}f_X(x)\norm{\int w(z)w'(z)K(z)dz}>0.$
Next because $f_X$ is bounded away from 0 on its support and $A\mapsto A^{-1}$ is Lipschitz on $\{A:\norm{A}\geq c\}$ for any $c>0$, we have 
\begin{align}
	\sup_{x\in \Supp{X}}\norm{\chi(x,h_n)}&<Ch_n,\label{eq:mean_sq_cont_8}
\end{align} 
for $\chi(x,h)=\Omega^{-1}(x,h)f_X(x)-\left[\int w(u)w'(u)K(u)du\right]^{-1}$ and some constant $C$. Lindeberg-Feller Theorem ensures
\begin{align}
	\frac{1}{\sqrt{n}}\sum_{i=1}^n\epsilon_{ji}e'_{(0)}\int_{zh_n\in X_i-\Supp(X)} \chi(X_i-zh_n,h_n)w\left(z\right)\widetilde{\lambda}_{j}(X_i-zh_n)K\left(z\right)dz&=o_P(1).\label{eq:mean_sq_cont_9}
	\end{align}
Because $x\mapsto \tilde{\lambda}_j(x)$ is Lipschitz on $\Supp(X)$ and $ \int ||w(z)||\times ||z|| K(z) dz<\infty$, we have: \begin{align}
\sup_{x\in \Supp(X)}\norm{\int_{x-zh\in \Supp(X)} w(z)\left(\tilde{\lambda}_j(x-zh)-\tilde{\lambda}_j(x)\right)K(z)dz}\leq C h_n. 
\label{eq:mean_sq_cont_10} 
\end{align}
From \eqref{eq:mean_sq_cont_7}, \eqref{eq:mean_sq_cont_9} and \eqref{eq:mean_sq_cont_10}, we deduce
\begin{align}
	&\sqrt{n}\int \underline{v}_{\gamma j}(x,s)(\widehat{\gamma}_j(x)-\gamma_{0j}(x))dP_{X,S}(x,s)\nonumber\\
	&=\frac{1}{\sqrt{n}}\sum_{i=1}^n\epsilon_{ji}\widetilde{\lambda}_j(X_i)e_{(0)}'\left[\int w(u)w(u)'K(u)du\right]^{-1}\int_{X_i-zh_n\in \Supp (X)}w(z)K(z)dz+o_p(1).
\label{eq:mean_sq_cont_11}\end{align}
Consider the Hilbert space of functions on $\mathbb{R}^{pT}$ equipped with the scalar product $\langle g, f\rangle=\int g(u) f(u) K(u) du$. Because the constant function 1 belongs to the linear space generated by $w(.)$, it is equal to its min-square projection on this space and next \begin{align}1=&e_{(0)}'\left[\int w(u)w(u)'K(u)du\right]^{-1}\int w(z)K(z)dz.\label{eq:mean_sq_cont_12} \end{align}
Because $\int_{X_i-zh_n\in \Supp (X)}w(z)K(z)dz$ converges almost-surely to $\int w(z)K(z)dz$, dominated convergence and Lindeberg-Feller theorems combined with \eqref{eq:mean_sq_cont_11} and \eqref{eq:mean_sq_cont_12} ensure:
 \begin{align*}
 	\sqrt{n}\int \underline{v}_{\gamma j}(x,s)(\widehat{\gamma}_j(x)-\gamma_{0j}(x))dP_{X,S}(x,s)&=\frac{1}{\sqrt{n}}\sum_{i=1}^n\epsilon_{ji}\widetilde{\lambda}_j(X_i)+o_p(1)
 \end{align*}
for all $j$. Since $\underline{v}_{\gamma j}(x,s)$ does not depend on $s$, $\widetilde{\lambda}_{j}(x)=\E\left(\underline{v}_{\gamma j}(X,S)|X=x\right) = \underline{v}_{\gamma j}(x,s)$ and Condition \ref{cond.mean_sq_cont} in Part 1, Step 3 of the proof of Theorem \ref{thm:as_normality} holds.

	\begin{lem}\label{lem:as_nor_triv}
		Let $\mathcal{P}'$ the set of probability distributions of $(Y,X,\alpha)$ such that Assumption \ref{hyp:model}-\ref{hyp:iid_bdX} hold with $\mathcal{I}_{0}>>\underline{A}$ and $\sigma^2\geq \underline{\sigma}^2$ for some definite positive matrix $\underline{A}$ and some $\underline{\sigma}>0$. Then $\widehat{\sigma}\convP \sigma$ and \eqref{eq:asnor_triv} hold uniformly on $\mathcal{P}'$.
	\end{lem}
	
	\textbf{Proof:} To show these results, it suffices to show that they hold along any sequence of probability distribution $(P_n)_{n\geq 1}$ in $\mathcal{P}'$. We use the same notation as in the other proofs but index parameters, variables and the expectation operator by $n$ to underline their dependence on $P_n$ when deemed necessary. Relatedly, we use $o_{P_n}(1)$ as a shortcut for a sequence of random variables (or vectors or matrices) $\eps_n$ satisfying $P_n(\norm{\eps_n}>\eta)\to 0$ for all $\eta>0$. First, let $W_i=(X_i,S_i)$. We have $\tilde{\delta}_n = \E_n[p(W_1,\beta_0)]$ and $\widehat{\tilde{\delta}}=\sum_{i=1}^n p(W_i,\widehat{\beta})/n$. Assumptions \ref{hyp:fct_obj_simple} and \ref{hyp:iid_bdX} ensure that $\norm{f}_{\infty}=\sup_{(w,b)\in \Supp{W}\times B}\norm{f(w,b)}<\infty$ for $f=p, \deriv{p}{\beta}, \deriv{{}^2p}{\beta\partial \beta'}$. Lemma \ref{lem:conv_unif_P} ensures
	\begin{equation}
		\sqrt{n}\left(\widehat{\beta}-\beta_0\right) =\frac{1}{n^{1/2}} \sum_{i=1}^n \phi_{n,i} + o_{P_n}(1).
		\label{eq:beta_infl_fun}
	\end{equation}

	\medskip
Since $\beta\mapsto p(w,\beta)$ is differentiable for all $w$, by the mean value theorem, there exists $\overline{\beta}_{n,i}=t_{n,i}\widehat{\beta}+(1-t_{n,i})\beta_{0n}$, with $t_{in}\in[0,1]$, such that $p(W_i,\widehat{\beta})-p(W_i,\beta_{0n})=\deriv{p}{\beta'}(W_i,\overline{\beta}_{n,i})(\widehat{\beta}-\beta_{0n})$. Let
$$\widehat{G} =  \frac{1}{n}\sum_{i=1}^n \Deriv{p}{\beta}(W_i,\overline{\beta}_{n,i})\text{ and }G_n=\E_n\left(\Deriv{p}{\beta}(W_1,\beta_{0})\right).$$
Then it follows
$$\sqrt{n}\left(\widehat{\tilde{\delta}} - \widetilde{\delta}_n\right)  = \frac{1}{n^{1/2}} \sum_{i=1}^n \left[\widehat{G} \phi_{n,i} + p(W_i,\beta_{0n}) - \widetilde{\delta}\right] +\widehat{G}  o_{P_n}(1).$$
Note that for any $i$, $\beta_{n,i}\in B$ and $\norm{\overline{\beta}_{n,i} -\beta_{0n}}\leq \norm{\widehat{\beta}-\beta_{0n}}$. Hence, $\norm{G_n}\leq \norm{\deriv{p}{\beta}}_{\infty}$ and
\begin{equation}
	\norm{\widehat{G}-G_n} \leq \norm{\deriv{{}^2 p}{\beta \partial \beta'}}_{\infty} \norm{\widehat{\beta} - \beta_{0n}} + \norm{\frac{1}{n}\sum_{i=1}^n \Deriv{p}{\beta}(W_i,\beta_{0n}) - G_n}.	
	\label{eq:bound_G_hat}
\end{equation}
By the first part of the proof, $\widehat{\beta} - \beta_{0n}=o_{P_n}(1)$. Next, because $\deriv{p}{\beta}(.,\beta_{0n})$ is bounded on $\Supp(W)$, the uniform integrability condition of \cite{gut1992weak} also holds for this variable. Then, by the weak LLN of \cite{gut1992weak}, the second term of \eqref{eq:bound_G_hat} is an $o_{P_n}(1)$. Thus, $\norm{\widehat{G}-G_n}=o_{P_n}(1)$ and $\norm{\widehat{G}}\leq \norm{\deriv{p}{\beta}}_{\infty}+o_{P_n}(1)$. As a result,
$$\sqrt{n}\frac{\widehat{\tilde{\delta}} - \widetilde{\delta}}{\sigma_n}  = \frac{1}{n^{1/2}} \sum_{i=1}^n \frac{G_n \phi_{n,i} + p(W_i,\beta_{0n}) - E_n[p(W_1,\beta_{0n})]}{\sigma_n}+ o_{P_n}(1).$$
Now, by the triangle and Cauchy-Schwarz inequalities, we have
\begin{equation}
	\left|G_n \phi_{n,i} + p(W_i,\beta_{0n}) - E_n[p(W_1,\beta_{0n})] \right| \leq \norm{G_n} \norm{\phi_{n,i}} + \left|p(W_i,\beta_{0n}) - E_n[p(W_1,\beta_{0n})]\right|.	
	\label{eq:bound_for_CLT}
\end{equation}
We have $\norm{G_n}\leq \norm{\deriv{p}{\beta}}_{\infty}$, $\left|p(W_i,\beta_{0n}) - E_n[p(W_i,\beta_{0n})]\right|\leq 2\norm{p}_{\infty}$ and
$$\norm{\phi_{n,i}}=\norm{\mathcal{I}_{n0}^{-1}\Deriv{\ell_c}{\beta}(W_i,\beta_{0n})}\leq \norm{\underline{A}^{-1} \Deriv{\ell_c}{\beta}(W_i,\beta_{0n})}\leq \underline{\rho}^{-1} \norm{\Deriv{\ell_c}{\beta}}_{\infty},$$
where $\underline{\rho}>0$ denotes the smallest eigenvalue of $\underline{A}$. Then, using \eqref{eq:bound_for_CLT} and $\sigma_n\geq \underline{\sigma}$, the variables $(G_n \phi_{n,i} + p(W_i,\beta_{0n})-E_n(p(W,\beta_{0n})))/\sigma_n$ are bounded by a constant independent of $n$. Thus, they satisfy the Lindeberg condition. Then, by the central limit theorem for triangular arrays,
$$\sqrt{n}\frac{\widehat{\tilde{\delta}} - \widetilde{\delta}}{\sigma_n} \convL \mathcal{N}(0,1),$$ ensuring that \eqref{eq:asnor_triv} holds uniformly on $\mathcal{P}'$.\\
We now show that $\widehat{\sigma}$ converges to $\sigma$ uniformly over $\mathcal{P}'$. First note for $\psi_{n,i}=p(W_i,\beta_{0n})-\E_n(p(W_1,\beta_{0n}))+G_n'\phi_{n,i}$, we have:
\begin{align}
	|\psi_{n,i}|&\leq 2 \norm{p}_{\infty}+\norm{\Deriv{p}{\beta}}_{\infty}\norm{\phi_{n,i}}\leq 2 \norm{p}_{\infty}+\norm{\Deriv{p}{\beta}}_{\infty}\underline{\rho}^{-1}\norm{\Deriv{\ell_c}{\beta}}_{\infty}\label{eq:bound_psi}
	\end{align}
and next, the LLN of \cite{gut1992weak} ensures \begin{equation}\frac{1}{n}\sum_{i=1}^n\psi_{n,i}^2=\sigma_n^2+o_{P_{n}}(1).\label{eq:sigma_convU}\end{equation}
We have by triangle inequality and bounded derivatives of $p$:
\begin{align*}
\left|\widehat{\psi}_{i}-\psi_{n,i}\right| &= \left|p(W_i,\widehat{\beta})-\frac{1}{n}\sum_{j=1}^n p(W_j,\widehat{\beta})+  \left[\frac{1}{n}\sum_{j=1}^n \Deriv{p}{\beta}(W_j,\widehat{\beta}) \right]' \widehat{\phi}_i-\psi_{n,i}\right|\\
&\leq 2\norm{\Deriv{p}{\beta}}_{\infty}\norm{\widehat{\beta}-\beta_0}+\norm{\frac{1}{n}\sum_{j=1}^n p(W_j,\beta_0)-\tilde{\delta}_n}+\left|\left[\frac{1}{n}\sum_{j=1}^n \Deriv{p}{\beta}(W_j,\widehat{\beta}) \right]' \widehat{\phi}_i-G_n'\phi_{n,i}\right|\\
&\leq 2\norm{\Deriv{p}{\beta}}_{\infty}\norm{\widehat{\beta}-\beta_0}+\norm{\frac{1}{n}\sum_{j=1}^n p(W_j,\beta_0)-\tilde{\delta}_n}+\left|\left[\frac{1}{n}\sum_{j=1}^n \Deriv{p}{\beta}(W_j,\widehat{\beta})-G_n \right]' \phi_{n,i}\right|\\
&~~~+\left|G_n' (\widehat{\phi}_i-\phi_{n,i})\right|+\left|\left[\frac{1}{n}\sum_{j=1}^n \Deriv{p}{\beta}(W_j,\widehat{\beta})-G_n \right]' (\widehat{\phi}_i-\phi_{n,i})\right|\end{align*}
Cauchy-Schwarz inequality and bound on $||G_n||$ ensure:
\begin{align*}
	\sup_{i=1,...,n}\left|\widehat{\psi}_{i}-\psi_{n,i}\right|&\leq2\norm{\Deriv{p}{\beta}}_{\infty}\norm{\widehat{\beta}-\beta_0}+\norm{\frac{1}{n}\sum_{j=1}^n p(W_j,\beta_0)-\tilde{\delta}_n}\\
	&+\norm{\frac{1}{n}\sum_{j=1}^n \Deriv{p}{\beta}(W_j,\widehat{\beta})-G_n } \sup_{i=1,...,n}\norm{\phi_{n,i}}+\norm{\Deriv{p}{\beta}}_{\infty} \sup_{i=1,...,n}\norm{\widehat{\phi}_i-\phi_{n,i}}\\
	&+\norm{\frac{1}{n}\sum_{j=1}^n \Deriv{p}{\beta}(W_j,\widehat{\beta})-G_n} \sup_{i=1,...,n}\norm{\widehat{\phi}_i-\phi_{n,i}} \end{align*}
Note that $\widehat{\beta}-\beta_{0n}=o_{P_n}(1)$, that the LLN of \cite{gut1992weak} implies that $\frac{1}{n}\sum_{j=1}^n p(W_j,\beta_0)-\tilde{\delta}_n=o_{P_n}(1)$. The norm of $=\frac{1}{n}\sum_{j=1}^n \deriv{p}{\beta}(W_j,\widehat{\beta}) -G_n$  could be bounded by the right hand side of \eqref{eq:bound_G_hat}, ensuring it is also an $o_{P_n}(1)$. Moreover $\sup_{i=1,...,n}\norm{\widehat{\phi}_i-\phi_{n,i}}=o_{P_n}(1)$ and $\sup_{i=1,...,n}\norm{\phi_{n,i}}\leq \underline{\rho}^{-1}\norm{\Deriv{\ell_c}{\beta}}_{\infty}$ (cf. arguments just before and after \eqref{eq:tau_conv} in the proof of Lemma \ref{lem:conv_unif_P}). This ensures that $\sup_{i=1,...,n}\left|\widehat{\psi}_{i}-\psi_{n,i}\right|=o_{P_n}(1)$. Combine \begin{align*}\left|\widehat{\sigma}^2 -\frac{1}{n}\sum_{i=1}^n \psi_{n,i}^2\right|&= \left|\frac{1}{n}\sum_{i=1}^n \left(\widehat{\psi}_i-\psi_{n,i}\right)^2+2\left(\widehat{\psi}_i-\psi_{n,i}\right)\psi_{n,i}\right|\\
	&\leq \sup_{i=1,...,n}\left|\widehat{\psi}_{i}-\psi_{n,i}\right|^2+2\sup_{i=1,...,n}\left|\widehat{\psi}_{i}-\psi_{n,i}\right|\sup_{i=1,...,n}\left|\psi_{n,i}\right|.\end{align*} with \eqref{eq:bound_psi} and \eqref{eq:sigma_convU} to conclude $\widehat{\sigma}^2=\sigma_n^2+o_{P_n}(1)$.

	\bigskip
	
	\begin{lem}\label{lem:conv_unif_P}
		Let $\mathcal{P}'$ the set of probability distributions of $(Y,X,\alpha)$ such that Assumptions \ref{hyp:model} and \ref{hyp:iid_bdX} hold with $\mathcal{I}_{0}>>\underline{A}$ for some definite positive matrix $\underline{A}$. Then for $\phi_i = \mathcal{I}_0^{-1}\deriv{\ell_c}{\beta}(Y_i|X_i;\beta_0)$ the influence function of $\widehat{\beta}$, we have:
		\begin{equation}
			\limsup_{n\to\infty} \sup_{P\in\mathcal{P}'} P\left(\norm{n^{1/2}(\widehat{\beta}-\beta_0) - \frac{1}{n^{1/2}}\sum_{i=1}^n \phi_i}>\eta\right)=0.	
			\label{eq:lineariz_unif}
		\end{equation}
		Moreover, for $\tau = \E(\phi_i \phi_i')$ and $\hat{\tau}$ its plug-in estimator, $\hat{\tau} \convP \tau $ holds uniformly over $\mathcal{P}'$. 
	\end{lem}
	
	\textbf{Proof:} To show these results, it suffices to show that they hold along any sequence of probability distribution $(P_n)_{n\geq 1}$ in $\mathcal{P}'$. We use the same notation as in the other proofs but index parameters, variables and the expectation operator by $n$ to underline their dependence on $P_n$ when deemed necessary. Relatedly, we use $o_{P_n}(1)$ as a shortcut for a sequence of random variable $\eps_n$ satisfying $P_n(\norm{\eps_n}>\eta)\to 0$ for all $\eta>0$.
	
	\medskip
	To prove the first point, let us first prove that $\widehat{\beta}-\beta_{0n}=o_{P_n}(1)$. To that end, consider the class of functions $\mathcal{L} := \{(y,x)\mapsto \ell_c(y|x;\beta); \, \beta \in B\}$. We apply a version of Glivenko-Cantelli theorem on $\mathcal{L}$ that is uniform over $P$. The functions $(y,x,\beta)\mapsto \ell_c(y|x;\beta)$ are $C^1$ on $\{0,1\}^T\times \Supp(X)\times B$, which is a compact set. The class $\mathcal{L}$ thus satisfies the Lipschitz requirement of Theorem 2.7.11 of \cite{VdV_Wellner}. Then, by that theorem and the fact that $B$ is compact,
	$$N(\epsilon \|F\|_{Q,1},\mathcal{L},L_1(Q)) \leq N_{[\;]}(\epsilon \|F\|_{Q,1},\mathcal{L},L_1(Q)) \leq N(\epsilon/2,B,\|.\|) <\infty,$$
	where $N_{[\;]}$ denotes bracketing numbers, $N$ denotes covering numbers and $F$ is the  envelope function defined in the same theorem. Hence,
	$$\sup_{Q} \log N(\epsilon \|F\|_{Q,1},\mathcal{L},L_1(Q))  <\infty.$$
	In view of the comment after its proof, we can then apply Theorem 2.8.1 of \cite{VdV_Wellner}. As a result,
	\begin{equation}
		\sup_{\beta\in B} \left|\frac{1}{n}\sum_{i=1}^n \ell_c(Y_i|X_i;\beta) - E_n[\ell_c(Y|X;\beta)]\right|=o_{P_n}(1).
		\label{eq:GCunif}
	\end{equation}
	We establish below a uniform version of the well-separation condition by proving that for all $\eta>0$, there exists $\nu>0$ such that for all $n\geq 1$,
	\begin{equation}
		\sup_{\beta:\norm{\beta-\beta_{0n}}>\eta} M_n(\beta) < M_n(\beta_{0n}) - \nu,	
		\label{eq:well_sep_unif}
	\end{equation}
	where $M_n(\beta)=E_n[\ell_c(Y|X;\beta)]$.
	
	\medskip
	Now, we prove that for any $\eta>0$, there exists $\nu>0$ such that \eqref{eq:well_sep_unif} holds. For any $\beta$ such that $\norm{\beta-\beta_{0n}}>\eta$, let
	$$\beta' =\frac{\eta}{\norm{\beta -\beta_{0n}}}\beta + \left(1 - \frac{\eta}{\norm{\beta -\beta_{0n}}}\right) \beta_{0n}.$$
	Then $\norm{\beta'-\beta_{0n}}=\eta$. Moreover, by concavity of $M_n$,
	$$M_n(\beta')  \geq  \frac{\eta}{\norm{\beta -\beta_{0n}}} M_n(\beta) + \left(1 - \frac{\eta}{\norm{\beta -\beta_{0n}}}\right)M_n(\beta_{0n}) \geq M_n(\beta) .$$
	Thus,
	$$\sup_{\beta:\norm{\beta-\beta_{0n}}>\eta} M_n(\beta) \leq \sup_{\beta\in S_{n,\eta}} M_n(\beta),$$
	where $S_{n,\eta}=\{\beta:\norm{\beta-\beta_{0n}}=\eta\}$.
	Next, for any $\beta\in S_{n,\eta}$ by a Taylor expansion of $M_n$ at $\beta_{0n}$,
	$$M_n(\beta) = M_n(\beta_{0n}) - \frac{1}{2}(\beta-\beta_{0n})' \mathcal{I}_{n,0} (\beta-\beta_{0n}) + \frac{1}{6}\frac{\partial^3 M_n}{\partial^{\otimes 3}\beta} (\tilde{\beta}) [\beta-\beta_{0n}],$$
	where $\tilde{\beta}=t\beta+(1-t)\beta_{0n}$ for some $t\in (0,1)$ and  $\frac{\partial^3 M_n}{\partial^{\otimes 3}\beta} (\tilde{\beta}) [\beta-\beta_{0n}]=\sum_{1\leq j_1,j_2,j_3\leq p}\frac{\partial^3M_n}{\partial \beta_{j_1}\partial \beta_{j_2}\partial \beta_{j_3}}(\tilde{\beta})\times \prod_{s=1}^3\left(\beta_{j_s}-\beta_{0nj_s}\right)$ is the third order differential of $M_n$ at $\tilde{\beta}$ evaluated at $\beta-\beta_{0n}$.
	We know that $\mathcal{I}_{0n} >>\underline{A}$, write $\underline{\rho}$ the smallest eigenvalue of $\underline{A}$.
	By Assumption \ref{hyp:iid_bdX}, there exists $D > 0$ such that $\left|\frac{1}{6} \Deriv{{}^3 M_n}{{}^{\otimes 3}\beta} (\tilde{\beta}) [\beta-\beta_{0n}]  \right| \leq D \eta^3$, which gives
	$$M_n(\beta) \leq M_n(\beta_{0n}) + \eta^2 \left(D \eta - \frac{1}{2}\underline{\rho}\right) \leq M_n(\beta_{0n}) - \eps \eta^2$$
	if $\eta \leq (\frac{1}{2}\underline{\rho} - \eps)/D$ for some $\eps > 0$. Taking $\eta$ small enough is without loss of generality, thus \eqref{eq:well_sep_unif} follows. By suitably modifying the proof of  Theorem 5.7 in \cite{vandervaart_2000} to the sequence $(P_n)$, we have $\left|\left|\widehat{\beta}-\beta_{0n}\right|\right|=o_{P_n}(1)$.
	
	\medskip
	
	Next, we prove \eqref{eq:lineariz_unif}. By a Taylor expansion, there exists $t_n \in (0,1)$ such that
	$$\frac{1}{n}\sum_{i=1}^n \Deriv{\ell_c}{\beta}(Y_i|X_i;\beta_{0n}) + \left[\frac{1}{n}\sum_{i=1}^n \Deriv{{}^2 \ell_c}{\beta\partial \beta'}(Y_i|X_i;\tilde{\beta}_n)\right] \left(\widehat{\beta}-\beta_{0n}\right)=0,$$
	where $\tilde{\beta}_n= t_n \widehat{\beta} + (1-t_n)\beta_{0n}$. Thus, by definition of $\phi_{n,i}$,
	\begin{equation}
		-\mathcal{I}_{0n}^{-1}\left[\frac{1}{n}\sum_{i=1}^n \Deriv{{}^2 \ell_c}{\beta\partial \beta'}(Y_i|X_i;\tilde{\beta}_n)\right] \sqrt{n}\left(\widehat{\beta}-\beta_{0n}\right)=\frac{1}{\sqrt{n}}\sum_{i=1}^n  \phi_{n,i}.	
		\label{eq:CPO_modif_beta}
	\end{equation}
	Now, by the triangle inequality and the fact that the third derivatives of $\ell_c$ are uniformly bounded, there exists $C>0$ such that
	\begin{align*}
		\norm{\frac{1}{n}\sum_{i=1}^n \Deriv{{}^2 \ell_c}{\beta\partial \beta'}(Y_i|X_i;\tilde{\beta}_n)+\mathcal{I}_{0n}} & \leq \norm{\frac{1}{n}\sum_{i=1}^n \Deriv{{}^2 \ell_c}{\beta\partial \beta'}(Y_i|X_i;\tilde{\beta}_n)- \Deriv{{}^2 \ell_c}{\beta\partial \beta'}(Y_i|X_i;\beta_{0n})} \notag \\
		& \; + \norm{\frac{1}{n}\sum_{i=1}^n \Deriv{{}^2 \ell_c}{\beta\partial \beta'}(Y_i|X_i;\beta_{0n}) + \mathcal{I}_{0n}} \notag \\
		& \leq C \norm{\widehat{\beta} - \beta_{0n}} + \norm{\frac{1}{n}\sum_{i=1}^n \Deriv{{}^2 \ell_c}{\beta\partial \beta'}(Y_i|X_i;\beta_{0n}) + \mathcal{I}_{0n}}.
	\end{align*}
	By what precedes, the first term is an $o_{P_n(1)}$. Moreover, for all $i$ and $n$, each element of the matrix $\deriv{{}^2 \ell_c}{\beta\partial \beta'}(Y_i|X_i;\beta_{0n})$ is bounded almost surely.  Thus, the uniform integrability condition of \cite{gut1992weak} holds for this variable. Then, by the weak LLN of \cite{gut1992weak}, the second term of the right-hand side above is also an $o_{P_n(1)}$. Thus, because $\mathcal{I}_{0n}^{-1} << \underline{A}^{-1}$ (since $P_n\in\mathcal{P}'$), we have
	$$\mathcal{I}_{0n}^{-1}\left[\frac{1}{n}\sum_{i=1}^n \Deriv{{}^2 \ell_c}{\beta\partial \beta'}(Y_i|X_i;\tilde{\beta}_n)\right]=-\text{Id} +o_{P_n}(1).$$
	Next, for all $i$ and $n$, we have
	\begin{equation}
		E_n[\phi_{n,i}]=0, \quad V_n(\phi_{n,i})=\mathcal{I}_{0n}^{-1} << \underline{A}^{-1}.	
		\label{eq:bound_var_phi}
	\end{equation}
	Hence, by Chebyshev's inequality, the right-hand side of \eqref{eq:CPO_modif_beta} is bounded in probability uniformly over $n$. Thus, this is also the case of $\sqrt{n}\left(\widehat{\beta}-\beta_{0n}\right)$. Hence,
	$$\sqrt{n}\left(\widehat{\beta}-\beta_{0n}\right)=\frac{1}{\sqrt{n}}\sum_{i=1}^n  \phi_{n,i} + o_{P_n}(1).$$
	In other words, \eqref{eq:lineariz_unif} holds.
	
	\medskip
	We now show that $\hat{\tau} $ converges uniformly  over $\mathcal{P}'$ to $\tau $. First, we have
	$$\norm{\phi_{n,i}}=\norm{\mathcal{I}_{n0}^{-1}\Deriv{\ell_c}{\beta}(W_i,\beta_{0n})}\leq \norm{\underline{A}^{-1} \Deriv{\ell_c}{\beta}(W_i,\beta_{0n})}\leq \underline{\rho}^{-1} \norm{\Deriv{\ell_c}{\beta}}_{\infty},$$
	and next the LLN of \cite{gut1992weak} ensures that \begin{equation}\frac{1}{n}\sum_{i=1}^{n}\phi_{n,i}\phi_{n,i}'=\tau_n+o_{P_n}(1).\label{eq:tau_conv}\end{equation}
	Moreover,
	\begin{align*}
		\widehat{\phi}_i -\phi_{n,i}& =-\left[ \frac{1}{n}\sum_{j=1}^n  \frac{\partial^2 \ell_c}{\partial \beta\partial \beta'}(Y_j|X_j;\widehat{\beta}) \right]^{-1}    \frac{\partial \ell_c}{\partial \beta}(Y_i|X_i;\widehat{\beta})-\mathcal{I}_{0n}^{-1}\frac{\partial \ell_c}{\partial \beta}(Y_i|X_i;\beta_{0n})\\
		& = \left\lbrace   -\left[\frac{1}{n}\sum_{j=1}^n  \frac{\partial^2 \ell_c}{\partial \beta \partial \beta'}(Y_j|X_j;\widehat{\beta}) \right]^{-1}-\mathcal{I}_{0n}^{-1}  \right\rbrace  \frac{\partial \ell_c}{\partial \beta}(Y_i|X_i;\widehat{\beta})\\
		& + \mathcal{I}_{0n}^{-1}  \left[\frac{\partial \ell_c}{\partial \beta}(Y_i|X_i;\widehat{\beta})-\frac{\partial \ell_c}{\partial \beta}(Y_i|X_i;\beta_{0n})\right].
	\end{align*}
	By the same argument as below Equation (\ref{eq:CPO_modif_beta}), using sequences of probability distributions and replacing $\tilde{\beta}_n$ with $\widehat{\beta}$, one can show that $\frac{1}{n}\sum_{j=1}^n  \deriv{{}^2 \ell_c}{\beta \partial \beta'}(Y_j|X_j;\widehat{\beta})=-\mathcal{I}_{0n}+o_{P_n}(1)$. And since  $\mathcal{I}^{-1}_{0n} << \underline{A}^{-1}$,
	$\left[\frac{1}{n}\sum_{j=1}^n  \deriv{{}^2 \ell_c}{\beta^2}(Y_j|X_j;\widehat{\beta})\right]^{-1}=-\mathcal{I}_{0n}^{-1}+o_{P_n}(1)$. It follows that
	$\sup_{i=1,...,n}||\widehat{\phi}_i-\phi_{n,i}||\leq o_{P_n}(1)\norm{\deriv{\ell_c}{\beta}}_{\infty}+\underline{\rho}^{-1}\norm{\widehat{\beta}-\beta_{0n}}\norm{\deriv{{}^2\ell_c}{\beta\partial \beta'}}_{\infty}$ where $\underline{\rho}>0$ is the smallest eigenvalue of $\underline{A}$. Because $\norm{\widehat{\beta}-\beta_{0n}}=o_{P_n}(1)$, we have $\sup_{i=1,...,n}||\widehat{\phi}_i-\phi_{n,i}||=o_{P_n}(1)$. Triangle inequality ensures
	\begin{align*}
		\norm{\widehat{\tau}-\frac{1}{n}\sum_{i=1}^n\phi_{n,i}\phi_{n,i}'}&\leq \norm{\frac{1}{n}\sum_{i=1}^n(\widehat{\phi}_{i}-\phi_{n,i})'\phi_{n,i}}+\norm{\frac{1}{n}\sum_{i=1}^n\phi_{n,i}'(\widehat{\phi}_{i}-\phi_{n,i})}+\norm{\frac{1}{n}\sum_{i=1}^n(\widehat{\phi}_{i}-\phi_{n,i})'(\widehat{\phi}_{i}-\phi_{n,i})}\\
		&\leq 2 \underline{\rho}^{-1}\norm{\deriv{\ell_c}{\beta}}_{\infty}\sup_{i=1,...,n}\norm{\widehat{\phi}_i-\phi_{n,i}}+\sup_{i=1,...,n}\norm{\widehat{\phi}_i-\phi_{n,i}}^2=o_{P_n}(1)
		\end{align*}
$\widehat{\tau}=\tau_n+o_{P_n}(1)$ follows from \eqref{eq:tau_conv}.

	\bigskip
	
	\begin{lem}\label{lem:conv_unif_R}
		\begin{enumerate}
			\item Suppose that Assumption \ref{hyp:model}-\ref{hyp:lambda2} hold. Then $\widehat{\overline{R}}$  converges to $\overline{R}$, and this convergence holds uniformly over $\mathcal{P}$.
			\item Suppose that Assumption \ref{hyp:model}-\ref{hyp:iid_bdX} hold. Then $\widehat{\overline{b}}$ converges to $\overline{b}$.
		\end{enumerate}		
	\end{lem}
	\textbf{Proof:} We start by showing Part 1. Define
	$$f(W_i,\beta) := \binom{T}{S_i}\frac{\rho(X_i,\beta) \exp(S_i v\left( X_{i},\beta\right)}{2\times 4^T \times C_{S_i}(X_i,\beta)}.$$
	Then, $\widehat{\overline{R}} = \sum_{i=1}^n |f(W_i,\widehat{\beta})|/n$.
	The function $w \mapsto f(w,.)$ is $C^1$ and $(w,\beta) \mapsto \deriv{f(w,\beta)}{\beta}$ is continuous over the compact set $\Supp(W)\times B'$. Hence, there exists $M>0$ such that $\beta \mapsto f(w,\beta)$ is Lipschitz with coefficient $M$ for all $\beta \in B'$ and $w\in \Supp(W)$.  The same property then holds for $\beta \mapsto|f(w,\beta)|$. Since $\widehat{\beta} \in B'$ with probability uniformly going to $1$ and $| f(W_i,\beta)|$ is bounded almost surely when $\beta \in B'$,  arguments and a decomposition similar to those below Equation (\ref{eq:CPO_modif_beta}) show that $\widehat{\overline{R}}$  converges to $\overline{R}$ and this convergence holds uniformly over $\mathcal{P}$.
	
	As for Part 2., note that $\widehat{\overline{b}} = \sum_{i=1}^n |\tilde{f}(W_i,\widehat{\beta})|/n$ where the difference between $f$ and $\tilde{f}$ is that $\rho$ is replaced with $\lambda_{T+1}$. Under Assumption \ref{hyp:lambda}, there exists $\tilde{M}>0$ such that $\beta \mapsto \tilde{f}(w,\beta)$ is Lipschitz with coefficient $\tilde{M}$ for all $\beta \in B'$ and $w\in \Supp(W)$ and the result thus holds.
	
	\begin{lem}\label{lem:q_alpha}
		Let $q_\alpha(b)$ denote the quantile of order $1-\alpha$ of a $|\mathcal{N}(b,1)|$. Then:
		\begin{enumerate}
			\item \label{lem:q_alpha_increasing} $x\mapsto q_{\alpha}(x)$ is increasing on $\R^+$.
			\item \label{lem:q_alpha_ineq} For all $x \in \R_+ $, $x+z_{1-\alpha}\le q_\alpha(x)\le x+z_{1-\alpha/2}$.	
		\end{enumerate}
	\end{lem}

	\textbf{Proof:} To prove the first point, take $h \geq 0$ and $x\ge 0$. Then:
	\begin{align*}
		& P\left(|Z+x+h| \leq q_\alpha(x)\right) - (1 - \alpha)  \\
		= & P\left(|Z+x+h| \leq q_\alpha(x)\right) - P\left(|Z+x| \le  q_\alpha(x)\right) \\
		= & \Phi(q_\alpha(x)-x - h) - \Phi(-x - h -q_\alpha(x)) - \Phi(q_\alpha(x)-x) + \Phi(-x-q_\alpha(x))\\
		= & \Phi(-x-q_\alpha(x))  - \Phi(-x - h -q_\alpha(x)) - [\Phi(q_\alpha(x)-x) - \Phi(q_\alpha(x)-x - h)]\\
		= &  \Phi( q_\alpha(x)+ x + h) - \Phi(q_\alpha(x)+x) - [\Phi(q_\alpha(x)-x) - \Phi(q_\alpha(x)-x - h)]\\
	   \le & \left[\phi(q_\alpha(x)+x)  - \phi(q_\alpha(x)-x)\right] h \\
	   \le & 0,
	\end{align*}
	where $\phi$ denotes the density of the standard normal distribution, which is decreasing on $\R_+$. The inequality above implies that $q_\alpha(x +h ) \geq q_\alpha(x)$.
	
	\medskip
	To prove the second point, note that  the first inequality comes from
	\begin{equation}
		\Phi(q_\alpha(x)-x) - \Phi(-x-q_\alpha(x))=1-\alpha	
		\label{eq:q_alpha_x}
	\end{equation}
	and $\Phi(-x-q_\alpha(x))\ge 0$. The second inequality comes from $-x-q_\alpha(x) \le x-q_\alpha(x)$ and thus, from \eqref{eq:q_alpha_x} again, $2\Phi(q_\alpha(x)-x) - 1 \le 1-\alpha$.
	
\medskip
	\begin{lem}\label{lem:random_func}
		Let $(X_i)_{i\geq 1}$ a sequence of i.i.d. random variables (resp. random vectors) with common distribution $P$. If $\mathcal{F}$ is a class of functions that is $P$-Donsker and $\hat{f}_n$ is a random function such that $\int \hat{f}^2_n(x)dP(x)$ converges to 0 in probability then for any $\varepsilon>0$, we have: \begin{align*}\limsup_n P\left(\norm{\sqrt{n}\left(\frac{1}{n}\sum_{i=1}^n\hat{f}_n(X_i)-\int \hat{f}_n(x)dP(x)\right)}>\varepsilon\right)\leq \limsup_n P(\hat{f}_n \notin \mathcal{F})\end{align*} 
		\end{lem}
	\textbf{Proof:}
	Because $P(\hat{f}_n \notin \mathcal{F}\cup\{0\})\leq P(\hat{f}_n \notin \mathcal{F})$, we can assume without loss of generality that the null function belongs to $\mathcal{F}$. Let $\hat{g}_n=\hat{f}_n\mathds{1}\{\hat{f}_n\in \mathcal{F}\}$, we have:  
	\begin{align*} &P\left(\norm{\sqrt{n}\left(\frac{1}{n}\sum_{i=1}^n\hat{f}_n(X_i)-\int \hat{f}_n(x)dP(x)\right)}>\varepsilon\right)\\
	&\leq P\left(\norm{\sqrt{n}\left(\frac{1}{n}\sum_{i=1}^n\hat{f}_n(X_i)-\int \hat{f}_n(x)dP(x)\right)}>\varepsilon, \hat{f}_n \in \mathcal{F}\right)+ P(\hat{f}_n \notin \mathcal{F})\\
&=P\left(\norm{\sqrt{n}\left(\frac{1}{n}\sum_{i=1}^n\hat{g}_n(X_i)-\int \hat{g}_n(x)dP(x)\right)}>\varepsilon\right)+ P(\hat{f}_n \notin \mathcal{F})\end{align*}
Moreover $\hat{g}_n\in \mathcal{F}$ and $\int \hat{g}^2_n(x)dP(x)$ converges to 0 in probability. Lemma 19.24 in \cite{vandervaart_2000} ensures that $\sqrt{n}\left(\frac{1}{n}\sum_{i=1}^n\hat{g}_n(X_i)-\int \hat{g}_n(x)dP(x)\right)$ converges in probability to 0.
		

\end{document}